\documentclass[article,12pt]{article} 

\usepackage{amscd,amsmath,amssymb,amsfonts,latexsym,mathrsfs,amsthm,mathtools}
\usepackage{graphicx} 
\usepackage{bigints}
\usepackage{setspace} 
\usepackage{graphics,epsfig}
\graphicspath{{figures/}} 
\usepackage{sectsty}
\usepackage[numbers]{natbib}
\usepackage[english]{babel}
\usepackage{bm}
\usepackage{multirow}
\usepackage{subfigure}
\usepackage[usenames]{color}
\usepackage{rotating}
\usepackage{url}
\usepackage{bbm}
\usepackage{float}
\usepackage{tikz}
\usepackage{pgf}
\usepackage{pdflscape}
\usepackage{xcolor}
\usepackage{colortbl}
\usepackage{bigints}

\def\checkmark{\tikz\fill[scale=0.4](0,.35) -- (.25,0) -- (1,.7) -- (.25,.15) -- cycle;}

\newtheorem{rem}{{\bf Remark}}

\usepackage{flushend}
\usepackage{verbatim}
\usepackage{tabularx}
\usepackage{arydshln}
\usepackage{hyperref}
\usepackage[toc,page]{appendix}
\usepackage{pdflscape}

\newcommand{\E}{\mathbb{E}}

\newcommand{\x}{\bm{\theta}}
\newcommand{\z}{{\bf z}}

\newcommand{\y}{{\bf y}}

\newcommand{\post}{P}
\newcommand{\norm}[1]{\left\lVert#1\right\rVert}

\definecolor{MYCOLOR0}{rgb}{0.92,0.92,0.92}
\definecolor{MYCOLOR}{rgb}{1,1,0}
\definecolor{MYCOLOR2}{rgb}{0.5,1,0.5}
\definecolor{MYCOLOR3}{rgb}{0.88,1,1}

\DeclarePairedDelimiter\floor{\lfloor}{\rfloor}

\usepackage{imakeidx}
\makeindex

\UseRawInputEncoding

\title{Marginal likelihood computation for model selection and hypothesis testing: an extensive review} 
 
\author{F. Llorente$^\star$, L. Martino$^{\star\star}$, D. Delgado$^\star$, J. Lopez-Santiago$^\star$  \\
{\small$^\star$  Universidad Carlos III de Madrid,  Legan\'es (Spain).}\\
{\small$^\star$$^\star$  Universidad Rey Juan Carlos,  Fuenlabrada (Spain).} \\
}

\date{}
 
\begin{document}

\maketitle

\begin{abstract}

This is an up-to-date introduction to, and overview of, marginal likelihood computation for model selection and hypothesis testing. Computing normalizing constants of probability
models (or ratio of constants) is a fundamental issue in many applications in statistics, applied mathematics, signal processing and machine learning. This article provides a comprehensive study of the state-of-the-art of the topic. We highlight limitations, benefits, connections and differences among the different techniques. Problems and possible solutions with the use of improper priors are also described. Some of the most relevant methodologies are compared through theoretical comparisons and numerical experiments.
\end{abstract}

\section{Introduction}


Marginal likelihood (a.k.a., Bayesian evidence) and Bayes factors are the core of the Bayesian theory for testing hypotheses and model selection \cite{Liu04b,Robert04}.  More generally, the computation of normalizing constants or ratios of normalizing constants has played an important role in statistical physics and numerical analysis \cite{stoltz2010free}.  In the Bayesian setting, the approximation of normalizing constants is also required in the study of the so-called double intractable posteriors \cite{Liang10}.  

Several methods have been proposed for approximating the marginal likelihood and normalizing constants in the last decades. Most of these techniques have been originally introduced in the field of statistical mechanics. Indeed, the marginal likelihood is the analogous of a central quantity in statistical physics known as the {\it partition function} which  is also closely related to another important quantity often called {\it free-energy}. The relationship  between statistical physics and Bayesian inference has been remarked in different works \cite{balasubramanian1997statistical, lamont2019correspondence}. 

The model selection problem has been also addressed from different points of view. Several criteria have been proposed to deal with the trade-off between the goodness-of-fit of the model and its simplicity. For instance,  the Akaike information criterion (AIC)  or the focused information criterion (FIC) are two examples of these approaches \cite{sakamoto1986akaike,claeskens2003focused}. The Bayesian-Schwarz information criterion (BIC) is related to the marginal likelihood approximation, as discussed in Section \ref{ExploitingFunctionalIdentity}.  The deviance information criterion (DIC) is a generalization of the AIC, which is often used in Bayesian inference  \cite{Spiegelhalter02,Spiegelhalter14}. It is particularly useful for hierarchical models and  it can be approximately computed when the outputs of a Markov Chain Monte Carlo (MCMC) algorithm are given. However, DIC is not directly related to the Bayesian evidence \cite{Pooley18}. Another different approach, also based on information theory, is the so-called minimum description length principle (MDL) \cite{Roos19}. MDL was originally
derived for data compression, and then was applied to model selection and hypothesis testing. Roughly speaking, MDL considers that the best explanation for a given set of data is provided by the {\it shortest description} of that data \cite{Roos19}.
\newline
In the Bayesian framework,  there are two main classes of sampling algorithms. The first one consists in approximating the marginal likelihood of different models or the ratio of two marginal likelihoods.  
In this work, we focus on this first approach.  
 The second  sampling approach extends the posterior space including a discrete indicator variable $m$, denoting the $m$-th model \cite{carlin1995bayesian,green1995reversible}. 
 For instance, in the well-known {\bf reversible jump MCMC}  \cite{green1995reversible}, a  Markov chain is generated in this extended space, allowing  jumps between models with possibly different dimensions. However, generally, these methods are difficult to tune and the mixing of the chain can be poor \cite{hastie2012model}. For further details, see also the interesting works  \cite{dellaportas2002bayesian,godsill2001relationship,congdon2006bayesian}. The average number of MCMC iterations when the chain jumps or stays into the $m$-th model is proportional to the marginal likelihood of the corresponding model. 
\newline
In this work, we provide an extensive review of computational techniques for the marginal likelihood computation. 
 The main contribution is to present jointly  numerous  computational schemes  (introduced independently in the literature) with a detailed description under the same  notation, highlighting their differences, relationships, limitations and strengths.   Most of them are based on the importance sampling (IS) approach and several of them are combination the MCMC and IS schemes.
 It is also important to remark that parts of the presented material are also novel, i.e., no contained in previous works.
We have widely studied, analyzed and jointly described with a unique notation and classification, the methodologies presented in a vast literature from 1990s to the recent proposed algorithms (see Table  \ref{SurveysTable}).    We also discuss issues and solutions when improper priors are employed.  Therefore, this survey provides an ample covering of the literature, where we highlight important details and comparisons in order to facilitate the understanding of the interested readers and practitioners. 
\newline
The problem statement and the main notation are introduced in the  Section \ref{ProSect0}. Relevant considerations regarding the marginal likelihood and other model selection strategies are given in Section \ref{ModelFitSect} and Section \ref{SectNuovaBella}. Specifically, a description of how the marginal likelihood handles the model fit and the model complexity is provided in Section \ref{ModelFitSect}.  The dependence on the prior selection and the possible choice of an improper prior are discussed in  Section \ref{SectNuovaBella}.
 The different techniques have been classified in four main families, as shown in Section \ref{GenOver}.  
  Sections \ref{ExploitingFunctionalIdentity}, \ref{ImportanceSamplingApproaches}, \ref{CombinationISandMCMC}, \ref{verticalLikelihoodApproach} are devoted to the detailed description of the computational schemes for approximating the Bayesian evidence. 
 Section \ref{NumSimu} contains some numerical experiments. In Section \ref{concluSect}, we conclude with a final summary and discussion. We provide also theoretical analyses of some of the experiments and other comparisons in the Supplementary Material. 

%
%

\setcounter{tocdepth}{2}
\tableofcontents






\section{Problem statement and preliminary discussions}\label{ProSect}

\subsection{Framework and notation}\label{ProSect0}
In many applications,  the goal is to make inference about a variable of interest,  
$\x=\theta_{1:D_\theta}=[\theta_1,\theta_2,\ldots,\theta_{D_\theta}]\in \Theta\subseteq\mathbb{R}^{D_\theta}$,
where $\theta_d\in \mathbb{R}$ for all $d=1,\ldots,D_\theta$, given a set of observed measurements, ${\bf y}=[y_1,\ldots,y_{D_y}]\in \mathbb{R}^{D_y}$. In the Bayesian framework, one complete model $\mathcal{M}$ is formed by a likelihood function $\ell({\bf y}|\x,\mathcal{M})$ and a prior probability density function (pdf) $g(\x|\mathcal{M})$. All the statistical information is summarized by the posterior pdf, i.e.,
\begin{equation}
\post(\x|\y,\mathcal{M})= \frac{\ell(\y|\x,\mathcal{M}) g(\x|\mathcal{M})}{p(\y|\mathcal{M})},
\label{eq:posterior}
\end{equation}
where 
\begin{equation}\label{MarginalLikelihood}
Z= p({\bf y}|\mathcal{M})=\int_{\Theta} \ell({\bf y}|\x,\mathcal{M}) g(\x|\mathcal{M}) d\x,
\end{equation}
is the so-called marginal likelihood, a.k.a., Bayesian evidence. This quantity is important for model selection purpose, as we show below.
However, usually $Z= p({\bf y}|\mathcal{M})$ is unknown and difficult to approximate, so that in many cases we are only able to evaluate the unnormalized target function,
\begin{equation}
\pi(\x|\y,\mathcal{M})=\ell({\bf y}|\x,\mathcal{M}) g(\x|\mathcal{M}).
\label{eq:target}
\end{equation}
Note that $\post(\x|\y,\mathcal{M})\propto \pi(\x|\y,\mathcal{M})$ \citep{Liu04b,Robert04}. For the sake of simplicity,  hereafter we use the simplified notation $\post(\x|\y)$ and $\pi(\x|\y)$. Thus, note that 
\begin{equation}\label{MarginalLikelihood2}
Z=\int_\Theta \pi(\x|\y) d\x.
\end{equation}
\newline
{\bf Model Selection and testing hypotheses.} Let us consider now $M$ possible models (or hypotheses), $\mathcal{M}_1,...,\mathcal{M}_M$,  with prior probability mass $p_m =\mathbb{P}\left(\mathcal{M}_m\right)$, $m=1,...,M$. Note that, we can have variables of interest $\x^{(m)}=[\theta_{1}^{(m)},\theta_{2}^{(m)},\ldots,\theta_{D_m}^{(m)}]\in \Theta_m\in \mathbb{R}^{D_m}$, with possibly different dimensions in the different models. The posterior of the $m$-th  model is given by
\begin{eqnarray}
p(\mathcal{M}_m|{\bf y})&=&\frac{p_m p({\bf y}|\mathcal{M}_m)}{p({\bf y})}\propto p_mZ_m
\end{eqnarray}
where $Z_m=p({\bf y}|\mathcal{M}_m)=\int_{\Theta} \ell({\bf y}|\x_m,\mathcal{M}_m) g(\x_m|\mathcal{M}_m) d\x_m$,
and $p({\bf y})=\sum_{m=1}^M p(\mathcal{M}_m)p({\bf y}|\mathcal{M}_m)$. 
Moreover, the ratio of two marginal likelihoods 
\begin{align}
	\frac{Z_m}{Z_{m'}} =\frac{ p({\bf y}|\mathcal{M}_m)}{p({\bf y}|\mathcal{M}_{m'})} =\frac{p(\mathcal{M}_m|\y)/p_m}{p(\mathcal{M}_{m'}|\y)/p_{m'}},
\end{align}
also known as {\it Bayes factors}, represents the posterior to prior odds of models $m$ and $m'$.  If some quantity of interest is common to all models, the posterior of this quantity can be studied via {\it model averaging} \cite{BMA99}, i.e., a complete posterior distribution as a mixture of $M$ partial posteriors linearly combined with weights proportionally to $p(\mathcal{M}_m|{\bf y})$  (see, e..g, \cite{Martino15PF,PetarCopy}). 
Therefore, in all these scenarios, we need the computation of $Z_m$ for all $m=1,...,M$. In this work, we describe different computational techniques for calculating $Z_m$, mostly based on Markov Chain Monte Carlo (MCMC) and Importance Sampling (IS) algorithms \cite{Robert04}. Hereafter, we assume proper prior $g(\x|\mathcal{M}_m)$. Regarding the use of {\it improper priors} see Section \ref{AppBayeFactImproperPrior}. Moreover, we usually denote $Z$, $\Theta$, $\mathcal{M}$, omitting the subindex $m$, to simplify notation.  It is important also to remark that, in some cases, it is also necessary to approximate normalizing constants (that are also functions of the parameters) in each iteration of an MCMC algorithm, in order to allow the study of the posterior density. For instance, this is the case of the so-called double intractable posteriors \cite{Liang10}. 
\begin{rem}
The evidence $Z$ is the normalizing constant of $\pi(\x|\y)$, hence most of the methods in this review can be used to approximate normalizing constants of generic pdfs. 
\end{rem}
\begin{rem}
Instead of approximating the single values $Z_m$ for all $m$, another approach consists in estimating directly the ratio of two marginal likelihoods $\frac{Z_m}{Z_{m'}}$, i.e., approximating directly the Bayes factors. For these reasons, several computational methods focus on estimating the ratio of two normalizing constants. However, they can be used also for estimating a single $Z_m$ provided that $Z_{m'}$ is known. 
\end{rem}

\begin{table}[H]
	\centering
	\caption{Main notation of the work.}
	\vspace{0.1cm}
	\begin{center}
	\begin{tabular}{|c|l||c|l|}
		\hline
		\cellcolor{MYCOLOR0} $D_\theta$ & \multicolumn{3}{l|}{dimension of the parameter space, $\x\in\Theta  \subset\mathbb{R}^{D_\theta}$.} \\ 
		\cellcolor{MYCOLOR0} $D_y$ & \multicolumn{3}{l|}{Total number of data.} \\
		\hline
		\hline
		\cellcolor{MYCOLOR0}$\x$ &\multicolumn{3}{l|}{parameters; $\x=[\theta_1,\ldots,\theta_{D_\theta}]$.}\\
		\cellcolor{MYCOLOR0}$\y$ &\multicolumn{3}{l|}{Data, $\y = [y_1,\dots,y_{D_y}]$. }\\
		\hline
		\hline
		\cellcolor{MYCOLOR0} $\ell(\y|\x)$ & \multicolumn{3}{l|}{Likelihood function.} \\ 
		\cellcolor{MYCOLOR0} $g(\x)$ & \multicolumn{3}{l|}{Prior pdf.} \\ 
		\cellcolor{MYCOLOR0} $\post(\x|\y)$ & \multicolumn{3}{l|}{Posterior pdf, {\small $\post(\x|\y)= \frac{\ell(\y|\x)g(\x)}{Z}$}.} \\
		\cellcolor{MYCOLOR0} $\pi(\x|\y)$ & \multicolumn{3}{l|}{Unnormalized posterior, {\small $\pi(\x|\y) = \ell(\y|\x)g(\x)\propto\post(\x|\y)$}.} \\
		\cellcolor{MYCOLOR0} $Z=p(\y)$ & \multicolumn{3}{l|}{Marginal likelihood, a.k.a., Bayesian evidence {\small $Z= \int_\Theta\pi(\x|\y)d\x$}.} \\
		\hline
		\hline
		\cellcolor{MYCOLOR0} $\bar{q}(\x)$ & \multicolumn{3}{l|}{Proposal pdf.} \\
		\cellcolor{MYCOLOR0} ${q}(\x)$ & \multicolumn{3}{l|}{Unnormalized proposal function, $q(\x) \propto \bar{q}(\x)$.} \\
		\hline
	\end{tabular}
	\end{center}
	\label{tab:notation}
\end{table}


\subsection{Model fit and model complexity}\label{ModelFitSect}

\subsubsection{Bounds of the evidence $Z$}

Let us denote the maximum and minimum value of the likelihood function as  $\ell_\text{min}=\ell(\y|\x_\text{min})= \min\limits_{\x\in\theta} \ell(\y|\x)$,  and $\ell_\text{max}=\ell(\y|\x_\text{max})=  \max\limits_{\x\in\theta}\ell(\y|\x)$, respectively. Note that
\begin{align*}
Z = \int_\Theta \ell(\y|\x)g(\x)d\x &\leq  \ell(\y|\x_\text{max})\int_\Theta g(\x)d\x = \ell(\y|\x_\text{max}).
\end{align*}
Similarly, we can obtain $ZÊ\geq  \ell(\y|\x_\text{min})$. The maximum and minimum value of $Z$ are reached with two degenerate choices of the prior, $g(\x) = \delta(\x-\x_\text{max})$ and $g(\x) = \delta(\x-\x_\text{min})$. Hence, for every other choice of $g(\x)$, we have 
\begin{align}
\ell(\y|\x_\text{min}) \leq Z \leq \ell(\y|\x_\text{max}).
\end{align}
Namely, depending on the choice of the prior $g(\x)$, we can have any value of Bayesian evidence contained in the interval $[\ell(\y|\x_\text{min}),\ell(\y|\x_\text{max})]$. 
For further discussion see Section \ref{SectNuovaBella}. 
\newline
  The two possible extreme values correspond to the worst and the best model fit, respectively. Below, we will see that if $Z=\ell(\y|\x_\text{min})$ the chosen prior, $g(\x) = \delta(\x-\x_\text{min})$, applies the greatest possible penalty to the model whereas, if $Z=\ell(\y|\x_\text{max})$, the chosen prior, $g(\x) = \delta(\x-\x_\text{max})$,  does not apply any penalization to the model complexity (we have the maximum overfitting). Namely,  the evidence $Z$ is an average of the likelihood values, weighted according to the prior.

\subsubsection{Occam factor and implicit/intrinsic complexity penalization in $Z$}
The marginal likelihood can be expressed as
\begin{align}\label{SuperIMPEqZ}
Z= \ell_\text{max}W,
\end{align}
where  $W \in[0,1]$ is the {\it Occam factor}  \cite[Sect. 3]{knuth2015bayesian}.  More specifically, the Occam factor is defined as 
\begin{align}
W = \frac{1}{\ell_\text{max}}\int_\Theta g(\x) \ell(\y|\x)d\x,
\end{align}
and it is $\frac{\ell_\text{min}}{\ell_\text{max}}\leq W \leq 1$. The factor $W$ measures  the penalty of the model complexity {\it intrinsically} contained in the marginal likelihood $Z$: this penalization depends on the chosen prior and the number of data involved. 
We show below that the Occam factor measures the ``overlap'' between likelihood and prior, i.e., how diffuse the prior is with respect to the likelihood function. Finally,  it is important to remark that, considering the posterior of the $m$-th model $p(\mathcal{M}_m|{\bf y})$, we have another possible penalization term due to the prior $p_m= P(\mathcal{M}_m)\in[0,1]$, i.e.,

$$
p(\mathcal{M}_m|{\bf y})\propto Zp_m=\ell_\text{max}Wp_m=\ell_\text{max}\widetilde{W},
$$
where we have defined the {\it posterior Occam factor} as $\widetilde{W}=Wp_m$. 
\subsubsection{Occam factor with uniform priors}

 {\bf One-dimensional case.} Let start with a single parameter, $\x = \theta$, and a uniform prior in $[a,b]$. We can define the amount of the likelihood mass is contained inside the prior bounds, 
\begin{align}
\Delta_{\ell} = \frac{1}{\ell_\text{max}}\int_{a}^{b}\ell(\y|\theta)d\theta, \quad \mbox{ where } \quad\ell_\text{max}=  \max\limits_{\theta\in\theta}\ell(\y|\theta).
\end{align}
Defining also  the width of the prior as $\Delta_\theta=|\Theta|=b-a$, 
note that $0\leq \Delta_\ell \leq \Delta_\theta$, where the equality $\Delta_\ell = \Delta_\theta$ is given when the likelihood is $\ell(\y|\theta)=\ell_\text{max}$ is constant.
 The Occam factor is given as the ratio of $\Delta_\ell$ and the width of a uniform prior $\Delta_\theta$ \cite{knuth2015bayesian},
\begin{align}
W = \frac{\Delta_\ell}{\Delta_\theta}.
\end{align}
If the likelihood function is integrable in $\mathbb{R}$, then there exists a finite upper bound for $\Delta_\ell$ when $\Delta_\theta \rightarrow  \infty$, that is $\Delta_\ell^*= \frac{1}{\ell_\text{max}}\int_{-\infty}^{+\infty}\ell(\y|\theta)d\theta$. Hence, in this scenario, we can see that an increase of $\Delta_\theta$ makes that $W$ approaches $0$.
\newline
\newline
 {\bf Multidimensional case.}  Consider now a multidimensional case, $\x=[\theta_1,\theta_2,\ldots,\theta_{D_\theta}]\in \Theta\subseteq\mathbb{R}^{D_\theta}$, where we can use the same uniform prior, with the same width  $\Delta_\theta=|\Theta|$,  for all the parameters. In this case,
 $\Delta_{\ell} = \frac{1}{\ell_\text{max}}\int_{\Theta}\ell(\y|\x)d\x \leq  (\Delta_\theta)^{D_\theta}$ is $D_\theta$-dimensional integral, and $\ell_\text{max}=  \max\ell(\y|\x)$.
  Then, for $D_\theta$ parameters, the Occam factor is 
\begin{align}\label{Wgen}
 W=\frac{\Delta_{\ell}}{(\Delta_\theta)^{D_\theta}}.
\end{align}
Usually, as $D_\theta$ grows, the fitting improves until reaching (or approaching) a maximum, possible overfitting. Then, with $D_\theta$ big enough,  $\ell_\text{max}$  tends to be virtually constant (reaching the maximum overfitting). If $\int_{\Theta}\ell(\y|\x)d\x$ grows slower than $(\Delta_\theta)^{D_\theta}$ as $D_\theta\to\infty$,
and assuming  for an illustrative purpose $\Delta_{\theta}> 1$,  then $W$ converges to $0$ as $D_\theta\to\infty$.
 That is, when we introduce more and more parameters, the increase in model fit will be dominated, at some point, by the model complexity penalization implicitly contained in the evidence $Z$. 
\subsubsection{Marginal likelihood and information criteria}\label{ObradeArteCriteria}
 Considering the expressions \eqref{SuperIMPEqZ} and \eqref{Wgen} and taking the logarithm, we obtain
\begin{align}
\log Z= \log\ell_{\max}+ \log W&=\log \ell_{\max}+\log \Delta_\ell-D_\theta \log \Delta_\theta,Ê  \nonumber \\
&=\log \ell_{\max}+\eta D_\theta,  \label{aquiEqDtheta}
\end{align}
where $\eta=\frac{\log \Delta_\ell}{D_\theta}-\log \Delta_\theta$ is a constant value, which also depends on the number of data $D_y$ and, generally, $\eta=\eta(D_y,D_\theta)$. Different model selection rules in the literature consider the simplification $\eta=\eta(D_y)$. Note that $\log \ell_{\max}$ is a fitting term whereas $\eta D_\theta$ is a penalty for the model complexity.  Instead of maximizing $Z$ (or $\log Z$) for model selection purposes, several authors consider the minimization of some cost functions derived by different information criteria.  To connect them with the marginal likelihood maximization, we consider the expression of $-2 \log Z=-2 I$ where $I=-\log Z$ resembles the Shannon information associated to $Z=p({\bf y})$ , i.e.,
\begin{eqnarray}
2 I=-2\log Z=-2\log \ell_{\max}- 2 \eta D_\theta.
\end{eqnarray}
The expression above encompasses several well-known information criteria proposed in the literature and shown in Table \ref{TablaIC}, which differ for the choice of $\eta$. In all these cases, $\eta$ is just a function of the number of data $D_y$.  
More details regarding these information criteria are given in Section \ref{ExploitingFunctionalIdentity}. 

{\rem The penalty term in the information criteria is the same for every parameter. The Bayesian approach allows the choice of different penalties, assuming different priors, one for each parameter. }

\begin{table}[!h]	
	 \caption{Different information criterion for model selection.}\label{TablaIC}
	 \vspace{-0.2cm}
	 \footnotesize
	\begin{center}
		\begin{tabular}{|c|c|} 
		\hline 
			 {\bf Criterion} & {\bf Choice - approximation of} $\eta$   \\ 
			\hline 
			\hline 
			Bayesian-Schwarz information criterion (BIC) \cite{schwarz1978estimating} &  $-\frac{1}{2}\log D_y$ \\
			\hline 
			  Akaike information criterion  (AIC) \cite{Spiegelhalter02} &  $-1$ \\
			\hline 
                         Hannan-Quinn information criterion (HQIC)  \cite{Hannan79}&  $-\log(\log(D_y))$ \\
			\hline 
		\end{tabular}
	\end{center}
\end{table}

\subsection{A general overview of the computational methods}\label{GenOver}

After a depth revision of the literature, we have recognized four main families of techniques, described below. We list them in order of complexity, from the simplest to the most complex underlying main idea. However, each class can contain  both simple and very sophisticated algorithms.
\newline
\newline
{\bf Family 1: {\it Deterministic approximations}.} These methods consider an analytical approximation of the function $\post(\x|\y)$. The Laplace method and the Bayesian Information Criterion (BIC), belongs to this family (see Section \ref{ExploitingFunctionalIdentity}).
\newline
\newline
{\bf Family 2: {\it Methods based on density estimation}.}  This class of algorithms uses the equality 
\begin{eqnarray}
\widehat{Z} = \dfrac{\pi(\x^*|\y)}{\widehat{\post}(\x^*|\y)},
\end{eqnarray}
where $\widehat{\post}(\x^*|\y) \approx \post(\x^*|\y)$ represents an estimation of the density $\post(\x|\y)$ at some point $\x^*$. Generally, the point $\x^*$ is chosen in a high-probability region.
The techniques in this family differ in the procedure employed for obtaining the estimation $\widehat{\post}(\x^*|\y)$. One famous example is the Chib's method \cite{chib1995marginal}. Section \ref{ExploitingFunctionalIdentity} is devoted to describe methods belonging to family 1 and family 2. 
\newline
\newline
 {\bf Family 3: {\it Importance sampling (IS) schemes.}}  The IS methods are based on rewriting Eq. \eqref{MarginalLikelihood} as an expected value w.r.t. a simpler  normalized density $\bar{q}(\x)$, i.e.,
 $Z = \int_\Theta \pi(\x|\y)d\x =  E_{\bar{q}}\left[\frac{\pi(\x|\y)}{\bar{q}(\x)}\right]$. This is the most considered class of methods in the literature,  containing numerous variants, extensions and generalizations. We devote Sections \ref{ImportanceSamplingApproaches}-\ref{CombinationISandMCMC} to this family of techniques. 
\newline
\newline
  {\bf Family 4: {\it Methods based on a vertical representation.}}  These schemes rely on  changing the expression of $Z = \int_\Theta\ell(\y|\x)g(\x)d\x$ (that is a multidimensional integral) to equivalent one-dimensional integrals \cite{polson2014vertical, weinberg2012computing, skilling2006nested}. Then, a quadrature scheme is applied to approximate this one-dimensional integral. The most famous example is the nested sampling algorithm \cite{skilling2006nested}. Section \ref{verticalLikelihoodApproach} is devoted to this class of methods.
\section{Methods based on deterministic approximations and density estimation}\label{ExploitingFunctionalIdentity}
In this section, we consider approximations of $\post(\x|\y)$, or its unnormalized version $\pi(\x|\y)$, in order to obtain an estimation $Z$. 
In a first approach, the methods consider $\post(\x|\y)$ or  $\pi(\x|\y)$ as a function, and try to obtain a good approximation given another parametric or non-parametric family of functions.  
Another approach consists in approximating  $\post(\x|\y)$ only at one specific point $\x^*$, i.e., $\widehat{\post}(\x^*|\y)\approx \post(\x^*|\y)$ ($\x^*$ is usually chosen in high posterior probability regions), and then using the identity
\begin{align}\label{TrivialIdentity}
	\widehat{Z} = \dfrac{\pi(\x^*|\y)}{\widehat{\post}(\x^*|\y)}.
\end{align}
The latter scheme is often called {\it candidate's estimation}.
\subsection{Laplace's method}\label{LaplaceSect}
Let us define $\widehat{\x}_{\text{MAP}}\approx \x_{\text{MAP}}=\mbox{arg} \max \post(\x|\y)$ (obtained by some optimization method), which is an approximation of the {\it maximum a posteriori} (MAP),  and consider a Gaussian approximation of $ \post(\x|\y)$ around $\widehat{\x}_{\text{MAP}}$, i.e.,
\begin{align}
\widehat{\post}(\x|\y) = \mathcal{N}(\x|\widehat{\x}_{\text{MAP}}, {\widehat{\bf \Sigma}}), 
\end{align}
with ${\widehat{\bf \Sigma}}{\approx}-{\bf H}^{-1}$,  which is an approximation of the negative inverse Hessian matrix of $\log\pi(\x|\y)$ at $\widehat{\x}_{\text{MAP}}$. Replacing in Eq. \eqref{TrivialIdentity}, with $\x^*=\widehat{\x}_{\text{MAP}}$, we obtain the Laplace approximation
\begin{align}
\widehat{Z} = \frac{\pi(\widehat{\x}_{\text{MAP}}|\y)}{\mathcal{N}(\widehat{\x}_{\text{MAP}}|\widehat{\x}_{\text{MAP}},{\widehat{\bf \Sigma}})} = (2\pi)^\frac{D_x}{2}\lvert {\widehat{\bf \Sigma}}  \rvert^\frac{1}{2}\pi(\widehat{\x}_{\text{MAP}|\y}).
\end{align}
\noindent This is equivalent to the classical derivation of Laplace's estimator, which is based on expanding the $\log \pi(\x|\y) = \log(\ell(\y|\x)g(\x))$ as quadratic around $\widehat{\x}_{\text{MAP}}$ and substituting in $Z = \int\pi(\x|\y)d\x$, that is,
\begin{align}
	Z &=  \int\pi(\x|\y)d\x =  \int\exp\{ \log \pi(\x|\y) \}d\x \\
	&\approx  \int\exp\left\{ \log \pi(\widehat{\x}_{\text{MAP}}|\y) - \frac{1}{2}(\x-\widehat{\x}_{\text{MAP}})^T{\widehat{\bf \Sigma}}^{-1}(\x-\widehat{\x}_{\text{MAP}}) \right \}d\x \\
	&= (2\pi)^\frac{D_x}{2}\lvert {\widehat{\bf \Sigma}}\rvert^\frac{1}{2}\pi(\widehat{\x}_{\text{MAP}}|\y).
\end{align}
\noindent In \cite{lewis1997estimating}, they propose to use samples generated by a Metropolis-Hastings algorithm to estimate the quantities $\widehat{\x}_\text{MAP}$  and ${\widehat{\bf \Sigma}}$ \cite{Robert04}. The resulting method is called Laplace-Metropolis estimator. The authors in \cite{diciccio1997computing} present different variants of the Laplace's estimator.  A relevant extension for Gaussian Markov random field models, is the so-called {\it integrated nested Laplace approximation} (INLA) \cite{rue2017bayesian}.
\subsection{Bayesian-Schwarz information criterion (BIC)}\label{BICSect} 
Let us define $\widehat{\x}_{\text{MLE}}\approx \x_{\text{MLE}}=\mbox{arg} \max \ell(\y|\x)$. 
The following quantity
\begin{align}
	\text{BIC} = D_\theta\log{D_y} - 2\log\ell(\y|\widehat{\x}_{\text{MLE}}),
\end{align}
was introduced by Gideon E. Schwarz in \cite{schwarz1978estimating}, where $D_\theta$ represents the number of parameters of the model ($\x \in \mathbb{R}^{D_\theta}$), $D_y$ is the number of data,\footnote{Note that, for simplicity, we are considering scalar observations $y_i$, so that the dimension $D_y$ of the data vector $\y$ coincides with the number of data.} and $\ell(\y|\widehat{\x}_\text{MLE})$ is the estimated maximum value of the likelihood function. The value of $\widehat{\x}_\text{MLE}$ can be obtained using samples generated by a MCMC scheme. The BIC expression can be derived similarly to the Laplace's method, but this time with a second-order Taylor expansion of the $\log Z$ around its maximum $\x_\text{MLE}$ and a first-order expansion of the prior around $\x_\text{MLE}$ \cite[Ch. 9.1.3]{konishi2008information}.  The derivation is given in the Supplementary Material. Then, the final approximation is
\begin{align}
	Z\approx \widehat{Z}= \exp\left( \log \ell(\y|\widehat{\x}_{\text{MLE}}) - \frac{D_\theta}{2}\log D_y\right)=\exp\left( -\frac{1}{2} \text{BIC} \right), \quad \text{as $D_y\rightarrow \infty$}, 
\end{align}
and $\text{BIC} \approx -2\log Z$, asymptotically as the number of data $D_y$ grows. Then, smaller BIC values are associated to better models. Note that BIC clearly takes into account the complexity of the model since higher BIC values are given to models with more number of parameters $D_\theta$. 
Namely the penalty $D_\theta\log{D_y} $ discourages overfitting,  since increasing the number of parameters generally improves the goodness of the fit. 
 Other criteria can be found in the literature, such as the well-known Akaike information criterion (AIC), 
$$
\text{AIC} =2 D_\theta - 2\log\ell(\y|\widehat{\x}_{\text{MLE}}).
 $$ 
However, they are not an approximation of the marginal likelihood $Z$ and  are usually founded on information theory derivations.  Generally, they have the form of $c_p- 2\log\ell(\y|\widehat{\x}_{\text{MLE}})$ where the penalty term $c_p$ of the model complexity changes in each different criterion (e.g., $c_p= D_\theta\log{D_y} $ in BIC and $c_p=2D_\theta$ in AIC).  Another example that uses MCMC samples  is the Deviance Information Criterion (DIC), i.e.,
\begin{equation}
\text{DIC} =- \frac{4}{N}\sum_{n=1}^N \log\ell(\y|\x_n) - 2\log\ell(\y|\bar{\x}), \quad \mbox{ where } \quad \bar{\x}=\frac{1}{N}\sum_{n=1}^N \x_n,
\end{equation}
and $\{\x_n\}_{n=1}^N$ are outputs of an MCMC algorithm \cite{Spiegelhalter02}. In this case, note that $c_p=- \frac{4}{N}\sum_{i=1}^N \log\ell(\y|\x_n)$. DIC is considered more adequate for  hierarchical models than AIC, BIC \cite{Spiegelhalter02}, but is not directly related to the marginal likelihood \cite{Pooley18}. See also related comments in Section \ref{ObradeArteCriteria}.
\subsection{Kernel density estimation (KDE)}\label{KDESect}  
KDE can be used to approximate the value of the posterior density at a given point $\x^*$, and then consider $Z \approx \frac{\pi(\x^*|\y)}{\widehat{\post}(\x^*|\y)}$. For instance, we can build a kernel density estimate (KDE) of $\post(\x|\y)$ based on $M$ samples distributed according to the posterior  (obtained via an MCMC algorithm, for instance) by using $M$ normalized kernel functions $k(\x|{\bm \mu}_m,h)$ (with $\int_{\Theta} k(\x|{\bm \mu}_m,h) d\x=1$ for all $m$) where ${\bm \mu}_m$ is a location parameter and $h$ is a scale parameter,
\begin{align}
	\widehat{\post}(\x^*|\y) = \frac{1}{M}\sum_{m=1}^M k(\x^*|{\bm \mu}_m,h), \quad \{\bm{\mu}_m \}_{m=1}^M \sim \post(\x|\y) \quad \mbox{(e.g., via MCMC)}.
\end{align} 
Generally, $\widehat{\post}(\x^*|\y)$ is  a biased estimation of $\post(\x^*|\y)$.
The estimator is $\widehat{Z}= \frac{\pi(\x^*|\y)}{\widehat{\post}(\x^*|\y)}$ where the point $\x^*$ can be chosen as $\widehat{\x}_{\text{MAP}}$. If we consider $N$ different points $\x_1, ...,\x_N$ (selected without any specific rule) we can also write a more general approximation,
\begin{equation}\label{aquiKDEcasiIS}
\widehat{Z}=\frac{1}{N} \sum_{n=1}^N \frac{\pi(\x_n|\y)}{\widehat{\post}(\x_n|\y)}.
\end{equation}
\begin{rem}
	The estimator above is generally biased and depends on the choices of {\bf(a)}  of the points $\x_1, ...,\x_N$, {\bf(b)} the scale parameter $h$, and {\bf (c)} the number of samples $M$ for building $\widehat{\post}(\x^*|\y)$.
\end{rem}  
\begin{rem}
	A improved version of this approximation can be obtained  by the {\it importance sampling} approach described in Sect. \ref{ImportanceSamplingApproaches},  where $\x_1,...,\x_N$ are drawn from the KDE mixture $\widehat{\post}(\x|\y)$. In this case, the resulting estimator is {\it unbiased}.
\end{rem}
\subsection{Chib's method}\label{ChibEstSect}
In \cite{chib1995marginal, chib2001marginal}, the authors present more sophisticated methods to estimate $\post(\x^*|\y)$ using outputs from Gibbs sampling and the Metropolis-Hastings (MH) algorithm respectively \cite{Robert04}. Here we only present the latter method, since it can be applied in more general settings. 
In \cite{chib2001marginal}, the authors propose to estimate the value of the posterior at one point $\x^*$,i.e.,  $\post(\x^*|\y)$, using the output from a MH sampler. More specifically, let us denote the current state as $\x$. A possible candidate as future state $\z \sim \varphi(\z|\x)$ (where $ \varphi(\z|\x)$ represents the proposal density used within MH), is accepted with probability
		$\alpha(\x,\z) = \min \left\{1, \frac{\pi(\z|\y)\varphi(\x|\z)}{\pi(\x|\y)\varphi(\z|\x)}\right\}$  \cite{Robert04,MHchap17}.
	This is just an example of $\alpha(\x,\z)$ that
	by construction the probability $\alpha$ satisfies the detailed balance condition \cite[Section 2.4]{MHchap17},\cite{MARTINO_REV_MTM}, i.e.,
	\begin{equation}
        \alpha(\x,\z)\varphi(\z|\x)\post(\x|\y) = \alpha(\z,\x)\varphi(\x|\z)\post(\z|\y).
   \end{equation}
 By integrating in $\x$ both sides, we obtain 
       \begin{align*}
	     	\int_ {\Theta} \alpha(\x,\z)\varphi(\z|\x)\post(\x|\y) d\x &= \int_ {\Theta} \alpha(\z,\x)\varphi(\x|\z)\post(\z|\y) d\x, \\
      		&= \post(\z|\y) \int_ {\Theta} \alpha(\z,\x)\varphi(\x|\z) d\x, 	
	\end{align*}
hence finally we can solve with respect to $\post(\z|\y)$ obtaining 
	\begin{align}
	\post(\z|\y) = \frac{\int_ {\Theta}\alpha(\x,\z)\varphi(\z|\x)\post(\x|\y)d\x}{\int_\Theta\alpha(\z,\x)\varphi(\x|\z)d\x}. 
	\end{align}
	 This suggests the following estimate of $\post(\x^*|\y)$ at a specific point $\x^*$ (note that $\x^*$ plays the role of $\z$ in the equation above),
	\begin{align}
	\widehat{\post}(\x^*|\y) = \dfrac{\dfrac{1}{N_1}\sum_{i=i}^{N_1}\alpha(\x_i,\x^*)\varphi(\x^*|\x_i)}{\dfrac{1}{N_2}\sum_{j=1}^{N_2}\alpha(\x^*,{\bf v}_j)}, \quad \enskip \{\x_i\}_{i=1}^{N_1}\sim \post(\x|\y),  \enskip \{{\bf v}_j\}_{j=1}^{N_2}\sim \varphi(\x|\x^*).
	\end{align}
	The same outputs of the MH scheme can be considered as $\{\x_i\}_{i=1}^{N_1}$. The final estimator is again $\widehat{Z}= \frac{\pi(\x^*|\y)}{\widehat{\post}(\x^*|\y)}$, i.e.,
	\begin{align}\label{eq:ChibsEst}
	\widehat{Z}= \dfrac{\pi(\x^*|\y) \dfrac{1}{N_2}\sum_{j=1}^{N_2}\alpha(\x^*,{\bf v}_j) }{\dfrac{1}{N_1}\sum_{i=i}^{N_1}\alpha(\x_i,\x^*)\varphi(\x^*|\x_i)   }, \quad \enskip \{\x_i\}_{i=1}^{N_1}\sim \post(\x|\y),  \enskip \{{\bf v}_j\}_{j=1}^{N_2}\sim \varphi(\x|\x^*).
	\end{align}	
	The point $\x^*$ is usually chosen in an high probability region.
	 Interesting discussions are contained in  \cite{mira2003bridge}, where the authors also show that this estimator is related to bridge sampling idea described in Section \ref{TwoDensitiesIS}. For more details, see Section \ref{ChibBridgeSect}.  
\subsection{Interpolative approaches}
Another possibility is to approximate $Z$ by substituting the true $\pi(\x|\y)$ with interpolation or a regression function $\widehat{\pi}(\x|\y)$ in the integral \eqref{MarginalLikelihood2}. For simplicity, we focus on the interpolation case, but all the considerations can be easily extended for a regression scenario.  Given a set of nodes $\{\x_1,\dots,\x_N \}\subset \Theta$ and $N$ nonlinear functions $k(\x,\x'): \Theta\times\Theta \rightarrow \mathbb{R}$ chosen in advance by the user (generally, centered around $\x'$), we can build  the interpolant of unnormalized posterior $\pi(\x|\y)$ as follows 
\begin{align}\label{Interpolant_of_pi(x)}
\widehat{\pi}(\x|\y) = \sum_{i=1}^N\beta_i k(\x,\x_i),
\end{align}
where  $\beta_i \in \mathbb{R}$ and the subindex $u$ denotes that is an approximation of the unnormalized function $\pi(\x|\y)$. The coefficients $\beta_i$ are chosen such that $\widehat{\pi}_u(\x|\y)$ interpolates the points $\{\x_n,\pi(\x_n|\y)\}$, that is, $\widehat{\pi}(\x_n|\y) = \pi(\x_n|\y)$. Then, we desire that 
\begin{align*}
 \sum_{i=1}^N\beta_i k(\x_n,\x_i) = \pi(\x_n|\y), 
\end{align*}
for all $n=1,...,N$. Hence, we can write a $N\times N$ linear system  where  the $\beta_i$ are the $N$ unknowns, i.e.,
\begin{align}\label{Computation of coeff beta}
\begin{pmatrix}
k(\x_1,\x_1) & k(\x_1,\x_2) & \dots & k(\x_1,\x_N) \\	
k(\x_2,\x_1) & k(\x_2,\x_2) & \dots & k(\x_2,\x_N) \\
\vdots &  & \ddots & \vdots \\
k(\x_N,\x_1) & k(\x_N,\x_2) & \dots & k(\x_N,\x_N)
\end{pmatrix}
\begin{pmatrix}
\beta_1 \\
\beta_2 \\
\vdots \\
\beta_N
\end{pmatrix}
&= \begin{pmatrix}
\pi(\x_1|\y) \\
\pi(\x_2|\y) \\
\vdots \\
\pi(\x_N|\y)
\end{pmatrix} 
\end{align}
In matrix form, we have 
\begin{equation}
{\bf K}\bm{\beta} = {\bf y},
\end{equation}
where $({\bf K})_{i,j}=k(\x_i,\x_j)$ and ${\bf y}= [\pi(\x_1|\y),\dots,\pi(\x_N|\y)]^{\top}$. Thus, the solution is $\bm{\beta} = {\bf K}^{-1}{\bf y}$. Now the interpolant $\widehat{\pi}_u(\x|\y)= \sum_{i=1}^N\beta_i k(\x,\x_i)$ can be used to approximate $Z$ as follows
\begin{align}\label{Z approx by substituting pi with interpolant}
\widehat{Z} = \int_\Theta\widehat{\pi}_u(\x|\y)d\x
=\sum_{i=1}^{N} \beta_i\int_\Theta k(\x,\x_i)d\x.
\end{align}
If we are able to compute analytically $\int_\Theta k(\x,\x_i)d\x$, we have an approximation $\widehat{Z}$. Some suitable choices of $k(\cdot,\cdot)$ are rectangular, triangular and Gaussian functions. More specifically, if all the nonlinearities $k(\x,\x_i)$ are normalized (i.e. $\int_\Theta k(\x,\x_i)d\x=1$), the approximation of $Z$ is $\widehat{Z} = \sum_{i=1}^N\beta_i$. 
This approach is related to the so-called Bayesian quadrature (using Gaussian process approximation) \cite{rasmussen2003bayesian} and the sticky proposal constructions within MCMC or rejection sampling algorithms \cite{Gilks92,Gilks95,Sticky13,MartinoA2RMS}. Adaptive schemes adding sequentially more nodes could be also considered, improving the approximation $\widehat{Z}$   \cite{Gilks95,Sticky13}. The quality of the interpolating approximation deteriorates as the dimension of $\x$ grows (see e.g. \cite{briol2019probabilistic} for explicit error bounds).

\section{Techniques based on IS}\label{ImportanceSamplingApproaches}
Most of the techniques for approximating the marginal likelihood are based on the importance sampling (IS) approach. Other methods are directly or indirectly related to the IS framework. In this sense, this section is the core of this survey. 
The standard IS scheme relies on the following equality,
\begin{align}\label{StandardISidentity}
Z = \int_\Theta \pi(\x|\y)d\x = 
\E_{\bar{q}}\left[\frac{\pi(\x|\y)}{\bar{q}(\x)}\right]
&=\int_\Theta \frac{\pi(\x|\y)}{\bar{q}(\x)}\bar{q}(\x)d\x \\
&= \int_\Theta\frac{\ell(\y|\x)g(\x)}{\bar{q}(\x)} \bar{q}(\x)d\x,
\end{align}
where $\bar{q}(\x)$ is a simpler normalized proposal density, $\int_\Theta \bar{q}(\x)d\x=1$.
\newline
\newline
{\bf IS version 1.}  Drawing $N$ independent samples from proposal $\bar{q}(\x)$, the {\it unbiased} IS estimator (denoted as IS vers-1) of $Z$ is 
\begin{align}\label{StandardIS}
\widehat{Z}_{IS1} &= \frac{1}{N}\sum_{i=1}^N\frac{\pi(\x_i|\y)}{\bar{q}(\x_i)}\\
&= \frac{1}{N}\sum_{i=1}^N w_i,\\
&= \frac{1}{N}\sum_{i=1}^N\frac{\ell(\y|\x_i)g(\x_i)}{\bar{q}(\x_i)}= \frac{1}{N}\sum_{i=1}^N\rho_i \ell(\y|\x_i), \qquad \{\x_i\}_{i=1}^N \sim \bar{q}(\x),
\end{align}
where $w_i=\frac{\pi(\x_i|\y)}{\bar{q}(\x_i)}$ are the standard IS weights and $\rho_i =\frac{g(\x_i)}{\bar{q}(\x_i)}$. 
\newline
\newline
{\bf Optimal proposal in IS vers-1.}  The optimal proposal, in terms of mean square error (MSE), in the standard IS scheme above  is $\bar{q}^{\text{opt}}({\bm \theta})=P(\x|\y)$.
\newline
\newline
{\bf IS version 2.} An alternative IS estimator (denoted as IS vers-2) is given by, considering a possibly unnormalized proposal pdf $q(\x)\propto \bar{q}(\x)$ (the case $q(\x)= \bar{q}(\x)$ is also included),
\begin{align}\label{StandardIS_2}
\widehat{Z}_{IS2} &=\frac{1}{\sum_{n=1}^N\frac{g(\x_n)}{q(\x_n)}} \sum_{i=1}^N\frac{g(\x_i)}{q(\x_i)} \ell(\y|\x_i), \\
&=\frac{1}{\sum_{n=1}^N\rho_n} \sum_{i=1}^N\rho_i \ell(\y|\x_i), \\
&= \sum_{i=1}^N\bar{\rho}_i \ell(\y|\x_i), \qquad \{\x_i\}_{i=1}^N \sim \bar{q}(\x).
\end{align} 
The estimator above is biased. However, it is a convex combination of likelihood values $\ell(\y|\x_i)$ since $\sum_{i=1}^N \bar{\rho}_i=1$. Hence, in this case  $\min\limits_{i} \ell(\y|\x_i) \leq\widehat{Z}\leq \max\limits_{i}  \ell(\y|\x_i)$, i.e., the estimator fulfills the bounds of $Z$, shown Section \ref{ModelFitSect}. Moreover, the estimator allows the use of an unnormalized proposal pdf $q(\x)\propto \bar{q}(\x)$ and $\rho_i =\frac{g(\x_i)}{q(\x_i)}$. For instance, one could consider $\bar{q}(\x)=\post(\x|\y)$, i.e., generate samples $ \{\x_i\}_{i=1}^N \sim \post(\x|\y)$ by an MCMC algorithm and then evaluate $\rho_i =\frac{g(\x_i)}{\pi(\x_i|\y)}$. 
\newline
\newline
{\bf Optimal proposal in IS vers-2. The optimal proposal, in terms of MSE, for the IS vers-2 is $\bar{q}^{\text{opt}}({\bm \theta})\propto|P(\x|\y)-g(\x)|$.} 
\newline
\newline
 Table \ref{TablaIS_0} summarizes the IS estimators and shows some important special cases that will be described in the next section.

\begin{table}[!h]	
	 \caption{IS estimators Eqs. \eqref{StandardIS}-\eqref{StandardIS_2} and relevant special cases. }\label{TablaIS_0}
	 \vspace{-0.2cm}
	\begin{center}
		\begin{tabular}{|c|c|c|c|c|c| } 
		\hline
		\multicolumn{6}{|c|}{ } \\
		\multicolumn{6}{|c|}{ $\widehat{Z}_{IS1} = \frac{1}{N} \sum_{i=1}^N\frac{g(\x_i)}{\bar{q}(\x_i)} \ell(\y|\x_i)=\frac{1}{N}\sum_{i=1}^N \rho_i \ell(\y|\x_i)$, \quad  $\rho_i=\frac{g(\x_i)}{\bar{q}(\x_i)}$}\\
		\multicolumn{6}{|c|}{ } \\ 
		\hline
		Name   & Estimator & $q(\x)$ & ${\bar q}(\x)$  & Need of MCMC & Unbiased\\ 
			\hline 
			 Naive Monte Carlo & $\frac{1}{N}\sum_{i=1}^N  \ell({\bf y}|\x_i)$ & $g(\x)$ &  $g(\x)$ & --- &  \checkmark \\
			\hline
			\hline
		\multicolumn{6}{|c|}{ } \\
		\multicolumn{6}{|c|}{ $\widehat{Z}_{IS2} = \frac{1}{\sum_{n=1}^N\frac{g(\x_n)}{q(\x_n)}} \sum_{i=1}^N\frac{g(\x_i)}{q(\x_i)} \ell(\y|\x_i)=\sum_{i=1}^N\bar{\rho}_i \ell(\y|\x_i)$}\\
		\multicolumn{6}{|c|}{ } \\ 
		\hline 
			 Name   & Estimator & $q(\x)$ & ${\bar q}(\x)$  & Need of MCMC & Unbiased\\ 
			\hline 
			 Naive Monte Carlo & $\frac{1}{N}\sum_{i=1}^N  \ell({\bf y}|\x_i)$ & $g(\x)$ &  $g(\x)$ & --- &  \checkmark \\
			 Harmonic mean  &  $\left(\frac{1}{N}\sum_{i=1}^N\frac{1}{\ell(\y|\x_i)}\right)^{-1}$  & $\pi(\x|\y)$& $\post(\x|\y)$ & \checkmark& --- \\ 
			\hline
		\end{tabular}
	\end{center}
\end{table}

\noindent 
Different sub-families of IS schemes are commonly used for computing normalizing constants \cite[chapter 5]{chen2012monte}. A first approach uses draws from a proposal density $\bar{q}(\x)$ that is completely known (i.e. direct sampling and evaluate). Sophisticated choices of $\bar{q}(\x)$ frequently imply the use of MCMC algorithms to sample from $\bar{q}(\x)$ and that we can only evaluate $q(\x)\propto \bar{q}(\x)$. The one-proposal approach is described in Section \ref{OneProp_IS_Sect}.  A second class is formed by methods which use more than one proposal density or a mixture of them (see Sections \ref{TwoDensitiesIS}, \ref{ISMultipleProposal} and  \ref{CombinationISandMCMC}). 
Moreover, {\bf adaptive importance sampling (AIS)} schemes are often designed, where the proposal (or the cloud of proposals)  is improved during some iterations, in some way such that  $\bar{q}_t(\x)$ (where $t$ is an iteration index) becomes closer and closer to  the optima proposal $q^{\text{opt}}(\x)$. For more details, see the reviews in \cite{bugallo2017adaptive}. Some AIS methods, obtained combining MCMC and IS approaches, are described in Section \ref{CombinationISandMCMC}.

\subsection{Techniques using draws from one proposal density}\label{OneProp_IS_Sect}
In this section, all the techniques are IS schemes which use a unique proposal pdf, and are based on the identity Eq. \eqref{StandardISidentity}. The techniques differ in the choice of $\bar{q}(\x)$. 
Recall that the optimal proposal choice for IS vers-1 is  $\bar{q}(\x)=\post(\x|\y)=\frac{1}{Z} \pi(\x|\y)$. This choice is clearly difficult for two reasons: (a) we have to draw from $\post$ and (b) we do not know $Z$, hence we cannot evaluate $\bar{q}(\x)$ but only  $q(\x)=\pi(\x|\y)$ (where $q(\x) \propto \bar{q}(\x)$). However, there are some methods based on this idea, as shown in the following. 
 The techniques below are enumerated in an increasing order of complexity. 
\newline
\newline
{\bf Naive Monte Carlo (arithmetic mean estimator).} It is straightforward to note that the integral above can be expressed as $Z=\E_{g}[\ell(\y|\x)]$, then we can draw $N$ samples $\{\x_i\}_{i=1}^N$ from the prior $g(\x)$ and compute the following estimator  
	\begin{align}\label{NaiveMC}
	\widehat{Z} = \frac{1}{N}\sum_{i=1}^N \ell(\y|\x_i),\qquad \{\x_i\}_{i=1}^N \sim g(\x).
	\end{align}
Namely a simple average of the likelihoods of a sample from the prior. Note that $\widehat{Z}$ will be very inefficient (large variance) if the posterior is much more concentrated than the prior (i.e., small overlap between likelihood and prior pdfs). Therefore, alternatives have been proposed, see below. It is a special case of the IS estimator with the choice $\bar{q}(\x)=g(\x)$ (i.e., the proposal pdf is the prior).
\newline
\newline
{\bf Harmonic mean (HM) estimators}. The HM estimator can be directly derived from the following expected value,
\begin{align}
 {\E_{\post}\left[\frac{1}{\ell(\y|\x)} \right]} &= \int_\Theta\frac{1}{\ell(\y|\x)}\post(\x|\y)d\x, \\
  &=\frac{1}{Z}\int_\Theta\frac{1}{\ell(\y|\x)}\ell(\y|\x) g(\x) d\x =\frac{1}{Z} \int_\Theta g(\x) d\x= \frac{1}{Z}.
\end{align}
The main idea is again to use the posterior itself as proposal. Since direct sampling from $\post(\x|\y)$ is generally impossible, this task requires the use of MCMC algorithms. Thus, the HM estimator is
\begin{align}\label{HarmonicMeanEst}
	\widehat{Z}=\left(\frac{1}{N}\sum_{i=1}^N\frac{1}{\ell(\y|\x_i)}\right)^{-1}=\frac{1}{\frac{1}{N}\sum_{i=1}^N\frac{1}{\ell(\y|\x_i)}}, \quad \{ \x_i \}_{i=1}^N \sim \post(\x|\y) \enskip (\text{via MCMC}).
\end{align}
The HM estimator converges almost surely to the correct value, but the variance of $\widehat{Z}$ is often high and possibly infinite. 
\footnote{See the comments of Radford Neal's blog, \url{https://radfordneal.wordpress.com/2008/08/17/the-harmonic-mean-of-the-likelihood-worst-monte-carlo-method-ever/}, where R. Neal defines the HM estimator as ``the worst estimator ever''.}  
The HM estimator is a special case of Reverse Importance Sampling (RIS) below.
\newline
\newline
 {\bf Reverse Importance Sampling (RIS).} 
 The RIS scheme \cite{gelfand1994bayesian}, also known as {\it reciprocal} IS, can be derived from the identity   
\begin{eqnarray}\label{ReverseISidentity}
 \frac{1}{Z} =\E_{\post}\left[ \frac{f(\x)}{\pi(\x|\y)} \right] &=&\int_\Theta\frac{f(\x)}{\pi(\x|\y)}\post(\x|\y)d\x \
\end{eqnarray}
where  we consider an auxiliary normalized function $f(\x)$, i.e., $\int_{\Theta} f(\x) d\x=1$. Then, one could consider the estimator
\begin{eqnarray}\label{ReverseIS}
 \widehat{Z} &=& \left(\frac{1}{N}\sum_{i=1}^N\frac{f(\x_i)}{\pi(\x_i|\y)}\right)^{-1} =\left(\frac{1}{N}\sum_{i=1}^N\frac{f(\x_i)}{\ell(\y|\x_i)g(\x_i)}\right)^{-1},\quad \x_i \sim \post(\x|\y) \enskip (\text{via MCMC}).
\end{eqnarray}
The estimator above is consistent but biased. Indeed, the expression $\frac{1}{N}\sum_{i=1}^N\frac{f(\x_i)}{\pi(\x_i|\y)}$ is a unbiased estimator of $1/Z$, but $ \widehat{Z}$ in the Eq. \eqref{ReverseIS}  is not an unbiased estimator of $Z$.  
Note that  $\post(\x|\y)$ plays the role of importance density from which we need to draw from. Therefore, another sampling technique must be used (such as a MCMC method) in order to generate samples from $\post(\x|\y)$. In this case,  we do not need samples from $f(\x)$, although its choice affects the precision of the approximation. Unlike in the standard IS approach, $f(\x)$ must have lighter tails than $\pi(\x|\y)=\ell(\y|\x)g(\x)$. For further details, see the example in Section \ref{Comparison IS vs RIS}.
Finally, note that the HM estimator is a special case of RIS when $f(\x)=g(\x)$ in Eq. \eqref{ReverseIS}. In \cite{robert2009computational}, the authors propose taking $f(\x)$ that is uniform in a high posterior density region whereas, in \cite{wang2018new}, they consider taking $f(\x)$ to be a piecewise constant function.

\subsubsection{The pre-umbrella estimators}

\noindent
All the estimators that we have seen so far can be unified within a common formulation, considering the more general problem of estimating a ratio of two normalizing constants $c_1/c_2$, where $c_i = \int q_i(\x)d\x$ and $\bar{q}_i(\x) = q_i(\x)/c_i$, $i=1,2$. Assuming we can evaluate both $q_1(\x)$, $q_2(\x)$, and draw samples from one of them, say $\bar{q}_2(\x)$, the importance sampling estimator of ratio $c_1/c_2$ is
\begin{align}\label{OnePropIS}
\frac{c_1}{c_2} = \E_{\bar{q}_2}\left[ \frac{q_1(\x)}{q_2(\x)} \right] \approx \frac{1}{N}\sum_{i=1}^N \frac{q_1(\x_i)}{q_2(\x_i)}, \quad \{ \x_i \}_{i=1}^N \sim \bar{q}_2(\x).
\end{align}
{\rem{The relative MSE (rel-MSE) of \eqref{OnePropIS}, in estimation of the ratio $r=\frac{c_1}{c_2}$, i.e.,
$\text{rel-MSE}= \frac{\mathbb{E}[(\widehat{r}-r)^2]}{r^2}$,
 is given by $\text{rel-MSE}= \frac{1}{N}\chi^2(\bar{q}_1||\bar{q}_2)$, where $\chi^2(\bar{q}_1||\bar{q}_2)$ is the Pearson divergence between $\bar{q}_1$ and $\bar{q}_2$ \cite{chen1997monte}. } }
\newline
\newline
This framework includes  almost all the estimators discussed so far in this section, as shown in Table \ref{Tabla2}. However, the IS vers-2 estimator is not a special case of Eq. \eqref{OnePropIS}. 

\begin{table}[!h]	
	 \caption{Summary of techniques considering the expression \eqref{OnePropIS}.}\label{Tabla2}
	 \vspace{-0.2cm}
	\begin{center}
		\begin{tabular}{|c|c| c|c|c|c|c|} 
		\hline 
			 Name & $q_1(\x)$  &$q_2(\x)$ & $c_1$ & $c_2$  & Proposal pdf $\bar{q}_2(\x)$ &  $c_1/c_2$ \\ 
			\hline 
			\hline 
			 IS vers-1 &  $\pi(\x|\y)$   & $\bar{q}(\x)$ & $Z$ &  $1$  & $\bar{q}(\x)$ &   $Z$ \\
			 Naive Monte Carlo & $\pi(\x|\y)$  & $g(\x)$ & $Z$ & $1$  & $g(\x)$ &  $Z$ \\
			 Harmonic mean  &  $g(\x)$  &$\pi(\x|\y)$ & $1$ & $Z$ &$\post(\x|\y)$  &   $1/Z$ \\ 
			 RIS   & $f(\x)$  & $\pi(\x|\y)$ & $1$ &  $Z$ & $\post(\x|\y)$ &  $1/Z$ \\ 
			\hline
		\end{tabular}
	\end{center}
\end{table}

Below we consider an extension of Eq. \eqref{OnePropIS} where an additional density $\bar{q}_3(\x)$ is employed for generating samples.


\subsubsection{Umbrella Sampling (a.k.a. ratio importance sampling)}\label{UmbrellaSect}

The IS estimator of $c_1/c_2$ given in Eq. \eqref{OnePropIS} may be inefficient when there is little overlap between $\bar{q}_1(\x)$ and $\bar{q}_2(\x)$, i.e., when $\int_\Theta\bar{q}_1(\x)\bar{q}_2(\x)d\x$ is small.
Umbrella sampling (originally proposed in the computational physics literature, \cite{torrie1977nonphysical}; also studied under the name ratio importance sampling in \cite{chen1997monte}) is based on the identity
\begin{align}\label{UmbrellaSamplingIdentity}
	\frac{c_1}{c_2} = \frac{c_1/c_3}{c_2/c_3} = \frac{\E_{\bar{q}_3}\left[\frac{q_1(\x)}{q_3(\x)}\right]}{\E_{\bar{q}_3}\left[\frac{q_2(\x)}{q_3(\x)}\right]} \approx \frac{\sum_{i=1}^N\frac{q_1(\x_i)}{q_3(\x_i)}}{\sum_{i=1}^N\frac{q_2(\x_i)}{q_3(\x_i)}}, \quad \{\x_i \}_{i=1}^N\sim \bar{q}_3(\x)
\end{align}
where $\bar{q}_3(\x)\propto q_3(\x)$ represents a {\it middle} density. A good choice of $\bar{q}_3(\x)$ should have large overlaps with both $\bar{q}_i(\x)$, $i=1,2$. The performance of umbrella sampling clearly depends on the choice of $\bar{q}_3(\x)$. 
Note that, when $\bar{q}_3=\bar{q}_2$, we recover Eq. \eqref{OnePropIS}.
\newline
\newline
{\bf Optimal umbrella proposal.}
The optimal umbrella sampling density $\bar{q}_3^\text{opt}(\x)$, that minimizes the asymptotic relative mean-square error, is
\begin{align}\label{eq:OptProUmbrella}
	\bar{q}_3^\text{opt}(\x) = \frac{|\bar{q}_1(\x) - \bar{q}_2(\x)|}{\int |\bar{q}_1(\x') - \bar{q}_2(\x')|d\x '} = \frac{|{q}_1(\x) - \frac{c_1}{c_2}{q}_2(\x)|}{\int |{q}_1(\x') - \frac{c_1}{c_2}{q}_2(\x')|d\x '}.
\end{align}
{\rem{\label{EsteRemark6} The  rel-MSE in estimation of the ratio $\frac{c_1}{c_2}$ of the optimal umbrella estimator, with $N$  great enough, is given by $\text{rel-MSE}\approx \frac{1}{N}L_1^2(\bar{q}_1,\bar{q}_2)$, where $L_1^2(\bar{q}_1,\bar{q}_2)$ denotes the $L_1$-distance  between $\bar{q}_1$ and $\bar{q}_2$ \cite[Theorem 3.2]{chen1997monte}. Moreover, since $L_1^2(\bar{q}_1,\bar{q}_2)\leq \chi^2(\bar{q}_1||\bar{q}_2)$, the optimal umbrella estimator is asymptotically more efficient than the estimator \eqref{OnePropIS} \cite[Sect. 3]{chen1994importance}. }} 
\newline
\newline
{\bf Two-stage umbrella sampling.} Since this $\bar{q}_3^\text{opt}(\x)$ depends on the unknown ratio $\frac{c_1}{c_2}$, it is not available for a direct use. The following two-stage procedure is often used in practice: 
\begin{enumerate}
	\item {\it Stage 1}: Draw $N_1$ samples from an arbitrary density $\bar{q}_3^{(1)}(\x)$ and use them to obtain 
	\begin{align}
	\widehat{r}^{(1)} = \frac{\sum_{i=1}^{N_1}\frac{q_1(\x_i)}{q_3^{(1)}(\x_i)}}{\sum_{i=1}^{N_1}\frac{q_2(\x_i)}{q_3^{(1)}(\x_i)}}, \enskip \{\x_i \}_{i=1}^{N_1} \sim \bar{q}_3^{(1)}(\x).
	\end{align}
	and define
	\begin{align}
	\bar{q}_3^{(2)}(\x) \propto |q_1(\x) - \widehat{r}^{(1)}q_2(\x)|.
	\end{align}
	
	\item {\it Stage 2}: Draw $N_2$ samples from $\bar{q}_3^{(2)}(\x)$ via MCMC and define the umbrella sampling estimator $\widehat{r}^{(2)}$ of $\frac{c_1}{c_2}$ as follows
	\begin{align}
	\widehat{r}^{(2)} = \frac{\sum_{i=1}^{n_2}\frac{q_1(\x_i)}{q_3^{(2)}(\x_i)}}{\sum_{i=1}^{n_2}\frac{q_2(\x_i)}{q_3^{(2)}(\x_i)}}, \enskip \{\x_i \}_{i=1}^{n_2} \sim \bar{q}_3^{(2)}(\x).
	\end{align}
\end{enumerate}
{\rem The number of stages could be increased considering, at each $t$-th stage, the proposal $\bar{q}_3^{(t)}(\x)\propto |q_1(\x) - \widehat{r}^{(t-1)}q_2(\x)|$ and obtaining a new estimation $\widehat{r}^{(t)}$. In this case, we have an umbrella scheme with adaptive proposal $\bar{q}_3^{(t)}(\x)$.}

\subsubsection{Umbrella for $Z$: the self-normalized Importance Sampling (Self-IS)}\label{UmbrellaForZsect}

Here, we describe an important special case of the umbrella sampling approach.
Considering  the umbrella identity \eqref{UmbrellaSamplingIdentity} an setting $q_1(\x)=\pi(\x|\y)$, $q_2(\x)=\bar{q}_2(\x)=f(\x)$, $c_1=Z$, $c_2=1$ and $c_3\in \mathbb{R}$, we obtain
\begin{equation}
\label{IS_self}
\widehat{Z} = \frac{1}{\sum_{i=1}^N\frac{f(\x_i)}{q_3(\x_i)}}\sum_{i=1}^N\frac{\pi(\x_i|\y)}{q_3(\x_i)}. \qquad \{\x_i\}_{i=1}^N \sim \bar{q}_3(\x).
\end{equation}
which is called the {\it self-normalized IS} (Self-IS) estimator. Note that  $f(\x)$ is an auxiliary normalized pdf, but we draw samples from $\bar{q}_3(\x)$.  In order to understand the reason of its name is interesting to derive it with standard IS arguments.  Let us consider that our proposal $q(\x)$ in the standard IS scheme is not normalized, and we can evaluate it up to a normalizing constant  $q(\x) \propto \bar{q}(\x)$. We also denote $c=\int_{\Theta} q(\x)d\x$. Note that this also occurs in the ideal case of using $\bar{q}(\x)=\post(\x|\y)=\frac{1}{Z} \pi(\x|\y)$ where $c=Z$ and $q(\x)=\pi(\x|\y)$. In this case, we have
\begin{align}\label{ESTASelfIS0}
\frac{\widehat{Z}}{c} &= \frac{1}{N}\sum_{i=1}^N\frac{\pi(\x_i|\y)}{q(\x_i)},\qquad \{\x_i\}_{i=1}^N \sim \bar{q}(\x).
\end{align}
Therefore, we need an additional estimation of $c$. We can also use IS for this goal, considering a new normalized reference function $f(\x)$, i.e., $\int_{\Theta} f(\x) d\x=1$. Now,
\begin{align}\label{ESTASelfIS}
\frac{1}{c} =E_{\bar q}\left[\frac{f(\x)}{q(\x)}\right]=\int_{\Theta}\frac{f(\x)}{q(\x)} \bar{q}(\x) d\x \approx \frac{1}{N}\sum_{i=1}^N\frac{f(\x_i)}{q(\x_i)},\qquad \{\x_i\}_{i=1}^N \sim \bar{q}(\x).
\end{align} 
Replacing  \eqref{ESTASelfIS} into   \eqref{ESTASelfIS0}, we obtain the self-normalized IS estimator in Eq. \eqref{IS_self}, i.e., 
$\widehat{Z}= \frac{1}{\sum_{i=1}^N\frac{f(\x_i)}{q(\x_i)}}\sum_{i=1}^N\frac{\pi(\x_i|\y)}{q(\x_i)}$ with $ \{\x_i\}_{i=1}^N \sim \bar{q}(\x)$.
\newline
The HM estimator is also a special case of Self-IS  setting  again $f(\x)=g(\x)$ and $\bar{q}(\x)= \post(\x|\y)$, so that $q(\x)=\pi(\x|\y)$. Moreover, the RIS estimator is a special case of  the Self-IS estimator above when $\bar{q}(\x)=\post(\x|\y)$ and $q(\x)=\pi(\x|\y)$.  
\newline
\newline
{\bf Optimal self-IS (O-Self-IS).} Since the Self-IS estimator is a special case of umbrella sampling, the optimal proposal in this case is 
$\bar{q}^\text{opt}(\x)\propto |\post(\x|\y) - f(\x)|$, and the optimal   estimator is
	\begin{align}
	\widehat{Z}_\text{O-Self-IS} = \frac{\sum_{i=1}^{N}\frac{\pi(\x_i)}{|\post(\x_i|\y) - f(\x_i)|}}{\sum_{i=1}^{N}\frac{f(\x_i)}{|\post(\x_i|\y) - f(\theta_i)|}},\quad \x_i \sim \bar{q}^\text{opt}(\x)\propto |\post(\x|\y) - f(\x)|.
	\end{align}
	Since the density cannot be evaluated (and also is not easy to draw from), this estimator is not of direct use and we need to resort to the two-stage procedure that we discussed above. Due to Remark \ref{EsteRemark6}, the O-Self-IS estimator is asymptotically more efficient than IS vers-1 estimator using $f(\x)$ as proposal, i.e., drawing samples from $\bar{q}(\x)=f(\x)$.


\subsubsection{Summary}

The more general expressions are the two identities \eqref{OnePropIS}-\eqref{UmbrellaSamplingIdentity} for estimating a ratio of normalizing constants $\frac{c_1}{c_2}$. The umbrella identity  \eqref{UmbrellaSamplingIdentity} is the more general since three densities are involved, and contains the Eq.  \eqref{OnePropIS} as special case when $q_3(\x)=q_2(\x)$. The Self-IS estimator coincides with the umbrella estimator when we approximate only one normalizing constant, $Z$ (i.e., for $\frac{c_1}{c_2}=Z$). Therefore, regarding the estimation of only one constant $Z$, the Self-IS estimator has the more general form and includes the rest of estimators as special cases. All these connections are summarized in  Table \ref{TablaUmbrella}. 
Finally, Table \ref{Tabla1} provides another summary of the one-proposal estimators of $Z$. Note that in the standard IS estimator the option $\bar{q}(\x)= \post(\x|\y)$ is not feasible, whereas it is possible for its second version.

\begin{table}[!h]	

	 \caption{Summary of techniques considering the umbrella sampling identity \eqref{UmbrellaSamplingIdentity} for computing $\frac{c_1}{c_2}=Z$. Note that Self-IS has the more general form and includes the rest of estimators as special cases.}\label{TablaUmbrella}
	 \vspace{-0.2cm}
	\begin{center}
		\begin{tabular}{|c||c|c|c|c|c|c|c|} 
	            \multicolumn{8}{c}{{\sl For estimating a generic ratio $c1/c2$}} \\
		\hline 
			 Umbrella & $q_1(\x)$  &$q_2(\x)$ & $q_3(\x)$ & $c_1$ & $c_2$ & $c_3$  & sampling from $\bar{q}_3(\x)$\\ 
			 \hline
			 \hline
			Eq. \eqref{OnePropIS} - ($q_3=q_2$) & $q_1(\x)$  &$q_2(\x)$ & $q_2(\x)$ & $c_1$ & $c_2$ & $c_2$  & sampling from $\bar{q}_2(\x)$\\ 
			\hline 
			 \multicolumn{8}{c}{{\sl For estimating $Z$}} \\
			\hline 
			 Self-IS & $\pi(\x|\y)$ & $f(\x)$ & $q(\x)$ & $Z$ &  $1$ & $c_3$  &  $\bar{q}(\x)$ \\
			 \hline
			  \multicolumn{8}{c}{{\sl Special cases of Self-IS}} \\ 
			 \hline
			 Naive Monte Carlo & $\pi(\x|\y)$ & $g(\x)$ & $g(\x)$ & \multirow{5}{*}{$Z$} & \multirow{5}{*}{$1$}  & $1$  &  $g(\x)$ \\
			 Harmonic Mean  &  $\pi(\x|\y)$   & $g(\x)$ &$\pi(\x|\y)$ &&  & $Z$  &$\post(\x|\y)$ \\ 
			 RIS   & $\pi(\x|\y)$  & $f(\x)$ & $\pi(\x|\y)$ &  &   & $Z$ & $\post(\x|\y)$  \\ 
			 IS vers-1; Eq. \eqref{StandardIS} &  $\pi(\x|\y)$   &  $\bar{q}(\x)$ & $\bar{q}(\x)$ & &  & $1$  & $\bar{q}(\x)$\\
			  IS vers-2; Eq. \eqref{StandardIS_2}  &  $\pi(\x|\y)$   &  $g(\x)$ & $\bar{q}(\x)$ & &  & $1$  & $\bar{q}(\x)$\\
			\hline
		\end{tabular}
	\end{center}

\end{table}

\begin{table}[!h]	
	 \caption{One-proposal estimators of $Z$}\label{Tabla1}
	 \vspace{-0.6cm}
	\begin{center}
		\begin{tabular}{|c|c| c|  c|c| } 
		\hline 
			 Name & Estimator & Proposal pdf  & Need of MCMC & Unbiased\\ 
			\hline 
			\hline 
			IS vers-1 &  $\frac{1}{N}\sum_{i=1}^N \rho_i \ell({\bf y}|\x_i)$ & Generic, $\bar{q}(\x)$ & --- &  \checkmark\\
			 IS vers-2 &  $\sum_{i=1}^N \bar{\rho}_i \ell({\bf y}|\x_i)$ & Generic, $\bar{q}(\x)$ & no, if $\bar{q}(\x)\neq \post(\x|\y)$ &  ---\\ 
			 Naive MC & $\frac{1}{N}\sum_{i=1}^N  \ell({\bf y}|\x_i)$ & Prior, $g(\x)$ & --- &  \checkmark \\
			 Harmonic mean  &  $\left(\frac{1}{N}\sum_{i=1}^N\frac{1}{\ell(\y|\x_i)}\right)^{-1}$  & Posterior, $\post(\x|\y)$ & \checkmark& --- \\ 
			 RIS   &  $\left(\frac{1}{N}\sum_{i=1}^N\frac{f(\x_i)}{\pi(\x_i|\y)}\right)^{-1} $ & Posterior, $\post(\x|\y)$ & \checkmark & --- \\ 
			  Self-IS     &$\left(\sum_{i=1}^N\frac{f(\x_i)}{q(\x_i)}\right)^{-1}\sum_{i=1}^N\frac{\pi(\x_i|\y)}{q(\x_i)}$ & Generic, $\bar{q}(\x)$ & no, if $\bar{q}(\x)\neq \post(\x|\y)$ & ---  \\ 
			\hline
		\end{tabular}
	\end{center}
\end{table}

In the next section, we discuss a generalization of Eq. \eqref{OnePropIS} for the case where we use samples from both $\bar{q}_1(\x)$ and $\bar{q}_2(\x)$. 

\subsection{Techniques using draws from two proposal densities} \label{TwoDensitiesIS}


In the previous section, we considered estimators of $Z$ that use samples drawn from a single proposal density. 
 More specifically, we have described several IS schemes using a generic pdf $\bar{q}(\x)$ {\it or} $\post(\x|\y)$ as proposal density. In this section,  we introduce schemes where  $\bar{q}(\x)$ {\it and} $\post(\x|\y)$ are employed jointly.
More generally, we consider estimators of a ratio of constants, $\frac{c_2}{c_1}$, that employ samples from two proposal densities,  denoted as $\bar{q}_i(\x)=\frac{q_i(\x)}{c_i}, i=1,2$.  Note that drawing $N_1$ samples from $\bar{q}_1(\x)$ and $N_2$   samples from $\bar{q}_2(\x)$ is equivalent to sampling by a {\it deterministic mixture} approach from the mixture $\bar{q}_{mix}(\x)= \frac{N_1}{N_1+N_2}\bar{q}_1(\x)+\frac{N_2}{N_1+N_2}\bar{q}_2(\x)$, i.e., a single density defined as mixture of two pdfs \cite{ElviraMIS15}. Thus, methods drawing from a mixture of two pdfs as $\bar{q}_{mix}(\x)$, are also considered in this section.    

\subsubsection{Bridge sampling identity } \label{PUENTE_sect}

\begin{figure}[!h]
	\centering
	\includegraphics[width=15cm]{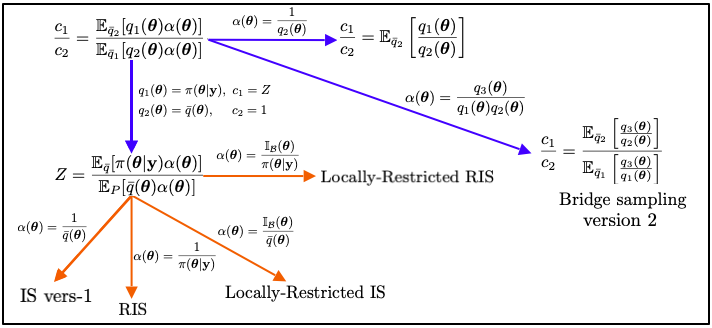}
	\caption{Graphical representation of the relationships among the Eqs. \eqref{OnePropIS} (pre-umbrella identity), \eqref{BridgeSamplingIdentity} (general bridge sampling identity), \eqref{BridgeSamplingIdentityForZ} (bridge sampling for $Z$) and the corresponding different methods, starting from bridge sampling identity  \eqref{BridgeSamplingIdentity}.}
	\label{fig1TEO}
\end{figure}

All the techniques, that we will describe below, are based on the following {\it bridge sampling} identity \cite{meng1996simulating}, 
\begin{align}\label{BridgeSamplingIdentity}
	\frac{c_1}{c_2} = \frac{\E_{\bar{q}_2}[q_1(\x)\alpha(\x)]}{\E_{\bar{q}_1}[q_2(\x)\alpha(\x)]}.
\end{align}
where $\alpha(\x)$ is an arbitrary function defined on the intersection of the supports of $\bar{q}_1$ and $\bar{q}_2$. 
Note that the expression above is an extension of the Eq. \eqref{OnePropIS}. Indeed, taking $\alpha(\x)=\frac{1}{q_2(\x)}$, we recover Eq. \eqref{OnePropIS}. The identity in Eq. \eqref{BridgeSamplingIdentity} and the umbrella identity in Eq. \eqref{UmbrellaSamplingIdentity} are both useful when $\bar{q}_1$ and $\bar{q}_2$ have little overlap, i.e.,  $\int_\Theta \bar{q}_1(\x)\bar{q}_2(\x)d\x$ is small. 
Moreover, If we set $q_1(\x)=\pi(\x|\y)$, $c_1=Z$, $q_2(\x)=\bar{q}(\x)$ and $c_2=1$, then the identity becomes
\begin{align}
\label{BridgeSamplingIdentityForZ}
Z=\frac{\E_{\bar{q}}\left[\pi(\x|\y)\alpha(\x)\right]}{\E_{\post}\left[\bar{q}(\x)\alpha(\x)\right]}.
\end{align}
The corresponding estimator employs samples from both $\bar{q}$ and $\post$, i.e., 
\begin{align}
\label{BridgeSamplingIdentityForZ_esy} 
\widehat{Z}=\frac{ \frac{1}{N_2}\sum_{j=1}^{N_2}  \alpha({\bf z}_j) \pi({\bf z}_j|\y) }{\frac{1}{N_1}\sum_{i=i}^{N_1} \alpha(\x_i) \bar{q}(\x_i)   }, \quad \enskip \{\x_i\}_{i=1}^{N_1}\sim \post(\x|\y),  \enskip \{{\bf z}_j\}_{j=1}^{N_2}\sim \bar{q}(\x).
\end{align}
Figure \ref{fig1TEO} summarizes the connections among the Eqs. \eqref{OnePropIS}, \eqref{BridgeSamplingIdentity}, \eqref{BridgeSamplingIdentityForZ} and the corresponding different methods. The standard IS and RIS schemes have been described in the previous sections, whereas the corresponding {\it locally-restricted} versions will be introduced below. 
\subsubsection{Relationship with Chib's method}\label{ChibBridgeSect}
The Chib estimator, described in Section \ref{ChibEstSect}, is
	\begin{align}\label{ChibAqui}
	\widehat{Z}=   \frac{\pi(\x^*|\y) \frac{1}{N_2}\sum_{j=1}^{N_2} \alpha(\x^*,{\bf v}_j) }{\frac{1}{N_1}\sum_{i=i}^{N_1}\alpha(\x_i,\x^*)\varphi(\x^*|\x_i)   }, \quad \enskip \{\x_i\}_{i=1}^{N_1}\sim \post(\x|\y),  \enskip \{{\bf v}_j\}_{j=1}^{N_2}\sim \varphi(\x|\x^*),
	\end{align}	
where $\varphi(\x|\x^*)$ is the proposal used inside an MCMC algorithm, $\alpha({\bf x},{\bf z}): \mathbb{R}^{D_\theta}\times \mathbb{R}^{D_\theta} \rightarrow \mathbb{R}^+$ represents acceptance probability of this MCMC scheme and the point $\x^*$ is usually chosen in a high probability region. 
Note that the balance condition involving the function $\varphi$, $\post$ and $\alpha$ must be satisfied,
\begin{equation*}
        \alpha(\x,\z)\varphi(\z|\x)\pi(\x|\y) = \alpha(\z,\x)\varphi(\x|\z)\pi(\z|\y).
   \end{equation*}
  Using the balance condition above, if we replace $ \alpha(\x^*,{\bf v}_j)=\frac{\alpha({\bf v}_j,\x^*)\varphi(\x^*|{\bf v}_j)\pi({\bf v}_j|\y)}{\varphi({\bf v}_j|\x^*)\pi(\x^*|\y)}$ inside the numerator of \eqref{ChibAqui}, we obtain
     \begin{align}
\widehat{Z}=  \frac{ \frac{1}{N_2}\sum_{j=1}^{N_2} 	
	\frac{\varphi(\x^*|{\bf v}_j)}{\varphi({\bf v}_j|\x^*)} \alpha({\bf v}_j,\x^*)\pi({\bf v}_j|\y)	
	 }{\frac{1}{N_1}\sum_{i=i}^{N_1} \alpha(\x_i,\x^*) \varphi(\x^*|\x_i)   },  
  \end{align}
and if we also assume a symmetric proposal $\varphi(\x|\x^*)=\varphi(\x^*|\x)$, we can finally write 
    \begin{align}\label{ChibAqui2}
\widehat{Z}=  \frac{ \frac{1}{N_2}\sum_{j=1}^{N_2}  \alpha({\bf v}_j,\x^*)\pi({\bf v}_j|\y)	
	 }{\frac{1}{N_1}\sum_{i=i}^{N_1} \alpha(\x_i,\x^*) \varphi(\x_i|\x^*)  },  \quad \enskip \{\x_i\}_{i=1}^{N_1}\sim \post(\x|\y),  \enskip \{{\bf v}_j\}_{j=1}^{N_2}\sim \varphi(\x|\x^*), 
  \end{align}
We can observe a clear connection between the estimators \eqref{BridgeSamplingIdentityForZ_esy} and \eqref{ChibAqui}. Clearly,  $\varphi(\x|\x^*)$ plays the role of $\bar{q}(\x)$ in \eqref{BridgeSamplingIdentityForZ_esy}, and the acceptance function $\alpha({\bf x},{\bf z})$ plays the role of the $\alpha$ function in \eqref{BridgeSamplingIdentityForZ_esy}.  However, in this case, $\varphi(\x|\x^*)$  participates also {\it inside} the MCMC used for generating $ \{\x_i\}_{i=1}^{N_1}\sim \post(\x|\y)$. The function $\alpha$ takes also part to the generation MCMC chain, $ \{\x_i\}_{i=1}^{N_1}$ (being the acceptance probability of the new states), and generally its evaluation involves the evaluation of $\varphi$ and $\post$. Note also that \eqref{ChibAqui} is more generic than \eqref{ChibAqui2}, being valid also for non-symmetric proposals $\varphi$. For further discussion see \cite{mira2003bridge}. 

\subsubsection{Locally-restricted IS and RIS}
In the literature, there exist variants of the estimators in Eqs. \eqref{NaiveMC}  and \eqref{HarmonicMeanEst}. These corrected estimators are attempts to improve the efficiency (e.g., remove the infinite variance cases, specially in the harmonic estimator) by restricting the integration to a smaller subset of $\Theta$ (usually chosen in high posterior/likelihood-valued regions) generally denoted by $\mathcal{B} \subset \Theta$. As an example,  $\mathcal{B}$
can be a rectangular or ellipsoidal region centered at the {\it maximum a posteriori} (MAP) estimate $\widehat{\x}_\text{MAP}$.
\newline
\newline
{\it Locally-restricted IS estimator.} Consider the posterior mass of subset $\mathcal{B}$ $\subset\Theta$,
\begin{align}
Z_\mathcal{B}=\int_\mathcal{B} \post(\x|\y)d\x = \int_\Theta \mathbb{I}_\mathcal{B}(\x) \frac{\ell(\y|\x)g(\x)}{Z}d\x,
\end{align}
where $ \mathbb{I}_\mathcal{B}(\x)$ is an indicator function, taking value $1$ for $\x \in \mathcal{B}$ and $0$ otherwise. It leads to the following representation
\begin{align}
Z = \frac{1}{Z_\mathcal{B}}\int_\Theta \mathbb{I}_\mathcal{B}(\x) {\ell(\y|\x)g(\x)}d\x=\frac{1}{Z_\mathcal{B}}\E_{\bar{q}}\left[\mathbb{I}_\mathcal{B}(\x)\frac{\ell(\y|\x) g(\x)}{\bar{q}(\x)} \right]. 
\end{align}
We can estimate $Z_\mathcal{B}$  considering $N_1$ samples from $\post(\x|\y)$ by taking the proportion of samples inside $\mathcal{B}$. The resulting locally-restricted IS estimator of $Z$ is
	\begin{align}
	\widehat{Z} = \frac{\frac{1}{N_1}\sum_{i=1}^{N_1}\frac{\mathbb{I}_\mathcal{B}(\z_i)\ell(\y|\z_i)g(\z_i)}{\bar{q}(\z_i)}}{\frac{1}{N_2}\sum_{i=1}^{N_2}\mathbb{I}_\mathcal{B}(\x_i)}, \quad \{\z_i \}_{i=1}^{N_1} \sim \bar{q}(\x), \enskip \{\x_i \}_{i=1}^{N_2} \sim \post(\x|\y) \quad \mbox{(via MCMC)}.	
	\end{align}
Note that the above estimator requires samples from two densities, namely the proposal $\bar{q}(\x)$ and the posterior density $\post(\x|\y)$ (via MCMC).
\newline
\newline
{\it Locally-restricted RIS estimator.} To derive the locally-restricted RIS estimator, consider the mass of $\mathcal{B}$ under $\bar{q}(\x)$,
\begin{align}
\bar{Q}(\mathcal{B})= \int_\mathcal{B} \bar{q}(\x)d\x = Z \cdot \E_{\post}\left[ \mathbb{I}_\mathcal{B}(\x)\frac{\bar{q}(\x)}{\ell(\y|\x)g(\x)} \right],
\end{align}
which leads to the following representation
\begin{align}
Z = \frac{\bar{Q}(\mathcal{B})}{\E_{\post}\left[ \frac{\mathbb{I}_\mathcal{B}(\x)\bar{q}(\x)}{\ell(\y|\x)g(\x)} \right]}.
\end{align}
$\bar{Q}(\mathcal{B})$ can be estimated using a sample from $\bar{q}(\x)$ by taking the proportion of sampled values inside $\mathcal{B}$. The locally-restricted RIS estimator is 
	\begin{align}
	\widehat{Z} = \frac{\frac{1}{N_1}\sum_{i=1}^{N_1}\mathbb{I}_\mathcal{B}(\z_i)}{\frac{1}{N_2}\sum_{i=1}^{N_2} \frac{\mathbb{I}_\mathcal{B}(\x_i)\bar{q}(\x_i)}{\ell(\y|\x_i)g(\x_i)} }, \quad \{\z_i \}_{i=1}^{N_1} \sim \bar{q}(\x), \qquad \{\x_i \}_{i=1}^{N_2} \sim \post(\x|\y).	
	\end{align}
Other variants, where $\mathcal{B}$ corresponds to highest density regions, can be found in \cite{robert2009computational}. 
\subsubsection{Optimal construction of bridge sampling}
 Identities as  \eqref{BridgeSamplingIdentity} are associated  to the bridge sampling approach. However, considering $\alpha(\x)=\frac{q_3(\x)}{q_2(\x) q_1(\x)}$ in Eq.  \eqref{BridgeSamplingIdentity}, bridge sampling can be also motivated from the expression
\begin{align}
\label{BSidentity}
	\frac{c_1}{c_2} = \frac{c_3/c_2}{c_3/c_1}= \frac{\E_{\bar{q}_2}\left[ \frac{q_3(\x)}{q_2(\x)} \right]}{\E_{\bar{q}_1}\left[ \frac{q_3(\x)}{q_1(\x)} \right]},
\end{align}
where the density $\bar{q}_3(\x)\propto q_3(\x)$ is in some sense ``in between'' $q_1(\x)$ and $q_2(\x)$. That is, instead of applying directly \eqref{OnePropIS} to $\frac{c_1}{c_2}$, we apply it to first estimate $\frac{c_3}{c_2}$ and $\frac{c_3}{c_1}$, and then take the ratio to cancel $c_3$.
The bridge sampling estimator of $\frac{c_1}{c_2}$ is then 
\begin{align}
	\frac{c_1}{c_2} \approx \frac{\frac{1}{N_2}\sum_{i=1}^{N_2}\frac{q_3({\bf z}_i)}{q_2({\bf z}_i)}}{\frac{1}{N_1}\sum_{i=1}^{N_1}\frac{q_3({\x}_i)}{q_1({\x}_i)}}, \quad \{{\x}_i \}_{i=1}^{N_1} \sim \bar{q}_1(\x), \quad \{ {\bf z}_i \}_{i=1}^{N_2} \sim \bar{q}_2(\x). 
\end{align}
{\rem We do not need to draw samples from $\bar{q}_3(\x)$, but only evaluate $q_3(\x)$. For a comparison with umbrella sampling see Table \ref{BridgeUmb}. 
}
\begin{table}[!h]	

	 \caption{Joint use of three densities: comparison between bridge and umbrella sampling.}\label{BridgeUmb}
	 \vspace{-0.2cm}
	\begin{center}
		\begin{tabular}{|c||c|c|c||c|} 
		\hline 
			 Method & $\bar{q}_1(\x)$  & $\bar{q}_3(\x)$  & $\bar{q}_2(\x)$ & Identity  \\ 
			\hline 
			\hline 
			 Umbrella sampling & evaluate & draw from & evaluate &   $\frac{c_1}{c_2} = \frac{c_1/c_3}{c_2/c_3}$ - \eqref{UmbrellaSamplingIdentity} \\
			 Bridge  sampling &  draw from & evaluate &draw from &   $\frac{c_1}{c_2} = \frac{c_3/c_2}{c_3/c_1}$ -  \eqref{BSidentity}\\
			\hline
		\end{tabular}
	\end{center}
	
\end{table}

\noindent
{\bf Optimal bridge density.} It can be shown that the optimal bridge density $\bar{q}_3(\x)$ can be expressed as a weighted harmonic mean of $\bar{q}_1(\x)$ and $\bar{q}_2(\x)$ (with weights being the sampling rates), 
\begin{eqnarray}
\bar{q}_3^{\text{opt}}(\x)&=& \frac{1}{\frac{N_2}{N_1+N_2}[\bar{q}_1(\x)]^{-1}+\frac{N_1}{N_1+N_2}[\bar{q}_2(\x)]^{-1}}Ê\nonumber \\
&=&\frac{1}{c_2}\cdot \frac{N_1 + N_2}{N_2\frac{c_1}{c_2}q_1^{-1}(\x)+N_1q^{-1}_2(\x)} \nonumber  \\
&\propto&q_3^{\text{opt}}(\x)= \frac{q_1(\x)q_2(\x)}{N_1q_1(\x) + N_2\frac{c_1}{c_2}q_2(\x)}. \label{eq:Optgamma}
\end{eqnarray}
This is an optimal bridge density if both $N_i$ are strictly positive, $N_i>0$, hence we draw from both $\bar{q}_i(\x)$. Note that $\bar{q}_3^{\text{opt}}(\x)$ depends on the unknown ratio $r=\frac{c_1}{c_2}$. 
Therefore, we cannot even evaluate  $q_3^{\text{opt}}(\x)$.
Hence, we need to resort to the following iterative procedure to approximate the optimal bridge sampling estimator. Noting that 
\begin{equation}
\frac{q_3^{\text{opt}}(\x)}{q_2(\x)}=  \frac{q_1(\x)}{N_1q_1(\x) + r N_2q_2(\x)}, \qquad \frac{q_3^{\text{opt}}(\x)}{q_1(\x)}=  \frac{q_2(\x)}{N_1q_1(\x) + r N_2q_2(\x)}.
\end{equation}
The iterative procedure is formed by the following steps:
\begin{enumerate}
	\item Start with an initial estimate  $\widehat{r}^{(1)} \approx \frac{c_1}{c_2}$ (using e.g. Laplace's).
	\item For $t=1,...,T:$
	\begin{enumerate}
	\item Draw $\{\x_i \}_{i=1}^{N_1}\sim \bar{q}_1(\x)$ and $\{{\bf z}_i \}_{i=1}^{N_2}\sim \bar{q}_2(\x)$ and iterate
	\begin{align}\label{iterOptBridge}
		\widehat{r}^{(t+1)} = \frac{  \frac{1}{N_2}\sum_{i=1}^{N_2}  \dfrac{q_1({\bf z}_i)}{N_1q_1({\bf z}_i) + N_2\widehat{r}^{(t)}q_2({\bf z}_i)} }{\frac{1}{N_1}\sum_{i=1}^{N_1}  \dfrac{q_2(\x_i)}{N_1q_1({\x}_i) + N_2\widehat{r}^{(t)}q_2({\x}_i)}}.
	\end{align}
\end{enumerate}
	\end{enumerate}
	{\rem{
In \cite[Theorem 3.3]{chen1997monte}, the authors show that the asymptotic error of optimal bridge sampling with $\bar{q}_3^\text{opt}$ in Eq. \eqref{eq:Optgamma} is always greater than the asymptotic error of optimal umbrella sampling using $\bar{q}_3^\text{opt}(\x) \propto |\bar{q}_1(\x) - \bar{q}_2(\x)|$ in Eq. \eqref{eq:OptProUmbrella}.}}
\newline
{\bf Optimal bridge sampling for $Z$.} Given the considerations above, an iterative bridge sampling estimator of $Z$ is obtained by setting $q_1(\x)=\pi(\x|\y)$, $c_1=Z$, $\bar{q}_2(\x)=\bar{q}(\x)$, so that  
\begin{align}
\label{IterativeBS}
\widehat{Z}^{(t+1)} = \frac{\frac{1}{N_2}\sum_{i=1}^{N_2}\dfrac{\pi(\z_i|\y)}{N_1\pi(\z_i|\y) + N_2Z^{(t)}\bar{q}(\z_i)}}{\frac{1}{N_1}\sum_{i=1}^{N_1}\dfrac{\bar{q}(\x_i)}{N_1\pi(\x_i|\y) + N_2Z^{(t)}\bar{q}(\x_i)}},\quad \{ \z_i \}_ {i=1}^{N_2} \sim \bar{q}(\x) \enskip \text{and} \enskip \{\x_i\}_{i=1}^{N_1} \sim \post(\x|\y). 
\end{align}
for $t=1,...,T$. Looking at Eqs. \eqref{eq:Optgamma} and \eqref{BSidentity}, when $N_1=0$, that is, when all samples are drawn from $\bar{q}(\x)$, the estimator above reduces to (non-iterative) standard IS scheme with proposal $\bar{q}(\x)$. When $N_2=0$, that is, when all samples are drawn from $\post(\x|\y)$, the estimator becomes the  (non-iterative)  RIS estimator. See \cite{chen1997monte} for a comparison of optimal umbrella sampling, bridge sampling and path sampling (described in the next section). An alternative derivation of the optimal bridge sampling estimator is given in \cite{robert2009computational}, by generating samples from a mixture of type $\psi(\x) \propto  \pi(\x|\y) + \upsilon\bar{q}(\x)$. However, the resulting estimator employs the same samples drawn from $\psi(\x)$  in the numerator and denominator, unlike in Eq. \eqref{IterativeBS}. 

\subsubsection{Other estimators drawing from a generic proposal and the posterior}

Let consider again the scenario where we have a set of samples $\{\x_i\}_{i=1}^{N_1}$ from the posterior $\post(\x|\y)$ and set $\{\z_i\}_{i=1}^{N_2}$ from some proposal $\bar{q}(\x)$, as in the bridge sampling case described above. However, here we consider that these two sets $\{\widetilde{\x}_i\}_{i=1}^{N_1+N_2} = \{\{\x_i\}_{i=1}^{N_1},\{\z_i\}_{i=1}^{N_2}\}$ are drawn from the mixture $\bar{q}_\text{mix}(\x) = \frac{N_1}{N_1+N_2}\post(\x|\y) + \frac{N_2}{N_1+N_2}\bar{q}(\x)$ considering a deterministic mixture sampling approach \cite{ElviraMIS15}, Thus, we can use the IS identities that use a {\it single} proposal, namely Eqs. \eqref{OnePropIS} and \eqref{UmbrellaSamplingIdentity}.
\newline
\newline
{\bf Importance sampling with mixture (M-IS).}  Setting $\bar{q}_1(\x)=\post(\x|\y)$ and $\bar{q}_2(\x) = \bar{q}_\text{mix}(\x)$ in Eq. \eqref{OnePropIS}, we have
\begin{align}
\widehat{Z}_\text{M-IS} =
\frac{1}{N_1+N_2}\sum_{i=1}^{N_1+N_2}\frac{\pi(\widetilde{\x}_i|\y)}{\bar{q}_\text{mix}(\widetilde{\x}_i)}
=\frac{1}{N_1+N_2}\sum_{i=1}^{N_1+N_2}\frac{\pi(\widetilde{\x}_i|\y)}{\frac{N_1}{N_1+N_2}\post(\widetilde{\x}_i|\y) + \frac{N_2}{N_1+N_2}\bar{q}(\widetilde{\x}_i)},
\end{align}
where $\widetilde{\x}_i\sim \bar{q}_\text{mix}(\x) = \frac{N_1}{N_1+N_2}\post(\x|\y) + \frac{N_2}{N_1+N_2}\bar{q}(\x)$ \cite{ElviraMIS15}. This estimator cannot be directly used since it requires the evaluation of $\post(\x|\y)=\frac{1}{Z} \pi(\x|\y)$. From an initial guess $\widetilde{Z}^{(0)}$, the following iterative procedure can be used
\begin{align}\label{EstoLucaIter}
\widehat{Z}^{(t)} =
\frac{1}{N_1+N_2}\sum_{i=1}^{N_1+N_2}\frac{\widehat{Z}^{(t-1)}\pi(\widetilde{\x}_i|\y)}{\frac{N_1}{N_1+N_2}\pi(\widetilde{\x}_i|\y) + \frac{N_2}{N_1+N_2}\widehat{Z}^{(t-1)}\bar{q}(\widetilde{\x}_i)}, \quad t\in \mathbb{N}.
\end{align}
\newline
{\bf Self-IS with mixture proposal (M-Self-IS).} Setting $\bar{q}_1(\x)=\post(\x|\y)$, $\bar{q}_2(\x) = \bar{q}(\x)$ and $\bar{q}_3(\x)=\bar{q}_\text{mix}$ in Eq. \eqref{UmbrellaSamplingIdentity}, we have
\begin{align}
\widehat{Z}_\text{M-Self-IS} =
 \frac{\sum_{i=1}^{N_1+N_2}\frac{\pi(\widetilde{\x}_i|\y)}{\bar{q}_\text{mix}(\widetilde{\x}_i)}}{\sum_{i=1}^{N_1+N_2}\frac{\bar{q}(\widetilde{\x}_i)}{\bar{q}_\text{mix}(\widetilde{\x}_i)}}
 =
\frac{\sum_{i=1}^{N_1+N_2}\frac{\pi(\widetilde{\x}_i|\y)}{\frac{N_1}{N_1+N_2}\post(\widetilde{\x}_i) + \frac{N_2}{N_1+N_2}\bar{q}(\widetilde{\x}_i)}}{\sum_{i=1}^{N_1+N_2}\frac{\bar{q}(\widetilde{\x}_i)}{\frac{N_1}{N_1+N_2}\post(\widetilde{\x}_i|\y) + \frac{N_2}{N_1+N_2}\bar{q}(\widetilde{\x}_i)}},
\end{align}
where $\widetilde{\x}_i\sim \bar{q}_\text{mix} = \frac{N_1}{N_1+N_2}\post(\x|\y) + \frac{N_2}{N_1+N_2}\bar{q}(\x)$ (drawn in a deterministic way). As above, this estimator is not of direct use, so we need to iterate
\begin{align}\label{EstoIterFer}
\widehat{Z}^{(t)} =
\frac{\sum_{i=1}^{N_1+N_2}\frac{\pi(\widetilde{\x}_i|\y)}{\frac{N_1}{N_1+N_2}\pi(\widetilde{\x}_i) + \frac{N_2}{N_1+N_2}\widehat{Z}^{(t-1)}\bar{q}(\widetilde{\x}_i)}}{\sum_{i=1}^{N_1+N_2}\frac{\bar{q}(\widetilde{\x}_i)}{\frac{N_1}{N_1+N_2}\pi(\widetilde{\x}_i|\y) + \frac{N_2}{N_1+N_2}\widehat{Z}^{(t-1)}\bar{q}(\widetilde{\x}_i)}}, \quad t\in \mathbb{N}.
\end{align}
This iterative estimator is very similar to the iterative optimal bridge sampling estimator in Eq. \eqref{IterativeBS}, but it uses both set of samples in numerator and denominator. This estimator is also related to the  reverse logistic regression method in \cite{geyer1994estimating} (for more details see \cite{chen1997monte,cameron2014recursive}, and the next section). 
 Furthermore, the iterative estimator \eqref{EstoIterFer} is also discussed for the case $\bar{q}(\x) = g(\x)$ in \cite{newton1994approximate}, in an attempt to exploit the advantages of the Naive Monte Carlo and the harmonic mean estimators, while removing their drawbacks. 

{\rem{Both iterative versions \eqref{EstoLucaIter}-\eqref{EstoIterFer} converge to the optimal bridge sampling estimator \eqref{IterativeBS}. See \cite{meng1996simulating}, for a related discussion. As we show in the simulation study, the speed of convergence of each iterative method is different. The iterative  bridge sampling estimator seems to be the quickest one.  }}

\subsubsection{Summary}

Several techniques described in the last two subsections, including both umbrella and bridge sampling, are encompassed by the generic formula 
\begin{equation}
\label{SuperGen}
\frac{c_1}{c_2} = \E_{\bar{\xi}}[q_1(\x)\alpha(\x)] \Big/ \E_{\bar{\chi}}[q_2(\x)\alpha(\x)] 
\end{equation}
as shown in Table \ref{TableBridgeUmbrella}. The techniques differ also for which densities are drawn from and which  densities are just evaluated.  

\begin{table}[!h]	
	 \caption{Summary of the IS schemes (with one or two proposal pdfs), using Eq. \eqref{SuperGen}.}\label{TableBridgeUmbrella}
	 \vspace{-0.2cm}
	\begin{center}
	\footnotesize
		\begin{tabular}{|c|c|c|c|c|c|c|c|c|} 
		\hline 
		\multicolumn{9}{|c|}{ } \\
		\multicolumn{9}{|c|}{ $\frac{c_1}{c_2} = \E_{\bar{\xi}}[q_1(\x)\alpha(\x)] \Big/ \E_{\bar{\chi}}[q_2(\x)\alpha(\x)]$}\\
		\multicolumn{9}{|c|}{ } \\ 
		\hline
		        \multicolumn{9}{c}{{\sl For estimating a generic ratio $c1/c2$}} \\
		\hline
			 Name & $\alpha(\x)$ &  $\bar{\xi}(\x)$ & $\bar{\chi}(\x)$ & $q_1(\x)$  &$q_2(\x)$  & $c_1$ & $c_2$ & sampling from \\ 
			\hline 
			\hline
			Bridge Identity - Eq. \eqref{BridgeSamplingIdentity} & $\alpha(\x)$ & $\bar{q}_2(\x)$ &$\bar{q}_1(\x)$  & \multirow{4}{*}{$q_1(\x)$} & \multirow{4}{*}{$q_2(\x)$} &   \multirow{4}{*}{$c_1$} &  \multirow{4}{*}{$c_2$}   &  $\bar{q}_1(\x)$,  $\bar{q}_2(\x)$  \\ 
				Bridge Identity - Eq. \eqref{BSidentity} & $\frac{q_3(\x)}{q_2(\x) q_1(\x)}$ &  $\bar{q}_2(\x)$ & $\bar{q}_1(\x)$  &  &     &  &  &  $\bar{q}_1(\x)$,  $\bar{q}_2(\x)$   \\ 					
	 Identity - Eq. \eqref{OnePropIS} & $\frac{1}{q_2(\x)}$ & $\bar{q}_2(\x)$ & $\bar{q}_1(\x)$ & &   &   &  &  $\bar{q}_2(\x)$   \\	
	 Umbrella - Eq. \eqref{UmbrellaSamplingIdentity} & $\frac{1}{q_3(\x)}$ & $\bar{q}_3(\x)$ & $\bar{q}_3(\x)$ &  &  &  &  &  $\bar{q}_3(\x)$   \\	
	 \hline	
 \multicolumn{9}{c}{{\sl For estimating $Z$, with one proposal}} \\	
 \hline  	 
Self-norm. IS - Eq. \eqref{IS_self}  & $\frac{1}{q_3(\x)}$ & $\bar{q}_3(\x)$ & $\bar{q}_3(\x)$ & $\pi(\x|\y)$ & $f(\x)$  &   \multirow{3}{*}{ $Z$} &  \multirow{3}{*}{$1$} &  $\bar{q}_3(\x)$   \\	
IS vers-1 & $1/\bar{q}(\x)$ & $\bar{q}(\x)$ &  $\post(\x|\y)$  & $\pi(\x|\y)$ & $\bar{q}(\x)$    &  & &  $\bar{q}(\x)$  \\ 	
RIS & $1/\pi(\x|\y)$ & $\bar{q}(\x)$ &  $\post(\x|\y)$  & $\pi(\x|\y)$ & $\bar{q}(\x)$    & &  & $\post(\x|\y)$    \\ 	
 \hline
  \multicolumn{9}{c}{{\sl For estimating $Z$, with two proposals, $\post(\x|\y)$ and  $\bar{q}(\x)$ }} \\	
 \hline 	 
	 Bridge Identity - Eq. \eqref{BridgeSamplingIdentityForZ} & $\alpha(\x)$ & $\bar{q}(\x)$ &  $\post(\x|\y)$  & $\pi(\x|\y)$ & $\bar{q}(\x)$  &  \multirow{3}{*}{$Z$} &  \multirow{3}{*}{$1$}   &  \multirow{3}{*}{$\post(\x|\y)$,  $\bar{q}(\x)$}  \\ 		 					
Locally-Restricted IS & $ \mathbb{I}_\mathcal{B}(\x)/\bar{q}(\x)$ & $\bar{q}(\x)$ &  $\post(\x|\y)$  & $\pi(\x|\y)$ & $\bar{q}(\x)$    &  &  &    \\ 	
Locally-Restricted RIS & $ \mathbb{I}_\mathcal{B}(\x)/\pi(\x|\y)$ & $\bar{q}(\x)$ &  $\post(\x|\y)$  & $\pi(\x|\y)$ & $\bar{q}(\x)$    &  &  &    \\	
			\hline
		\end{tabular}
	\end{center}
\end{table}

\subsection{IS based on multiple proposal densities}\label{ISMultipleProposal}

In this section we consider estimators of $Z$ using samples drawn from more than two proposal densities. These schemes are usually based on the so-called tempering and/or annealing approach.
\newline
\newline
{\bf Reasons for tempering.} The idea is again to consider densities that are in some sense ``in the middle'' between the posterior $\post(\x|\y)$ and an easier-to-work-with density (e.g. the prior $g(\x)$ or some other proposal density). These densities are usually scaled version of the posterior.
Generally, the scale parameter is called {\it temperature}.\footnote{The data tempering is also possible: the tempered posteriors contain less data than the complete posterior.} For this reason, the resulting pdfs are usually named  tempered posteriors and correspond to flatter, more diffuse distributions than the standard posterior.  The use of the tempered pdfs usually improve the mixing of the MCMC algorithms and foster the exploration of the space $\Theta$. Generally, it helps the Monte Carlo methods (as MCMC and IS) to find the regions of posterior high probability.  
The number of such middle densities is specified by the user, and in some cases, it is equivalent to the selection of a temperature schedule for linking the prior $g(\x)$ and $\post(\x|\y)$.  This idea is shared by the several methods, such as path sampling, power posterior methods and stepping-stone sampling described below. 
\newline
\newline
 First of all, we start with a general IS scheme considering different proposals $\bar{q}_n(\x)$'s.  Some of them could be tempered posteriors and the generation would be performed by an MCMC method in this case.  
\subsubsection{Multiple Importance Sampling (MIS) estimators}
 Here, we consider to generate samples from different proposal densities, i.e.,
\begin{equation}
\x_n \sim \bar{q}_n(\x), \qquad n=1,...,N.
\end{equation}
In this scenario, different proper importance weights can be used \cite{ElviraMIS15,HereticalMIS,EfficientMIS}. The most efficient MIS scheme considers the following weights
\begin{equation}
w_n=\frac{\pi(\x_n|\y)}{\frac{1}{N} \sum_{i=1}^N \bar{q}_i(\x_n)}=\frac{\pi(\x_n|\y)}{\psi(\x_n)},
\end{equation}
where $\psi(\x_n)=\frac{1}{N} \sum_{i=1}^N \bar{q}_i(\x_n)$. Indeed, considering the set of samples $\{\x_n\}_{n=1}^N$ drawn in a deterministic order, $\x_n \sim \bar{q}_n(\x)$, and given a sample $\x^*\in \{\x_1,...,\x_N\}$ uniformly chosen in $\{\x_n\}_{n=1}^N$, then we can write $\x^* \sim \psi(\x_n)$. The standard MIS estimator is 
\begin{eqnarray}
\widehat{Z}=\frac{1}{N} \sum_{n=1}^N w_n&=&\frac{1}{N} \sum_{n=1}^N \frac{\pi(\x_n|\y)}{\psi(\x_n)} \\
&=&\frac{1}{N} \sum_{n=1}^N \frac{g(\x_n) \ell({\bf y}|\x_n) }{\psi(\x_n)}, \\
&=&\frac{1}{N} \sum_{n=1}^N  \eta_n  \ell({\bf y}|\x_n), \qquad \x_n \sim \bar{q}_n(\x), \qquad n=1,...,N.
\end{eqnarray}
where $\eta_n =\frac{g(\x_n)  }{\psi(\x_n)}$. The estimator is unbiased \cite{ElviraMIS15}. As in the standard IS scheme, an alternative biased estimator is 
\begin{eqnarray}
\label{SuperMIS}
\widehat{Z}= \sum_{n=1}^N  \bar{\eta}_n  \ell({\bf y}|\x_n),  \qquad \x_n \sim \bar{q}_n(\x), \qquad n=1,...,N,
\end{eqnarray}
where $\bar{\eta}_n=\frac{\eta_n}{\sum_{i=1}^N \eta_i}$, so that $\sum_{i=1}^N \bar{\eta}_i=1$ and we have a convex combination of likelihood values $\ell({\bf y}|\x_n)$'s. It is a generalization of the estimator in Eq. \eqref{StandardIS_2} and recalled below in Eq. \eqref{key}.

\subsubsection{Tempered posteriors as proposal densities}\label{TemperedLuca}

Let  recall  the IS vers-2 estimator of $Z$ in Eq. \eqref{StandardIS_2}, which involves a weighted sum of likelihood evaluations at points $\{\x_i \}_{i=1}^N$ drawn from importance density $\bar{q}(\x)$ (but we can evaluate only $q(\x))\propto \bar{q}(\x)$), 
\begin{align}\label{key} 
\widehat{Z} = \sum_{i=1}^N \bar{\rho}_i\ell(\y|\x_i),\quad \bar{\rho}_i = \frac{\frac{g(\x_i)}{q(\x_i)}}{\sum_{n=1}^N \frac{g(\x_n)}{q(\x_n)}} \propto \frac{g(\x_i)}{q(\x_i)},
\end{align}
where $\sum_{i=1}^N\bar{\rho}_i=1$.
Let us consider
$$
\bar{q}(\x) = \post(\x|\y,\beta)\propto q(\x)=\pi(\x|\y,\beta)= g(\x) \ell(\y|\x)^{\beta},
$$
with $\beta\in[0,1]$. Namely, we use a tempered posterior as importance density.  Note that we can evaluate only the unnormalized density $q(\x)$. The IS estimator version 2 can be employed in this case, and we obtain  $\bar{\rho}_i \propto \frac{g(\x_i)}{g(\x_i)\ell(\y|\x_i)^\beta}=\frac{1}{\ell(\y|\x_i)^\beta}$. The resulting IS estimator version 2 is
\begin{align}
	\widehat{Z} &= \frac{\sum_{i=1}^N\frac{1}{ \ell(\y|\x_i)^{\beta}}\ell(\y|\x_i)}{\sum_{i=1}^N\frac{1}{ \ell(\y|\x_i)^{\beta}}} \\
	&= \frac{\sum_{i=1}^N\ell(\y|\x_i)^{1-\beta}}{\sum_{i=1}^N\ell(\y|\x_i)^{-\beta}} \qquad \{\x_i\}_{i=1}^N \sim \post(\x|\y,\beta) \quad \mbox{(via MCMC)}.
\end{align}
This method is denoted below as IS with a tempered posterior as proposal (IS-P).
Table \ref{PowerPost_as_ProposalPDF_Table} shows that this technique includes different schemes for different values of $\beta$. Different possible MIS schemes can be also considered, i.e., using Eq. \eqref{SuperMIS} for instance \cite{ElviraMIS15,EfficientMIS}.  
\begin{table}[!h]
	\caption{Different estimators of $Z$ using $\bar{q}(\x) \propto g(\x)\ell(\y|\x)^{\beta}$ as importance density, with  $\beta\in[0,1]$.}
	\label{PowerPost_as_ProposalPDF_Table}
	\begin{center}
	\footnotesize
		\begin{tabular}{|c|c|c|c|}  
			\hline
		Name	& Coefficient $\beta$ &  Weights $\bar{\rho}_i$ & Estimator $\widehat{Z} = \sum_{i=1}^N \bar{\rho}_i \ell(\y|\x_i)$ \\
			\hline
			\hline
		Naive Monte Carlo & $\beta=0$ & $\frac{1}{N}$ &$ \frac{1}{N}\sum_{i=1}^N\ell(\y|\x_i)$ \\
			\hline
		 Harmonic Mean Estimator & $\beta=1$ & $\frac{\frac{1}{\ell(\y|\x_i)}}{\sum_j \frac{1}{\ell(\y|\x_j)}}$ &$\widehat{Z} = \frac{1}{\frac{1}{N}\sum_{i=1}^N\frac{1}{\ell(\y|\x_i)}} $ \\ 
			\hline
		Power posterior	& & & \\
		as proposal pdf & $0<\beta<1$ & $\frac{\frac{1}{\ell(\y|\x_i)^\beta}}{\sum_j \frac{1}{\ell(\y|\x_j)^\beta}}$ & $\widehat{Z} = \frac{\sum_i \ell(\y|\x_i)^{1-\beta}}{\sum_i \ell(\y|\x_i)^{-\beta}}$ \\
			\hline
		\end{tabular}
	\end{center}
\end{table}

{\rem One could consider also  to draw samples from $N$ different tempered posteriors, $ \x_n \sim \post(\x|\y,\beta_n)\propto g(\x)\ell(\y|\x)^{\beta_n}$, with $n=1,...,N$, and then apply deterministic mixture idea in \eqref{SuperMIS}. However, in this case, we cannot evaluate properly the mixture 
$$
\psi(\x_n)=\frac{1}{N} \sum_{i=1}^N  \post(\x_n|\y,\beta_i)= \frac{1}{N} \sum_{i=1}^N  \frac{1}{Z(\beta_i)}\pi(\x_n|\y,\beta_i).
$$ 
Here, the issue is a not just a global unknown normalizing constant (as usual): in this case, we do not know the weights of the mixture since  all $Z(\beta)=\int_\Theta g(\x)\ell(\y|\x)^{\beta}d\x$ are unknown. This problem can be solved using the techniques described in the next sections. }
\newline
\newline
{\bf Reverse logistic regression (RLR).} In RLR, the idea is to apply IS with the mixture $\psi(\x_n)$ in the remark above. The normalizing constants $Z(\beta_i)$ are iteratively obtained by  maximizing of a suitable log-likelihood, built with the samples from each tempered posterior $\post(\x|\y,\beta_n)$ \cite{geyer1994estimating,liu2015estimating,cameron2014recursive}.
\newline
\newline
In the next section, we describe an alternative to RLR for employing different tempered posteriors as proposals.

\subsubsection{Stepping-stone (SS) sampling} \label{SSsect}

\begin{figure}[!h]
	\centering
	\includegraphics[width=10cm]{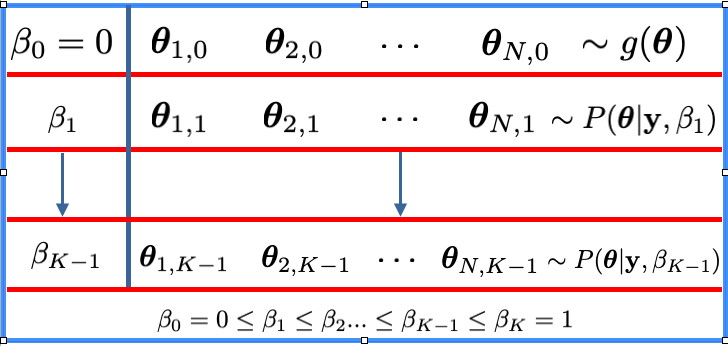}
	\caption{Sampling procedure in the SS method.  Note that samples from $\post(\x|\y)$ ($\beta_K=1$) are not considered. It is relevant to compare this figure with Figures \ref{FigAnIS}-\ref{FigSMC} in the next section. }
	\label{fig_stepping}
\end{figure}

Consider again $\post(\x|\y,\beta)\propto g(\x) \ell(\y|\x)^{\beta}$ and $Z(\beta)=\int_\Theta g(\x)\ell(\y|\x)^{\beta}d\x$.
The goal is to estimate $Z=\frac{Z(1)}{Z(0)}$, which can be expressed as the following product, with $\beta_0=0$ and $\beta_K=1$,
\begin{align}
Z=\frac{Z(1)}{Z(0)} = \prod_{k=1}^K \frac{Z(\beta_k)}{Z(\beta_{k-1})},
\end{align}
where $\beta_k$ are often chosen as $\beta_k=\frac{k}{K}$, $k=1,\dots,K$, i.e., with a uniform grid in $[0,1]$. Note that generally $Z(0)=1$, since it is normalizing constant of the prior. The SS method is based on the following identity,
\begin{align*}
\E_{\post(\x|y,\beta_{k-1})}\left[\dfrac{\pi(\x|\y,\beta_{k})}{\pi(\x|\y,\beta_{k-1})}\right]&=\int_{\Theta}\dfrac{\pi(\x|\y,\beta_{k})}{\pi(\x|\y,\beta_{k-1})} \post(\x|y,\beta_{k-1}) d\x, \\
&=\frac{1}{Z(\beta_{k-1})}\int_{\Theta}\pi(\x|\y,\beta_{k}) d\x =\frac{Z(\beta_k)}{Z(\beta_{k-1})}.
\end{align*}
Then, the idea of SS sampling is to estimate each ratio $r_k=\frac{Z(\beta_k)}{Z(\beta_{k-1})}$ by importance sampling as
\begin{align}
r_k& =\frac{Z(\beta_k)}{Z(\beta_{k-1})} =	\E_{\post(\x|{\bf y},\beta_{k-1})}\left[\dfrac{\pi(\x|\y,\beta_{k})}{\pi(\x|\y,\beta_{k-1})}\right] \label{MandaFinlandes_Cojonen}\\
&=\E_{\post(\x|{\bf y},\beta_{k-1})}\left[\dfrac{\ell(\y|\x)^{\beta_k}}{\ell(\y|\x)^{\beta_{k-1}}}\right]\\
&\approx \widehat{r}_k = \dfrac{1}{N}\sum_{i=1}^N\ell(\y|\x_{i,k-1})^{\beta_k-\beta_{k-1}}, \quad \{\x_{i,k-1} \}_{i=1}^N \sim \post(\x|\y,\beta_{k-1}).
\end{align}
Multiplying all ratio estimates yields the final estimator of $Z$
\begin{align}\label{Z_SS}
\widehat{Z} = \prod_{k=1}^K\widehat{r}_k= \prod_{k=1}^K \left( \frac{1}{N}\sum_{i=1}^N\ell(\y|\x_{i,k-1})^{\beta_k-\beta_{k-1}}\right), \quad \{\x_{i,k-1} \}_{i=1}^N \sim \post(\x|\y,\beta_{k-1}).
\end{align}
For $K=1$, we come back to the Naive MC estimator.
 The sampling procedure of the SS method is graphically represented in Figure \ref{fig_stepping}.
{\rem{The SS estimator is unbiased, since it a product of  unbiased estimators. }      }
\newline 
\newline 
The two following methods, path sampling and power posteriors, estimate $\log Z$ instead of $Z$.

\subsubsection{Path sampling (a.k.a., thermodynamic integration)}\label{PSsect}

More specifically, the method of path sampling for estimating $\frac{c_1}{c_2}$ relies on the idea of building and drawing samples from a sequence of distributions linking $\bar{q}_1(\x)$ and $\bar{q}_2(\x)$ (a continuous path). 
 For the purpose of estimating only one constant, the marginal likelihood $Z$, we set $\bar{q}_2(\x)=g(\x)$ and $\bar{q}_1(\x) = \post(\x|\y)$ and we link them by a univariate path with parameter $\beta$. Let 
\begin{align}
	\pi(\x|\y,\beta), \enskip \beta \in [0,1],
\end{align}
denote a sequence of (probably unnormalized except for $\beta=0$) densities such $\pi(\x|\y,\beta=0)=g(\x)$ and $\pi(\x|\y,\beta=1)=\pi(\x|\y)$. More generally, we could consider  $\pi(\x|\y,\beta=0)=\bar{q}(\x)$ where $\bar{q}(\x)$ is a generic normalized proposal density, possibly closer to the posterior than $g(\x)$.    
The path sampling method for estimating the marginal likelihood is based on expressing $\log Z$ as 
\begin{align}\label{PathSamplingIdentity}
\log Z  = \E_{p(\x,\beta|\y)}\left[\frac{U(\x,\beta)}{p(\beta)}\right],\quad \text{with} \enskip U(\x,\beta) = \dfrac{\partial}{\partial\beta}\log\pi(\x|\y,\beta), 
\end{align}
where the expectation is w.r.t. the joint $p(\x,\beta|\y)=\frac{1}{Z(\beta)}\pi(\x|\y,\beta)p(\beta)$, being $Z(\beta)$ the normalizing constant of $\pi(\x|\y,\beta)$ and $p(\beta)$ represents a density for $\beta \in [0,1]$. Indeed, we have 
\begin{align}\label{PathSamplingIdentity2}
\E_{p(\x,\beta{|\y})}\left[\frac{U(\x,\beta)}{p(\beta)}\right] &= \int_{\Theta}\int_{0}^1 \frac{1}{p(\beta)} \left[\frac{\partial}{\partial\beta}\log\pi(\x|\y,\beta)\right] \frac{\pi(\x|\y,\beta)}{Z(\beta)}p(\beta) d\x d\beta,Ê\nonumber \\
&= \int_{\Theta}\int_{0}^1 \frac{1}{\pi(\x|\y,\beta)} \left[\frac{\partial}{\partial\beta}\pi(\x|\y,\beta)\right] \frac{\pi(\x|\y,\beta)}{Z(\beta)} d\x d\beta, \nonumber \\
&= \int_{\Theta}\int_{0}^1\frac{1}{Z(\beta)}\dfrac{\partial}{\partial\beta}\pi(\x|\y,\beta) d\x d\beta, \nonumber \\
&= \int_{0}^1\frac{1}{Z(\beta)}\dfrac{\partial}{\partial\beta}\left(\int_{\Theta}\pi(\x|\y,\beta) d\x\right)d\beta, \nonumber \\
&=
\int_{0}^1\frac{1}{Z(\beta)}\dfrac{\partial}{\partial\beta}Z(\beta)d\beta \nonumber \\
&=
\int_{0}^1\dfrac{\partial}{\partial\beta}\log Z(\beta)d\beta= \log Z(1) - \log Z(0) = \log Z,
\end{align}
where we substituted $Z(\beta=1) = Z(1) = Z$ and $Z(\beta=0)=Z(0)=1$. Thus, using a sample $\{\x_i,\beta_i\}_{i=1}^N  \sim p(\x,\beta|\y)$, we can write the path sampling estimator for $\log Z$ 
\begin{align}\label{PathSampEst} 
\widehat{\log Z} =  \dfrac{1}{N}\sum_{i=1}^N\dfrac{U(\x_i,\beta_i)}{p(\beta_i)}, \quad \{\x_i,\beta_i \}_{i=1}^N \sim p(\x,\beta{|\y}).
\end{align}
The samples from $p(\x,\beta|\y)$ may be obtained by first drawing $\beta'$($\beta_i$) from $p(\beta)$ 
and then applying some MCMC steps to draw from $\post(\x|\y,\beta')$$\propto\pi(\x|\y,\beta')$ given $\beta'$. Therefore, in path sampling, we have to choose (a) the path and (b) and the prior $p(\beta)$. A discussion regarding the optimal choices of the path and $p(\beta)$, see \cite{gelman1998simulating}. The optimal path for linking any two given densities is impractical as it depends on the normalizing constants being estimated. The  geometric path described below, although suboptimal, is generic and simple to implement.
\newline
\newline
{\bf Geometric path.} Often a geometric path is employed,
\begin{align}Ê
	\pi(\x|\y,\beta) &= g(\x)^{1-\beta} \pi(\x|\y)^{\beta} \nonumber \\
	&= g(\x) \ell(\y|\x)^{\beta}, \enskip \beta\in[0,1]. \label{yamete_yamete}
\end{align}
Note that $\pi(\x|\y,\beta)$ is the posterior with a powered, ``less informative'' -``wider'' likelihood (for this reason, $\pi(\x|\y,\beta)$ is often called a ``power posterior''). 
In this case, we have
\begin{align*}
	 U(\x,\beta) &= \dfrac{\partial}{\partial\beta}\log\pi(\x|\y,\beta)=\log \ell(\y|\x),
\end{align*}
so the path sampling identity becomes
\begin{align}\label{PathSamplingAndPowerPosteriors}
	\log Z = \E_{p(\x,\beta{|\y})}\left[\frac{\log\ell(\y|\x)}{p(\beta)}\right],
\end{align}
which is also used in the power posterior method of \cite{friel2008marginal}, described in Section \ref{SectionPP}. 

\subsubsection{Connections among path sampling, bridge sampling and stepping-stones} \label{SectpathbridgeSS}
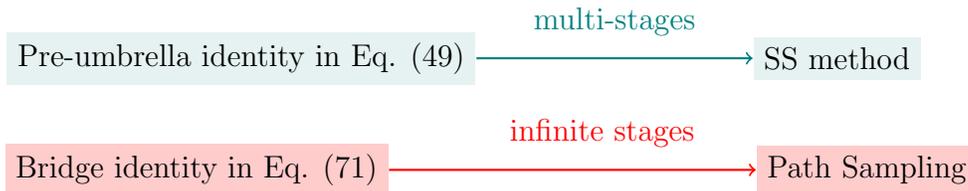
\begin{figure}[h!]
\begin{center}
\begin{tikzpicture}[->,scale=0.99,transform shape,node distance=3cm,thick]
  \tikzstyle{every state}=[fill=teal!10,draw=none,text=black]
  \node (RET) [fill=teal!10] {Pre-umbrella identity in Eq. \eqref{OnePropIS}};
   \node (SS)  [right of=RET,fill=teal!10,yshift=0cm,xshift=5cm] {SS method};
  \node (RTM)  [below of=RET,fill=red!20,yshift=+1.5cm,xshift=-0.6cm] {Bridge identity in Eq. \eqref{BSidentity}};
 \node (OBS)  [right of=RTM,fill=red!20,yshift=0cm,xshift=6cm] {Path Sampling};
  \path[solid, bend left=0, thick, color=red] (RTM)  edge node[yshift=+0.5cm,xshift=0.4cm] {infinite stages} (OBS);
    \path[solid, bend left=0, thick, color=teal] (RET)  edge node[yshift=+0.5cm,xshift=0cm] {multi-stages} (SS);
\end{tikzpicture}
\caption{Relationships among the identities \eqref{OnePropIS}-\eqref{BSidentity} and their multi-stages extensions: the SS method and path sampling scheme, respectively.}
\label{fig:RelSSpathBridge}
\end{center}
\end{figure}

The path sampling method can be motivated from bridge sampling by applying the bridge sampling identity in \eqref{BSidentity} in a chain fashion. Assume we have $K+1$ densities $\post(\x|\y,\beta_k)=\pi(\x|\y,\beta_k)/Z(\beta_k),\ k=0,\dots,K$ from which we can draw samples, with endpoints $\post(\x|\y,\beta_0=0)=g(\x)$ and $\post(\x|\y,\beta_K=1)=\post(\x|\y)$. 
 We can express $Z = Z(\beta_K)=Z(1)$ as follows
	\begin{align}
	Z = \prod_{k=1}^K \frac{Z(\beta_k)}{Z(\beta_{k-1})} = \prod_{k=1}^K\frac{\mathbb{E}_{\post(\x|\y,\beta_{k-1})}\left[\frac{\pi(\x|\y,\beta_{k-\frac{1}{2}})}{\pi(\x|\y,\beta_k)}\right]}{\mathbb{E}_{\post(\x|\y,\beta_{k})}\left[\frac{\pi(\x|\y,\beta_{k-\frac{1}{2}})}{\pi(\x|\y,\beta_{k})}\right]}.
	\end{align}
	Note that we have applied the bridge sampling identity in Eq. \eqref{BSidentity} to each ratio $\frac{Z(\beta_k)}{Z(\beta_{k-1})}$, using $K-1$ middle densities $\pi(\x|\y,\beta_{k-\frac{1}{2}})$. We can approximate the $k$-th term by using samples from $\post(\x|\y,\beta_{k-1})$ and $\post(\x|\y,\beta_{k})$, and take the product to obtain the final estimator of $Z$. Taking the logarithm of the above expression, as $K\to\infty$, results in the basic identity of path sampling for estimating $Z$ in Eq. \eqref{PathSamplingIdentity} \cite{gelman1998simulating}. 
	In this sense, path sampling can be interpreted as a continuous application of bridge sampling steps. 
	The difference with SS method is that it employs another identity, in \eqref{MandaFinlandes_Cojonen}, for estimating the ratios $\frac{Z(\beta_{k})}{Z(\beta_{k-1})}$.  Figure \ref{fig:RelSSpathBridge} summarizes the relationships among the identities \eqref{OnePropIS}-\eqref{BSidentity} and their multi-stages extensions: the SS method and path sampling scheme, respectively.


\subsubsection{Method of Power Posteriors }\label{SectionPP}

The previous expression  \eqref{PathSamplingAndPowerPosteriors} can also be  converted into an integral in $[0,1]$ as follows
\begin{align}
	\log Z &= \E_{p(\x,\beta{|\y})}\left[\frac{\log\ell(\y|\x)}{p(\beta)}\right], \nonumber\\
	&= \int_0^1d\beta   \int_\Theta\frac{\log\ell(\y|\x)}{p(\beta)} \frac{\pi(\x|\y,\beta)}{Z(\beta)}  p(\beta) d\x,   \nonumber \\
	&= \int_0^1d\beta   \int_\Theta\log \ell(\y|\x) \frac{\pi(\x|\y,\beta)}{Z(\beta)} d\x,   \nonumber \\
	&= \int_0^1 \E_{\post(\x|\y,\beta)}\left[ \log \ell(\y|\x)\right]d\beta, \label{PowerOneDim}
\end{align}
where $\post(\x|\y,\beta) = \frac{\pi(\x|\y,\beta)}{Z(\beta)}$ is a power posterior. 
The power posterior method aims at estimating the integral above by applying a quadrature rule. For instance, choosing a discretization $0=\beta_0<\beta_1<\dots<\beta_{K-1}<\beta_K=1$, leads to approximations of order 0, 
\begin{align}\label{PowerPostTrapRule0}  
\widehat{\log Z} = \sum_{k=1}^{K}(\beta_{k}-\beta_{k-1})\E_{\post(\x|\y,\beta_{k-1})}\left[\log \ell(\y|\x)\right],
\end{align}
or order 1 (trapezoidal rule),
\begin{align}\label{PowerPostTrapRule}
\widehat{\log Z} = \sum_{k=1}^{K}(\beta_{k}-\beta_{k-1})
\frac{\E_{\post(\x|\y,\beta_{k})}\left[\log \ell(\y|\x)\right]+\E_{\post(\x|\y,\beta_{k-1})}\left[\log \ell(\y|\x)\right]}{2},
\end{align}
where the expected values w.r.t. the power posteriors can be independently approximated via MCMC,
\begin{align}\label{PowerPostEstimExp}
\E_{\post(\x|\y,\beta_k)}\left[\log \ell(\y|\x)\right] \approx \frac{1}{N}\sum_{i=1}^N \log \ell(\y|\x_{i,k}),\quad \{ \x_{i,k} \}_{i=1}^N \sim \post(\x|\y,\beta_k),\quad k=0,\dots,K.
\end{align}
{\rem{ The identity \eqref{PowerOneDim} of method of power posteriors is derived by the path sampling  identity with a geometric path, as shown in \eqref{yamete_yamete}-\eqref{PathSamplingAndPowerPosteriors}. In this sense, the method of power posteriors is a special case of path sampling. However, unlike in path sampling,  the final approximation \eqref{PowerPostTrapRule} is based on a deterministic quadrature.  
}}
{
\rem{
Note that the approximation in Eq. \eqref{PowerPostTrapRule} is biased due to using a deterministic quadrature, unlike the  path sampling approximation in Eq. \eqref{PathSampEst} which is unbiased.} 
}
{
\rem{The need of using several values $\beta_i$ (i.e., several tempered posteriors) seems apparent in the estimator \eqref{PathSampEst}-\eqref{PowerPostTrapRule}. For instance, in \eqref{PowerPostTrapRule}, the choice a small value of $K$ yields a poor approximation of the integral \eqref{PowerOneDim}.  This is not the case in the SS method.
}
} 
\newline
\newline
{\bf Extensions.} Several improvements of the method of power posterior have been proposed in the literature \cite{friel2014improving,oates2016controlled}. In \cite{friel2014improving}, the authors note that the derivative of the integrand in \eqref{PowerOneDim} corresponds to 
\begin{align}
\frac{d}{d\beta}\E_{\post(\x|\y,\beta)}[\log\ell(\y|\x)] = 	\mbox{var}_{P(\x|\y,\beta)}[\log\ell(\y|\x)]	
\end{align}
so they propose to use this information to refine the trapezoidal rule in \eqref{PowerPostTrapRule} by adding additional terms
\begin{align}
\widehat{\log Z} = \sum_{k=1}^{K}(\beta_{k}-\beta_{k-1})
\frac{\E_{\post(\x|\y,\beta_{k})}\left[\log \ell(\y|\x)\right]+\E_{\post(\x|\y,\beta_{k-1})}\left[\log \ell(\y|\x)\right]}{2} - \\ \sum_{k=1}^{K}\frac{(\beta_{k}-\beta_{k-1})^2}{12}\left[
\mbox{var}_{P(\x|\y,\beta_{k})}[\log\ell(\y|\x)]- \mbox{var}_{P(\x|\y,\beta_{k-1})}[\log\ell(\y|\x)]
\right], \label{PowerPostTrapRuleMejorada}
\end{align}
This improvement comes at no extra cost since the same MCMC samples, used to estimate the expectations in \eqref{PowerPostEstimExp}, can be also used to estimate the variances in \eqref{PowerPostTrapRuleMejorada}. They also propose constructing the temperature ladder recursively, starting from $\beta_0=0$ and $\beta_K=1$, by leveraging the estimates of 
$\E_{\post(\x|\y,\beta_{k})}\left[\log \ell(\y|\x)\right]$ and $\mbox{var}_{P(\x|\y,\beta_{k})}[\log\ell(\y|\x)]$ (for further details see \cite[Sect. 2.2]{friel2014improving}).
In \cite{oates2016controlled}, they propose the use of control variates, a variance reduction technique, in order to improve the statistical efficiency of the estimator \eqref{PowerPostTrapRule}. However, this can only be applied in settings where $\nabla_{\x}\log P(\x|\y,\beta)$ is available.

\subsubsection{On the selection of $\beta_k$ }

The method of power posteriors and SS sampling require setting an increasing sequence of $\beta$'s.
Some strategies for selecting the sequence of values $\beta_k$'s, with $\beta_0=0$ and $\beta_K=1$, are discussed, e.g., in  \cite{friel2008marginal, friel2014improving, xie2010improving}. A uniform sequence $\beta_k = \frac{k}{K}$ for $k=0,\dots,K$ can be considered, although \cite{friel2008marginal} recommends putting more values near $\beta=0$, since it is where $\post(\x|\y,\beta)$ is changing more rapidly.  More generally, we can consider $\beta_k = (\frac{k}{K})^{1/\alpha}$. For choice of $\alpha \in [0,1]$, the values $\beta_k$ are evenly-spaced quantiles of a Beta($\alpha$,1), concentrating more and more near $\beta=0$ as $\alpha$ decreases to 0 \cite{xie2010improving}.

The path sampling method requires defining a prior density $p(\beta)$ from which samples are drawn. It can be shown that, for any given path, the optimal choice of $p(\beta)$ is a generalized local Jeffreys prior \cite[Sect. 4.1]{gelman1998simulating}.  

\subsubsection{Connection between stepping-stone and power posteriors methods}
Taking the logarithm of the SS estimator \eqref{Z_SS}, we obtain
\begin{align*}
 \log\widehat{Z}_{\text{SS}} =  \sum_{k=1}^K \log \left( \frac{1}{N}\sum_{i=1}^N\ell(\y|\x_{i{,k-1}})^{\beta_k-\beta_{k-1}}\right). 
\end{align*}
Applying the Jensen inequality and property of the logarithm, we can write 
\begin{align*}
 \log\widehat{Z}_{\text{SS}} &\geq  \sum_{k=1}^K \left( \frac{1}{N}\sum_{i=1}^N  \log\ell(\y|\x_{i,k-1})^{\beta_k-\beta_{k-1}}\right),\\
  &\geq  \sum_{k=1}^K  (\beta_k-\beta_{k-1}) \left( \frac{1}{N}\sum_{i=1}^N  \log\ell(\y|\x_{i,k-1})\right).
\end{align*}
The last expression is the estimator of the power posteriors method of order 0, i.e., replacing Eq. \eqref{PowerPostEstimExp} into \eqref{PowerPostTrapRule0}. If we denote here this estimator here as $\widehat{\log Z}_{PP}$, then we have
$\log\widehat{Z}_{\text{SS}} \geq  \widehat{\log Z}_{PP}$. Recall also the SS estimator is unbiased.

%

\section{Advanced schemes combining MCMC and IS} \label{CombinationISandMCMC}


%
%
%
%

In the previous sections, we have already introduced several methods which require the use of MCMC algorithms in order to draw from complex proposal densities. The RIS estimator, path sampling, power posteriors and the SS sampling schemes are some examples. All these previous schemes could be assigned to the family of ``MCMC-within-IS'' techniques. In this section, we describe more sophisticated schemes for estimating the evidence, which combine MCMC and IS techniques:  Annealed Importance Sampling (An-IS) in Section \ref{An-ISSect0}, Sequential Monte Carlo (SMC) in Section \ref{SMCsect}, Multiple Try Metropolis (MTM) in Section \ref{MTMsect}, and Layered Adaptive importance Sampling (LAIS) in Section \ref{LAISsect}. An-IS and SMC can be also considered ``MCMC-within-IS'' techniques.
They provide alternative ways to employ tempered posteriors and are related to SS method, described in the previous section. 
We also discuss the use of MCMC transitions and resampling steps for design efficient AIS schemes.  The MTM algorithm described here is an MCMC method, which belongs to the  family of ``IS-within-MCMC'' techniques. Indeed, internal IS steps are used for proposing good candidates as new state of the chain. LAIS is an AIS scheme driven by MCMC transitions. Since the the adaptation and sampling parts can be completely separated,  LAIS can be considered as a  ``IS-after-MCMC'' technique.

\subsection{MCMC-within-IS: weighted samples after MCMC iterations}\label{An-ISSect0}


In this section, we will see how to {\it properly} weight samples obtained by different MCMC iterations.  We denote as  $K({\bf z}|\x)$ the  transition kernel which summarizes all the steps of the employed MCMC algorithm. Note that  generally  $K({\bf z}|\x)$ cannot be evaluated. However, we can use MCMC kernels $K({\bf z}|\x)$ in the same fashion as proposal densities, considering the concept of the so-called {\it proper weighting} \cite{Liu04b,GIS18}.

\subsubsection{Weighting a sample after one MCMC iteration}
\label{weightafterMCMCsect}

Let us consider the following procedure:
\begin{enumerate}
\item Draw $\x_0\sim q(\x)$ (where $q(\x)$ is normalized, for simplicity). 
\item Draw $\x_1\sim K(\x_1|\x_0)$, where the kernel $K$ leaves invariant density  ${\bar \eta}(\x)=\frac{1}{c}\eta(\x)$, i.e., 
\begin{equation}
\int_{\Theta} K(\x'|\x) {\bar \eta}(\x) d\x={\bar \eta}(\x').
\end{equation}
\item Assign to $\x_1$ the weight
\begin{equation}
\rho(\x_0,\x_1)=\frac{\eta(\x_0)}{q(\x_0)}\frac{\pi(\x_1|\y)}{\eta(\x_1)}.
\end{equation}
\end{enumerate}
This weight is {\it proper} in the sense that can be used for building unbiased estimator $Z$ (or other moments $\post(\x|\y)$), as described in the Liu's definition \cite[Section 14.2]{Robert04}, \cite[Section 2.5.4]{Liu04b}.
Indeed, we can write
\begin{eqnarray}
\mathbb{E}[\rho(\x_0,\x_1) ]&=&\int_{\Theta}\int_{\Theta}  \rho(\x_0,\x_1) K(\x_1|\x_0) q(\x_0) d\x_0 d\x_1, \nonumber \\
&=&\int_{\Theta}\int_{\Theta}  \frac{\eta(\x_0)}{q(\x_0)}\frac{\pi(\x_1)}{\eta(\x_1)} K(\x_1|\x_0) q(\x_0) d\x_0d\x_1,  \nonumber \\
&=&\int_{\Theta} \frac{\pi(\x_1)}{\eta(\x_1)}\left[ \int_{\Theta} \eta(\x_0) K(\x_1|\x_0) d\x_0 \right] d\x_1,  \nonumber  \\
&=&\int_{\Theta} \frac{\pi(\x_1)}{c\bar{\eta}(\x_1)} c\bar{\eta}(\x_1) d\x_1=\int_{\Theta}  \pi(\x_1|\y) d\x_1=Z. 
\end{eqnarray}
Note that if $\eta(\x)\equiv \pi(\x|\y)$ then $\rho(\x_1)=\frac{\pi(\x_0|\y)}{q(\x_0)}$, i.e., the IS weights remain unchanged after an MCMC iteration with invariant density $\pi(\x|\y)$. Hence, if we repeat the procedure above $N$ times generating $\{\x_0^{(n)},\x_1^{(n)}\}_{n=1}^{N}$, we can build the following unbiased estimator of the $Z$,
\begin{equation}
\widehat{Z}=\frac{1}{N}\sum_{n=1}^N \rho(\x_0^{(n)},\x_1^{(n)})=\frac{1}{N}\sum_{n=1}^N \frac{\eta(\x_0^{(n)})}{q(\x_0^{(n)})}\frac{\pi(\x_1^{(n)}|\y)}{\eta(\x_1^{(n)})}
\end{equation}
 In the next section, we extend this idea where different MCMC updates are applied, each one addressing a different invariant density.

\subsubsection{Annealed Importance Sampling (An-IS)}\label{An-ISSect}

\begin{figure}[!h]
	\centering
	\includegraphics[width=11cm]{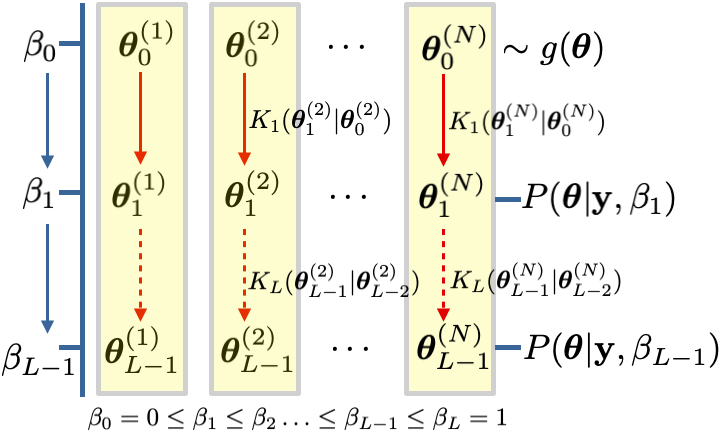}
	\caption{Sampling procedure in the An-IS method.} 
	\label{FigAnIS}
\end{figure}

In the previous section, we have considered the application of one MCMC kernel $K(\x_1|\x_0)$ (that could be formed by different MCMC steps). Below, we consider the application of several MCMC kernels addressing different target pdfs, and show their consequence in the weighting strategy. We consider again a sequence of tempered versions of the posterior, $\pi_1(\x|\y), \pi_2(\x|\y)$, $\ldots$, $\pi_L(\x|\y)\equiv \pi(\x|\y)$,
where the $L$-th version, $\pi_L(\x|\y)$,  coincides with the target function $\pi(\x|\y)$. One possibility is to considered  $\pi_i(\x|\y)=\left[\pi(\x|\y)\right]^{\beta_i}=g(\x)^{\beta_i} \ell({\bf y}|\x)^{\beta_i}$ or tempered posteriors, 
\begin{eqnarray}
\label{aquiTemperedEq2}
 \pi_i(\x|\y)= g(\x)\ell({\bf y}|\x)^{\beta_i}\quad \mbox{ where } \quad 0\leq \beta_1\leq \beta_2\leq \ldots \leq \beta_L=1.
\end{eqnarray}
as in path sampling and power posteriors. In any case, smaller $\beta$ values correspond to flatter distributions.\footnote{Another alternative is to use the so-called {\it data tempering} \cite{Chopin02}, for instance, setting
$ \pi_i(\x|\y)\propto p(\x|y_1,\ldots,y_{d+i})$, 
where $d\geq 1$ and $d+L=D_y$ (recall that ${\bf y}=[y_1,\ldots,y_{D_y}]\in \mathbb{R}^{D_y}$). } 
The use of the tempered sequence of target pdfs usually improve the mixing of the algorithm and foster the exploration of the space $\Theta$.
Since only the last function is the true target, $\pi_L(\x|\y)=\pi(\x|\y)$, different schemes have been proposed for suitable weighting the final samples. 
\newline
Let us consider conditional $L-1$ kernels $K_i({\bf z}|\x)$ (with $L\geq 2$), representing the probability of different MCMC updates of jumping from the state $\x$ to the state ${\bf z}$ (note that each $K_i$ can summarize the application of several MCMC steps), each one leaving invariant a different tempered target,  $\post_i(\x|\y)\propto \pi_i(\x|\y)$. 
The Annealed Importance Sampling (An-IS) is given in Table \ref{An-ISTable}.
 \begin{table}[!h]
\caption{\normalsize Annealed Importance Sampling (An-IS)}
\label{An-ISTable}
\begin{tabular}{|p{0.95\columnwidth}|}
   \hline
   \vspace{-0.7cm}
\begin{enumerate}
\item Draw $N$ samples $\x_0^{(n)}\sim \post_0(\x|\y)$ (usually $g(\x)$)   for $n=1,...,N$.
\item For $k=1,\ldots,L-1:$ 
\begin{enumerate}
\item  Draw $\x_{k}^{(n)}\sim K_{k}(\x|\x_{k-1}^{(n)})$ leaving invariant $\post_k(\x|\y)$ for $n=1,...,N$, i.e., we generate  $N$ samples using an MCMC with  invariant  distribution $\post_k(\x|\y)$ (with different starting points $\x_{k-1}^{(n)}$).
\item  Compute the weight associated to the sample $\x_{k}^{(n)}$, for $n=1,...,N$,
\begin{eqnarray}
\rho_{k}^{(n)}=\prod_{i=0}^{k}\frac{\pi_{i+1}(\x_{i}^{(n)}|\y)}{\pi_{i}(\x_i^{(n)}|\y)}=\rho_{k-1}^{(n)} \frac{\pi_{k+1}(\x_{k}^{(n)}|\y)}{\pi_{k}(\x_k^{(n)}|\y)}.
\end{eqnarray}
\end{enumerate}  
\item Return the weighted sample $\{\x_{L-1}^{(n)}, \rho_{L-1}^{(n)}\}_{n=1}^N$. The estimator of the marginal likelihood is 
$$
\widehat{Z}=\frac{1}{N}\sum_{n=1}^N \rho_{L-1}^{(n)}.
$$
 Combinations of An-IS with path sampling and power posterior methods can be also considered, employing the information of the rest of intermediate densities. \vspace{-0.3cm}
\end{enumerate}  \\
\hline 
\end{tabular}
\end{table}
Note that, when $L=2$, we have $\rho_{1}^{(n)}=\frac{\pi_1(\x_0^{(n)}|\y)}{q(\x_0^{(n)})}\frac{\pi(\x_1^{(n)}|\y)}{\pi_1(\x_1^{(n)}|\y)}$. If, $\pi_1=\pi_2 =\ldots =\pi_{L-1}=\eta \neq \pi$, then the weight is $\rho_{L-1}=\frac{\eta(\x_0^{(n)})}{\post_0(\x_0^{(n)}|\y)}\frac{\pi(\x_{L-1}^{(n)}|\y)}{\eta(\x_{L-1}^{(n)})}$. 
\newline
\newline
The method above can be modified by incorporating an additional MCMC transition $\x_{L}\sim K_{L}(\x|\x_{L-1})$, which leaves invariant $\post_L(\x|\y)=\post(\x|\y)$. However, since $\post_L(\x|\y)$ is the true target pdf, as we have seen above the weight remains unchanged (see the case $\bar{\eta}(\x)=\post(\x|\y)$ in the previous section). Hence, in this scenario, the output would be  $\{\x_{L}^{(n)}, \rho_{L}^{(n)}\}=\{\x_{L}^{(n)}, \rho_{L-1}^{(n)}\}$, i.e., $\rho_{L}^{(n)}=\rho_{L-1}^{(n)}$.  This method has been proposed in \cite{Neal01} but similarly schemes can be found in \cite{Chopin02, GilksBerzuini01}.
\newline
\newline
\begin{rem}
	The stepping-stones (SS) sampling method described in Section \ref{ISMultipleProposal} is strictly connected to an Ann-IS scheme. 
	 See Figures \ref{fig_stepping} and \ref{FigAnIS} for a comparison of the sampling procedures.
\end{rem}

\paragraph{Interpretation as Standard IS.} For the sake of simplicity, here we consider  {\it reversible} kernels, i.e., each kernel satisfies the detailed balance condition
\begin{eqnarray}
\pi_i(\x|\y) K_i({\bf z}|\x)=\pi_i(\z|\y) K_i(\x|{\bf z}) \quad \mbox{so that} \quad \frac{K_i({\bf z}|\x)}{K_i(\x|{\bf z})}=\frac{\pi_i(\z|\y)}{\pi_i(\x|\y)}.
\end{eqnarray}
We show that the weighting strategy suggested by An-IS can be interpreted as a standard IS weighting considering the following extended target density, defined in the extended space $\Theta^L$,
\begin{eqnarray}
\pi_g(\x_0,\x_1,\ldots,\x_{L-1}|\y) 
=\pi(\x_{L-1}|\y)\prod_{k=1}^{L-1} K_{k}(\x_{k-1}|\x_{k}).
\end{eqnarray}
Note that $\pi_g$ has the true target $\pi$ as a marginal pdf. Let also consider an extended proposal pdf defined as
\begin{eqnarray}
q_g(\x_0,\x_1,\ldots,\x_{L-1})  
=\post_0(\x_{0}|\y)\prod_{k=1}^{L-1} K_{k}(\x_{k}|\x_{k-1}).
\end{eqnarray}
The standard IS weight of an extended sample $[\x_0,\x_1,\ldots,\x_{L-1}]$ in the extended space $\Theta^L$ is
\begin{eqnarray}
w(\x_0,\x_1,\ldots,\x_{L-1})=\frac{\pi_g(\x_0,\x_1,\ldots,\x_{L-1}|\y)}{q_g(\x_0,\x_1,\ldots,\x_{L-1})}=\frac{\pi(\x_{L-1}|\y)\prod_{k=1}^{L-1} K_{k}(\x_{k-1}|\x_{k})}{\post_0(\x_{0}|\y)\prod_{k=1}^{L-1} K_{k}(\x_{k}|\x_{k-1})}. \label{W_An-IS}
\end{eqnarray}
Replacing the expression  $\frac{K_i({\bf z}|\x)}{K_i(\x|{\bf z})}=\frac{\pi_i(\z|\y)}{\pi_i(\x|\y)}$ in \eqref{W_An-IS}, we obtain the Ann-IS weights 
\begin{eqnarray}
w(\x_0,\x_1,\ldots,\x_{L-1})&=&\frac{\pi(\x_{L-1}|\y)}{\post_0(\x_{0}|\y)} \prod_{k=1}^{L-1} \frac{\pi_k(\x_{k-1}|\y)}{\pi_k(\x_{k}|\y)}, \\
&=&\frac{\pi_1(\x_{0}|\y)}{\post_0(\x_{0}|\y)} \prod_{k=1}^{L-1} \frac{\pi_{k+1}(\x_{k}|\y)}{\pi_k(\x_{k}|\y)}= \prod_{k=0}^{L-1} \frac{\pi_{k+1}(\x_{k}|\y)}{\pi_k(\x_{k}|\y)}=\rho_{L-1}, \label{W_An-IS2} 
\end{eqnarray}
where we have used $\pi_L(\x|\y)=\pi(\x|\y)$ and just rearranged the numerator.
The sampling procedure in An-IS is graphically represented in Figure \ref{FigAnIS}. 


 \subsection{Weighted samples after MCMC and resampling steps}\label{SMCsect}
In this section, we consider also the use of resampling steps jointly with MCMC transitions. The resulting algorithm is quite sophisticated (formed by several components that should by chosen by the user) but it is a very general technique, which includes the classical particle filters, several adaptive IS (AIS) schemes and the An-IS method as special case \cite{Moral06}.

\subsubsection{Generic Sequential Monte Carlo}
In this section, we describe a sequential IS scheme which encompasses the previous Ann-IS algorithm as a special case.
The method described here uses jointly MCMC transitions and, additionally, resampling steps as well.  It is called Sequential Monte Carlo (SMC), since we have a sequence of target pdfs $\pi_k(\x|\y)$, $k=1,\ldots,L$  \cite{Moral06}. This sequence of target densities can be defined by a state-space model as in a classical particle filtering framework (truly sequential scenario, where the goal is to track dynamic parameters).  Alternatively, we can also consider  a static scenario as in the previous sections, i.e., the resulting algorithm is an iterative importance sampler where we consider a  sequence of {\it tempered} densities $\pi_k(\x|\y)= g(\x)\ell({\bf y}|\x)^{\beta_k}$,  where  $0\leq \beta_1\leq \ldots \leq \beta_L=1$,  as in Eq.\eqref{aquiTemperedEq2}, so that $\pi_L(\x|\y)=\pi(\x|\y)$  \cite{Moral06}. Let us again define an extended proposal density in the domain $\Theta^k$,
\begin{equation}
\widetilde{q}_k(\x_1,\dots, \x_k)=q_1(\x_1) \prod_{i=2}^k F_i(\x_{i}|\x_{i-1}): \mbox{ } \mbox{ } \Theta^k\rightarrow \mathbb{R},
\end{equation}
where $q_1(\x_1)$ is a marginal proposal and $F_i(\x_{i}|\x_{i-1})$ are generic forward transition pdfs, that will be used as partial proposal pdfs. Extending the space from $\Theta^k$ to $\Theta^{k+1}$ (increasing its dimension), note that we can write the recursive equation  
$$
\widetilde{q}_{k+1}(\x_1,\dots, \x_k,\x_{k+1})=
 F_{k+1}(\x_{k+1}|\x_{k})\widetilde{q}_k(\x_1,\dots, \x_k):  \mbox{ } \mbox{ }   \Theta^{k+1}\rightarrow \mathbb{R}.
$$
 The marginal proposal pdfs are
\begin{eqnarray}
q_k(\x_k)&=& \int_{\Theta^{k-1}} \widetilde{q}_k(\x_1,\dots, \x_k) d\x_{1:k-1} \nonumber\\
&=& \int_{\Theta^{k-1}} q_1(\x_1) \prod_{i=2}^k F_i(\x_{i}|\x_{i-1}) d\x_{1:k-1},  \label{FirstQkEq}\\
&=&  \int_{\Theta} \left[\int_{\Theta^{k-2}}  q_1(\x_1) \prod_{i=2}^k F_i(\x_{i}|\x_{i-1}) d\x_{1:k-2} \right]  F_{k}(\x_{k}|\x_{k-1}) d\x_{k-1} , \nonumber \\
&=& \int_{\Theta} q_{k-1}(\x_{k-1})  F_{k}(\x_{k}|\x_{k-1}) d\x_{k-1},
\end{eqnarray} 
Therefore, we would be interested in computing the {\it marginal} IS weights, $w_{k}=\frac{\pi_k(\x_k|\y)}{q_k(\x_k)}$, for each $k$. However note that, in general, the marginal proposal pdfs $q_k(\x_k)$ cannot be computed and then cannot be evaluated. A suitable alternative approach is described next. Let us consider the extended target  pdf defined as
\begin{equation}
\widetilde{\pi}_k(\x_1,\dots, \x_k|\y)=\pi_k(\x_k|\y) \prod_{i=2}^k B_{i-1}(\x_{i-1}|\x_{i}): \mbox{ } \mbox{ } \Theta^k\rightarrow \mathbb{R},
\end{equation}
$B_{i-1}(\x_{i-1}|\x_{i})$ are arbitrary backward transition pdfs. Note that the space of $\{\widetilde{\pi}_k\}$ increases as $k$ grows, and $\pi_k$ is always a marginal pdf of $\widetilde{\pi}_k$. Moreover, writing the previous equation for $k+1$  
$$
\widetilde{\pi}_{k+1}(\x_1,\dots, \x_k,\x_{k+1}|\y)=\pi_{k+1}(\x_{k+1}|\y) \prod_{i=2}^{k+1} B_{i-1}(\x_{i-1}|\x_{i}),
$$
and writing the ratio of both, we get 
\begin{equation}
\frac{\widetilde{\pi}_{k+1}(\x_1,\dots, \x_k,\x_{k+1}|\y)}{\widetilde{\pi}_k(\x_1,\dots, \x_k|\y)}=\frac{\pi_{k+1}(\x_{k+1}|\y)}{\pi_k(\x_k|\y)}B_{k}(\x_{k}|\x_{k+1}).
\end{equation}
Therefore, the IS weights in the extended space $\Theta^k$ are
\begin{eqnarray}
w_k&=&\frac{\widetilde{\pi}_{k}(\x_1,\dots, \x_k|\y)}{\widetilde{q}_k(\x_1,\dots, \x_k)} \\
&=&\frac{\widetilde{\pi}_{k-1}(\x_1,\dots, \x_{k-1}|\y)}{\widetilde{q}_{k-1}(\x_1,\dots, \x_{k-1})}  \frac{\frac{\pi_{k}(\x_{k}|\y)}{\pi_{k-1}(\x_{k-1}|\y)}B_{k-1}(\x_{k-1}|\x_{k})}{F_{k}(\x_{k}|\x_{k-1})},  \\
&=& w_{k-1}  \frac{\pi_{k}(\x_{k}|\y)B_{k-1}(\x_{k-1}|\x_{k})}{\pi_{k-1}(\x_{k-1}|\y)F_{k}(\x_{k}|\x_{k-1})} \label{IMP_weightRec}.
\end{eqnarray}
where we have replaced $w_{k-1}=\frac{\widetilde{\pi}_{k-1}(\x_1,\dots, \x_{k-1}|\y)}{\widetilde{q}_{k-1}(\x_1,\dots, \x_{k-1})}$. The recursive formula in Eq. \eqref{IMP_weightRec} is the key expression for several sequential IS techniques. The SMC scheme summarized in Table \ref{SMCTable} is a general framework which contains different algorithms as a special cases  \cite{Moral06}. In Table \ref{SMCTable}, we have used the notation $\x_{1:k}=[\x_1,...,\x_k]$.

 \begin{table}[!h]
\caption{\normalsize Generic Sequential Monte Carlo (SMC)}
\label{SMCTable}
\begin{tabular}{|p{0.95\columnwidth}|}
   \hline
   \vspace{-0.7cm}
\begin{enumerate}
\item Draw $\x_1^{(n)} \sim q_1(\x)$, $n=1,\ldots,N$. 
\item For $k=2,\ldots,L:$
\begin{enumerate}
\item Draw $N$ samples $\x_{k}^{(n)} \sim F_k(\x|\x_{k-1}^{(n)})$. 
\item Compute the weights
\begin{eqnarray}
\label{WrecSIR}
w_k^{(n)}&=& w_{k-1}^{(n)}  \frac{\pi_{k}(\x_{k}^{(n)}|\y)B_{k-1}(\x_{k-1}^{(n)}|\x_{k}^{(n)})}{\pi_{k-1}(\x_{k-1}^{(n)}|\y)F_{k}(\x_{k}|\x_{k-1}^{(n)})}, \\
&=& w_{k-1}^{(n)}  \gamma_{k}^{(n)}, \qquad, k=1,\ldots, L,
\end{eqnarray}
 where we set $ \gamma_{k}^{(n)}=\frac{\pi_{k}(\x_{k}^{(n)}|\y)B_{k-1}(\x_{k-1}^{(n)}|\x_{k}^{(n)})}{\pi_{k-1}(\x_{k-1}^{(n)}|\y)F_{k}(\x_{k}|\x_{k-1}^{(n)})}$.
\item  Normalize the weights $\bar{w}_k^{(n)}=\frac{w_k^{(n)}}{\sum_{j=1}^N w_k^{(j)}}$, for $n=1,...,N$.
\item If $\widehat{ESS}\leq \epsilon N$:
\newline
{\small (with $0\leq \epsilon \leq 1$ and  $\widehat{ESS}$ is a effective sample size measure \cite{ESSarxiv16}, see section \ref{MLinSeqScenario}) }
\begin{enumerate}
\item Resample $N$ times $\{\x_{1:k}^{(1)},\ldots,\x_{1:k}^{(N)}\}$ according to $\{\bar{w}_k^{(n)}\}_{n=1}^N$, obtaining $\{{\bar \x}_{1:k}^{(1)},\ldots,{\bar \x}_{1:k}^{(N)}\}$. 
\item Set $\x_{1:k}^{(n)}={\bar \x}_{1:k}^{(n)}$,  $\widehat{Z}_k=\frac{1}{N}\sum_{n=1}^N w_k^{(n)}$ and $w_{k}^{(n)}=\widehat{Z}_k$ for all $n=1,\ldots,N$  \cite{GISssp16,GIS18,NSMC,Martino15PF}.
\end{enumerate}
 \end{enumerate}
\item Return the cloud of weighted particles and 
$$\widehat{Z}=\widehat{Z}_L=\frac{1}{N}\sum_{n=1}^N w_L^{(n)},$$
 if a proper  weighting of the resampled particles is used (as suggested in the step 2(d)-ii above).  Otherwise, you can use another estimator  $\widehat{Z}_L$, as shown in Section \ref{MLinSeqScenario} and the Supplementary Material.   \vspace{-0.3cm}
\end{enumerate}  \\ 
\hline 
\end{tabular}
\end{table}


\paragraph{Choice of the forward functions.}  
One  possible choice is to use independent proposal pdfs, i.e., $F_k(\x_k|\x_{k-1})=F_k(\x_k)$ or random walk proposal $F_k(\x_k|\x_{k-1})$, where $F_k$ represents standard distributions (e.g., Gaussian or t-Student). An alternative is to choose $F_k(\x_k|\x_{k-1})=K_k(\x_k|\x_{k-1})$, i.e., an MCMC kernel with invariant pdf $\post_k(\x_k|\y)$.

\paragraph{Choice of backward functions.} It is possible to show that  the optimal  backward transitions $\{B_k\}_{k=1}^L$ are \cite{Moral06}
\begin{equation}
B_{k-1}(\x_{k-1}|\x_{k}) =\frac{q_{k-1}(\x_{k-1})}{q_{k}(\x_{k})} F_k(\x_{k}|\x_{k-1}).
\end{equation}
This choice reduces the variance of the weights \cite{Moral06}.
However, generally, the marginal proposal $q_{k}$ in  Eq. \eqref{FirstQkEq} cannot be computed (are not available), other possible $\{B_k\}$ should be considered.
  For instance, with the choice
\begin{equation}\label{HereBAn}
B_{k-1}(\x_{k-1}|\x_{k}) =\frac{\pi_{k}(\x_{k-1}|\y)}{\pi_{k}(\x_{k}|\y)} F_k(\x_{k}|\x_{k-1}),
\end{equation}
we obtain 
\begin{eqnarray}
w_k &=& w_{k-1}   \frac{\pi_{k}(\x_{k}|\y)\frac{\pi_{k}(\x_{k-1}|\y)}{\pi_{k}(\x_{k}|\y)} F_k(\x_{k}|\x_{k-1}) }{\pi_{k-1}(\x_{k-1}|\y)F_{k}(\x_{k}|\x_{k-1})}  \\
&=& w_{k-1}  \frac{\pi_{k}(\x_{k-1}|\y)}{\pi_{k-1}(\x_{k-1}|\y)},
\end{eqnarray}
which is exactly the update rule for the weights in An-IS. 

\begin{figure}[!h]
	\centering
	\includegraphics[width=10cm]{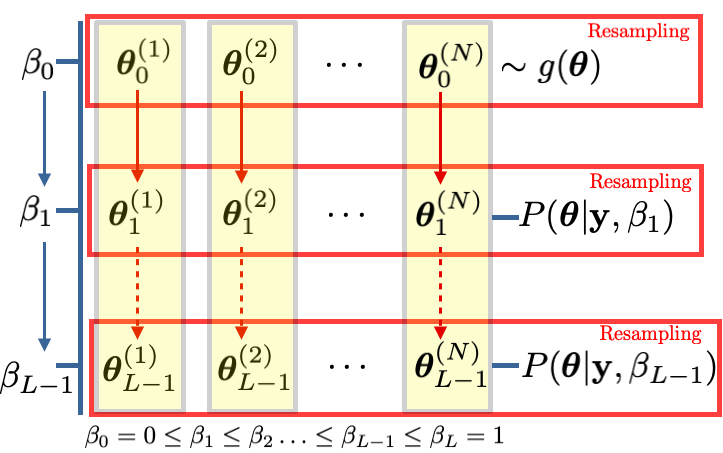}
	\caption{Sampling procedure in SMC. In this figure, we have considered resampling steps at each iteration $(\epsilon=1$).} 
	\label{FigSMC}
\end{figure}

\begin{rem}
 With the choice of  $B_{k-1}(\x_{k-1}|\x_{k})$ as in Eq. \ref{HereBAn}, and if $F_k(\x_{k}|\x_{k-1})=K_k(\x_{k}|\x_{k-1})$ is an MCMC kernel with invariant $\post_k(\x_k|\y)$, then we come back to An-IS algorithm \cite{Neal01,Chopin02, GilksBerzuini01}, described in  Table \ref{An-ISTable}. Hence,  the An-IS scheme is a special case of SMC method. 
\end{rem}
\noindent
 Several other methods are contained as special cases of algorithm in Table \ref{SMCTable}, with specific choice of $\{B_k\}$, $\{K_k\}$ and $\{\pi_{k}\}$, e.g., the Population Monte Carlo (PMC) method \cite{Cappe04}, that is a well-known AIS scheme. The sampling procedure in SMC is graphically represented in Figure \ref{FigSMC}.

\subsubsection{Evidence computation in a sequential framework with resampling steps} \label{MLinSeqScenario}
The generic algorithm in Table \ref{SMCTable} employs also resampling steps. Resampling consists in drawing particles from the current cloud according to the normalized importance weights $\bar{w}_k^{(n)}$, for $n=1,....,N$.  The resampling steps are applied only in certain iterations taking into account an ESS approximation, such as 
$\widehat{ESS}=\frac{1}{\sum_{n=1}^N (\bar{w}_k^{(n)})^2}$, or $\widehat{ESS}=\frac{1}{\max_n \bar{w}_k^{(n)}}$  \cite{Kong92,ESSarxiv16}. Generally,%
 if $\frac{1}{N}\widehat{ESS}$ is smaller than a pre-established threshold $\epsilon \in[0,1]$, all the particles are resampled. Thus, the condition for the adaptive resampling can be expressed as $\widehat{\textrm{ESS}} < \epsilon N$. When $\epsilon=1$, the resampling is applied at each iteration \cite{Djuric03,Doucet08tut}.
If $\epsilon= 0$, no resampling steps are applied, and we have a simple sequential importance sampling (SIS) method. There are two possible estimators of $Z_k$ in a sequential scenario: 
\begin{align}
\widehat{Z}_{k}^{(1)}&=\frac{1}{N} \sum\limits_{n=1}^Nw_{k}^{(n)}=\frac{1}{N} \sum\limits_{n=1}^N w_{k-1}^{(n)}  \gamma_{k}^{(n)}  =\frac{1}{N} \sum\limits_{n=1}^N \left[\prod_{j=1}^{k} \gamma_{j}^{(n)}\right],
\end{align}
and
\begin{equation}
\widehat{Z}_{k}^{(2)}=\prod_{j=1}^{k}\left[\sum_{n=1}^N\bar{w}_{j-1}^{(n)}\gamma_j^{(n)}\right].
\end{equation}
These two estimators are equivalent in SIS ($\epsilon= 0$, i.e., SMC without resampling), i.e., they are the same estimator, $\widehat{Z}_{k}^{(1)}=\widehat{Z}_{k}^{(2)}$. In SMC with $\epsilon>0$ and a proper weighting of the resampled particles, as used in Table \ref{SMCTable}, the two estimators are equivalent as well  \cite{GISssp16,GIS18,Martino15PF}.  If the proper weighting of the resampled particles  is not employed,  $\widehat{Z}_{k}^{(2)}$ is the only valid option.  See Table \ref{ProperWeightingTabla} for a summary and the Supp. Material for more details.

\begin{table}[!h]
	\caption{Possible estimators of the evidence in a sequential scenario.}
	\label{ProperWeightingTabla}
	\vspace{-0.3cm}
	\begin{center}
	\footnotesize
		\begin{tabular}{|c|c|c|c|c|c|}  
			\hline
		           {\bf Scenario} & {\bf Resampling } & {\bf Proper Weighting} \cite{GISssp16} & $\widehat{Z}_{k}^{(1)}$ &  $\widehat{Z}_{k}^{(2)}$ & {\bf Equivalence}  \\	        
			\hline
			\hline
			SMC - $\epsilon=0$ (SIS) &  x  & --- &   \checkmark & \checkmark &   \checkmark  \\
			\hline
			SMC - $\epsilon>0$ &   \checkmark   & x &   x & \checkmark &   x  \\
			\hline
			SMC - $\epsilon>0$ &   \checkmark   & \checkmark &   \checkmark & \checkmark &   \checkmark  \\
			\hline
		\end{tabular}
	\end{center}
\end{table}

\begin{figure}[!h]
	\centering
	\includegraphics[width=11cm]{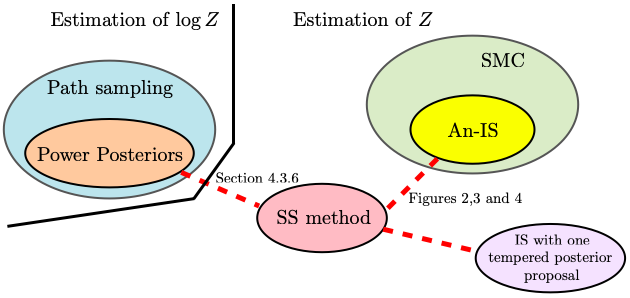}
	\caption{Graphical summary of the methods using tempered posteriors.} 
	\label{FigTotalRes}
\end{figure}

\subsection{IS-within-MCMC: Estimation based on Multiple Try MCMC  schemes} \label{MTMsect}
The Multiple Try Metropolis (MTM) methods are advanced MCMC algorithms which consider different candidates as possible new state of the chain \cite{MARTINO_REV_MTM,LucaJesse2,LucaJesse1}. More specifically,  at each iteration different samples are generated and compared by using some proper weights. Then one of them is selected and tested as possible future state. The main advantage of these algorithms is that they foster the exploration of a larger portion of the sample space, decreasing the correlation among the states of the generated chain. Here, we consider the use of importance weights for comparing the different candidates, in order to provide also an estimation of  the marginal likelihood \cite{LucaJesse2}.
More specifically, we consider the Independent Multiple Try Metropolis type 2 (IMTM-2) scheme \cite{MARTINO_REV_MTM} with an adaptive proposal pdf. 
The algorithm is given in Table \ref{AIMTM_Table}.
The mean vector  and covariance matrix are adapted using the empirical estimators yielded by all the weighted candidates drawn so far, i.e., $\{{\bf z}_{n,\tau},w_{n,\tau}\}$ for all $n=1,...,N$ and $\tau=1,...,T$. Two possible estimators of  the marginal likelihood can be constructed, one based on a standard adaptive importance sampling argument $\widehat{Z}^{(2)}$ \cite{AIS_SIG_PRO, Bugallo15} and other based on a group importance sampling idea provided in \cite{GIS18}.
\newline
For the sake of simplicity, we have described an independent MTM scheme, with the additional adaptation of the proposal. Random walk proposal pdfs can be also employed in an MTM algorithm \cite{MARTINO_REV_MTM}. In that case, the adaptation of the proposal could be not needed. However, in this scenario, the MTM algorithm requires the sampling (and weighting) of $N-1$ additional auxiliary points. Hence, the total number of weighted samples at each iterations are $2N-1$. These additional samples are just required for ensuring the ergodicity of the chain (including them in the acceptance probability $\alpha$), but are not included as states of the Markov chain. But, for our purpose, they can be employed in the estimators of $Z$, as we suggest for the $N$ candidates, $\{{\bf z}_{n,t},w_{n,t}\}$, in Table \ref{AIMTM_Table}. Note that the use of a random walk proposal in an MTM scheme of type in Table \ref{AIMTM_Table}, could be considered as ``MCMC-driven IS'' method, similar to the method introduced in the next section.   

 \begin{table}[!h]
\caption{\normalsize Adaptive Independent Multiple Try Metropolis type 2 (AIMTM-2)}
\label{AIMTM_Table}
\begin{tabular}{|p{0.95\columnwidth}|}
   \hline
   \vspace{-0.7cm}
\begin{enumerate}
\item Choose the initial parameters ${\bm \mu}_t$, ${\bf C}_t$ of the proposal $q$, an initial state $\x_0$ and a first estimation of the marginal likelihood $\widehat{Z}_0$.
\item For $t=1,...,T$:
\begin{enumerate}
\item Draw ${\bf z}_{1,t},....,{\bf z}_{N,t} \sim q({\bf z}|{\bm \mu}_t,{\bf C}_t)$.
\item Compute the importance weights 
$w_{n,t}=\frac{\pi({\bf z}_{n,t}|\y)}{q({\bf z}_{n,t}|{\bm \mu}_t,{\bf C}_t)}$, for  $n=1,...,N$. 
\item Normalize them ${\bar w}_{n,t}=\frac{w_{n,t}}{N \widehat{Z}'}$ where 
\begin{equation}
\widehat{Z}'=\frac{1}{N}\sum_{i=1}^N w_{i,t}, \quad \mbox{ and  set } \quad R_t=\widehat{Z}'. 
\end{equation}
\item Resample $\x' \in \{{\bf z}_{1,t},....,{\bf z}_{N,t} \}$ according to  ${\bar w}_n$, with $n=1,...,N$.
\item Set $\x_t=\x'$ and $\widehat{Z}_t=\widehat{Z}'$ with probability
\begin{eqnarray}
\alpha=\min\left[1, \frac{\widehat{Z}'}{\widehat{Z}_{t-1}}\right]
\end{eqnarray}
otherwise set $\x_t=\x_{t-1}$ and $\widehat{Z}_t=\widehat{Z}_{t-1}$.  
\item Update ${\bm \mu}_t,$ ${\bf C}_t$ computing the corresponding empirical estimators using $\{{\bf z}_{n,\tau},w_{n,\tau}\}$ for all $n=1,...,N$ and $\tau=1,...,T$. 
\end{enumerate}
\item Return the chain $\{\x_{t}\}_{t=1}^T$, $\{\widehat{Z}_t\}_{t=1}^T$ and $\{R_t\}_{t=1}^T$. Two possible estimators of $Z$ can be constructed: 
\begin{equation}
\widehat{Z}^{(1)}=\frac{1}{T} \sum_{t=1}^T \widehat{Z}_t, \qquad  \widehat{Z}^{(2)}=\frac{1}{T} \sum_{t=1}^T R_t
.\end{equation}

\end{enumerate} \\
\hline 
\end{tabular}
\end{table}

\subsection{IS-after-MCMC: Layered Adaptive Importance Sampling (LAIS)}\label{LAISsect}
The LAIS algorithm consider the use of $N$ parallel (independent or interacting) MCMC chains with invariant pdf $\post(\x|\y)$ or a tempered version $\post(\x|\beta)$ \cite{LAIS17,AIS_SIG_PRO}.  Each MCMC chain can address a different tempered version $\post(\x|\y,\beta)$ (or simply the posterior $\post(\x|\y)$) without jeopardizing the consistency of final estimators. After $T$ iterations of the $N$ MCMC schemes (upper layer), the resulting $NT$ samples, $\{{\bm \mu}_{n,t}\}$, for $n=1,...,N$ and $t=1,...,T$ are used as location parameters of $NT$  proposal densities  $q(\x|{\bm \mu}_{n,t}, {\bf C})$. Then, these proposal pdfs are employed within a MIS scheme (lower layer), weighting the generated samples $\x_{n,t}$'s with the generic weight $w_{n,t}=\frac{\pi(\x_{n,t}|\y)}{\Phi(\x_{n,t})}$ \cite{ElviraMIS15,EfficientMIS}. In the numerator of these weights in the lower layer,  we have always the unnormalized posterior $\pi(\x_{n,t}|\y)$.   The denominator $\Phi(\x_{n,t})$ is a mixture of (all or a subset of) proposal densities  which specifies the type of MIS scheme applied \cite{ElviraMIS15,EfficientMIS}.  The algorithm, with different possible choices of  $\Phi(\x_{n,t})$, is  shown in Table \ref{LAISTable}. The first choice in \eqref{PHI1} is the most costly since we have to evaluate all the proposal pdfs in all the generated samples $\x_{n,t}$'s, but provides the best  performance in terms of efficiency of the final estimator. The second and third choices are temporal and spatial mixtures, respectively. The last choice corresponds to standard importance weights given in Section \ref{ImportanceSamplingApproaches}.

 \begin{table}[!h]
\caption{\normalsize Layered Adaptive Importance Sampling (LAIS)}
\label{LAISTable}
\begin{tabular}{|p{0.95\columnwidth}|}
   \hline
   \vspace{-0.7cm}
\begin{enumerate}
\item Generate $NT$ samples, $\{{\bm \mu}_{n,t}\}$, using $N$ parallel MCMC chains of length $T$, each MCMC method using a proposal pdf $\varphi_n({\bm \mu}|{\bm \mu}_{t-1})$, with invariant distributions a power posterior $\post_n(\x|\y)=\post(\x|\y,\beta_n)$ (with $\beta_n>0$) or a posterior pdf with a smaller number of data.
\item Draw $NT$ samples $\x_{n,t} \sim q(\x|{\bm \mu}_{n,t}, {\bf C})$ where ${\bm \mu}_{n,t}$ plays the role of the mean, and  $ {\bf C}$ is a covariance matrix.
\item Assign to $\x_{n,t}$ the weights 
\begin{equation}
w_{n,t}=\frac{\pi(\x_{n,t}|\y)}{\Phi(\x_{n,t})}.
\end{equation}
There are different possible choices for $\Phi(\x_{n,t})$, for instance:
\begin{eqnarray}
\Phi(\x_{n,t})&=&\frac{1}{NT}\sum_{k=1}^T\sum_{i=1}^Nq_{i,k}(\x_{n,t}|{\bm \mu}_{i,k}, {\bf C}), \label{PHI1}\\
 \Phi(\x_{n,t})&=&\frac{1}{T}\sum_{k=1}^Tq(\x_{n,t}|{\bm \mu}_{n,k}, {\bf C}),  \label{PHI2}\\
 \Phi(\x_{n,t})&=&\frac{1}{N}\sum_{i=1}^Nq(\x_{n,t}|{\bm \mu}_{i,t}, {\bf C}),   \label{PHI3}\\
 \Phi(\x_{n,t})&=& q(\x_{n,t}|{\bm \mu}_{n,t}, {\bf C}), 
\end{eqnarray} 
\item Return all the pairs  $\{\x_{n,t},w_{n,t}\}$,  and $\widehat{Z}=\frac{1}{NT}\sum_{t=1}^T\sum_{n=1}^N w_{n,t}$.
\end{enumerate} \\ \\
\hline 
\end{tabular}
\end{table}

\noindent 
Let assume $\post_n(\x|\y)=\post(\x|\y)$ for all $n$ in the upper layer. Considering also standard parallel Metropolis-Hastings chains in the upper layer, the number of posterior evaluations in LAIS is $2NT$. Thus,  if only one chain $N=1$ is employed in the  upper layer, the number of posterior evaluations  is $2T$.
\newline
{\bf Special case with recycling samples.} The method in \cite{schuster2018markov} can be considered as a special case of LAIS when $N=1$, and $\{{\bm \mu}_{t}=\x_{t}\}$ i.e.,  all the samples $\{\x_{t}\}_{t=1}^T$ are generated by the unique MCMC chain with random walk proposal $\varphi(\x|\x_{t-1})=q(\x|\x_{t-1})$ with invariant density $\post(\x|\y)$.  In this scenario, the two layers of LAIS are collapsed in a unique layer, so that $\{{\bm \mu}_{t}=\x_{t}\}$. Namely, no additional generation of samples are needed in the lower layer, and the samples generated in the upper layer (via MCMC)  are recycled. Hence, the number of posterior evaluations is only $T$.  The denominator for weights used in \cite{schuster2018markov} is in Eq. \eqref{PHI2}, i.e., a temporal mixture as in \cite{CORNUET12}.  The resulting estimator is
 $$
  \widehat{Z}=\frac{1}{T}\sum_{t=1}^T \frac{\pi(\x_{t}|\y)}{\frac{1}{T}\sum_{k=1}^T\varphi(\x_k|\x_{k-1})}, \quad  \{\x_t\}_{t=1}^T \sim \post(\x|\y) \mbox{ (via MCMC with a proposal $\varphi(\cdot |\cdot)$)}.
 $$
 {\bf Relationship with KDE method.} LAIS can be interpreted as an extension of the KDE method in Section \ref{ExploitingFunctionalIdentity}, where the KDE function is also employed as a proposal density in the MIS scheme. Namely, the points used in Eq. \eqref{aquiKDEcasiIS}, in LAIS they are drawn from the KDE function using the deterministic mixture procedure \cite{ElviraMIS15,HereticalMIS,EfficientMIS}. 
\newline
{\bf  Compressed LAIS (CLAIS).} Let us consider the $T$ or $N$ is large (i.e., either large chains or several parallel chains; or both). Since $NT$ is large, the computation of the denominators Eqs. \eqref{PHI1}- \eqref{PHI2}- \eqref{PHI3} can be expensive. A possible solution is to use a partitioning or clustering procedure \cite{CMC} with $K<< NT$ clusters considering the $NT$ samples, and then employ as denominator the function
\begin{equation}\label{CLAISeq}
 \Phi(\x)=\sum_{k=1}^K {\bar a}_k \mathcal{N}(\x|{\bar {\bm \mu}}_{k}, {\bf C}_k), 
\end{equation}
where ${\bar {\bm \mu}}_k$ represents the centroid of the $k$-th cluster, the normalized weight ${\bar a}_k$ is proportional to the number  of elements in the $k$-th cluster ($\sum_{k=1}^K {\bar a}_k=1$),  and ${\bf C}_k={\bf \Sigma}_k+h {\bf I}$ with ${\bf \Sigma}_k$ the empirical covariance matrix  of $k$-th cluster and $h>0$. 
\newline
{\bf Relationship with other methods using tempered posteriors.} In the upper layer of LAIS, we can use non-tempered versions of the posterior, i.e., $\post_n(\x|\y)=\post(\x|\y)$ for all $n$, or tempered versions of the posterior  $\post_n(\x|\y)=\post(\x|\y,\beta_n)= \ell(\y|\x)^{\beta_n}g(\x)$. However, unlike in SS and/or power posterior methods, these samples are employed only as location parameters ${\bm \mu}_{n,t}$ of the proposal pdfs $q_{n,t}(\x|{\bm \mu}_{n,t}, {\bf C})$, and they are not included in the final estimators. Combining the tempered posteriors idea and the approach in  \cite{schuster2018markov}, we could recycle $\x_{n,t}={\bm \mu}_{n,t}$ and use $q_{n,t}(\x|{\bm \mu}_{n,t})=\varphi_{n,t}(\x|{\bm \mu}_{n,t})$ where we denote as $\varphi_{n,t}$ the proposal pdfs employed in the MCMC chains. Another difference is that, in LAIS, the use of an ``anti-tempered'' posteriors with $\beta_n>1$ is allowed and can be shown that is beneficial for the performance of the estimators (after the chains reach a good mixing) \cite{AntiTemperedLAIS17}. More generally, one can consider a time-varying $\beta_{n,t}$ (where $t$ is the iteration of the $n$-th chain). In the first iterations, one could use $\beta_{n,t}<1$ for fostering the exploration of the state space and helping the mixing of the chain. Then, in the last iterations, one could use  $\beta_{n,t}>1$ which increases the efficiency of the resulting IS estimators \cite{AntiTemperedLAIS17}.

\section{Vertical likelihood representations} \label{verticalLikelihoodApproach}
In this section, we introduce a different approach based on Lebesgue representations of the integral expressing the marginal likelihood $Z$. First of all, we derive two one-dimensional integral representations of $Z$, and then we describe how it is possible to use these alternative representations by applying one-dimensional quadratures. However, the application of these quadrature rules is not straightforward.  A possible final solution is the so-called nested sampling method.  

\subsection{Lebesgue representations of the marginal likelihood}
\subsubsection{First one-dimensional representation }
The $D_x$-dimensional integral $Z=\int_\Theta\ell(\y|\x)g(\x)d\x$ can be turned into a one-dimensional integral using an extended space representation.  Namely, we can write 
\begin{align}
Z &= \int_\Theta\ell(\y|\x)g(\x)d\x \\
&= \int_{\Theta}g(\x)d\x \int_0^{\ell(\y|\x)}d\lambda \quad \text{(extended space representation)} \\
&= \int_{\Theta}g(\x)d\x \int_0^\infty \mathbb{I}\{0<\lambda<\ell(\y|\x)\}d\lambda  
\end{align}
where $\mathbb{I}\{0<\lambda<\ell(\y|\x)\}$ is an indicator function which is $1$ if $\lambda \in [0,\ell(\y|\x)]$ and $0$ otherwise. Switching the integration order, we obtain 
\begin{align}
Z &= \int_0^\infty d\lambda\int_{\Theta}g(\x) \mathbb{I}\{0<\lambda<\ell(\y|\x)\}d\x \\
&= \int_0^\infty d\lambda\int_{\ell(\y|\x)>\lambda}g(\x)d\x \\
&= \int_0^\infty Z(\lambda)d\lambda = \int_0^{{\sup \ell(\y|\x)}} Z(\lambda)d\lambda,  \label{aquiHlam}
\end{align}
where we have set
\begin{align}\label{aquiHlam2}
Z(\lambda) = \int_{\ell(\y|\x)>\lambda}g(\x)d\x.
\end{align}
In Eq. \eqref{aquiHlam}, we have also assumed that $\ell(\y|\x)$ is bounded so the limit of integration is ${\sup \ell(\y|\x)}$.
\newline
Below, we define several variables and sampling procedures required for the proper understanding of the nested sampling algorithm.

\subsubsection{The survival function $ Z(\lambda)$ and related sampling procedures}\label{ZandSampling}

The function above $Z(\lambda): \mathbb{R}^{+}\rightarrow [0,1]$ is the mass of the prior restricted to the set $\{\x: \ell(\y|\x)>\lambda\}$.  Note also that
\begin{align}
Z(\lambda)=\mathbb{P}\left(\lambda<\ell(\y|\x)\right),\quad \mbox{ where } \x \sim g(\x).
\end{align}
Moreover, we have that  $Z(\lambda) \in [0,1]$ with $Z(0)=1$ and $Z(\lambda')=0$ for all  $\lambda' \geq {\sup \ell(\y|\x)}$,  and it is also an non-increasing function. Therefore, $Z(\lambda)$  is a {\it survival function}, i.e.,  
\begin{eqnarray}
F(\lambda)=1-Z(\lambda)=\mathbb{P}\left(\ell(\y|\x)<\lambda\right)=\mathbb{P}\left(\Lambda<\lambda\right),
\end{eqnarray} 
is the cumulative distribution of the random variable $\Lambda=\ell(\y|\x)$ with $\x \sim g(\x)$ \cite{martino2018independent,Robert04}.  
 \newline
  \newline
{\bf Sampling according to} $\bm{F(\lambda)=1-Z(\lambda)}$. Since  $\Lambda=\ell(\y|\x)$ with $\x \sim g(\x)$, the following procedure generates samples $\lambda_n$ from $\frac{dF(\lambda)}{d\lambda}$: 
\begin{enumerate}
\item Draw $\x_n\sim g(\x)$, for $n=1,...,N$.   
\item Set $\lambda_n=\ell(\y|\x_n)$, , for all $n=1,...,N$.
\end{enumerate}
Recalling the inversion method \cite[Chapter 2]{martino2018independent}, note also that the corresponding values 
\begin{eqnarray}
b_n=F(\lambda_n)\sim \mathcal{U}([0,1]),
\end{eqnarray}
i.e., they are uniformly distributed in $[0,1]$. Since $Z(\lambda)=1-F(\lambda)$, and since $V=1-U$ is also uniformly distributed $\mathcal{U}([0,1])$ if $U\sim \mathcal{U}([0,1])$, then
\begin{eqnarray} \label{AquiAn}
a_n=Z(\lambda_n)\sim \mathcal{U}([0,1]).
\end{eqnarray}
 In summary, finally we have that
 \begin{eqnarray}  \label{AquiAn2}
 \mbox{ if }  \x_n \sim g(\x),   \mbox{ and }  \lambda_n=\ell(\y|\x_n) \sim F(\lambda) \quad \mbox{ then }  \quad  a_n=Z(\lambda_n) \sim \mathcal{U}([0,1]).
\end{eqnarray}
\subsubsection{The truncated prior pdf $g(\x|\lambda)$ and other sampling procedures}\label{ZandSampling2} 
\begin{figure}[!h]
	\centering
\subfigure[]{\includegraphics[width=5cm]{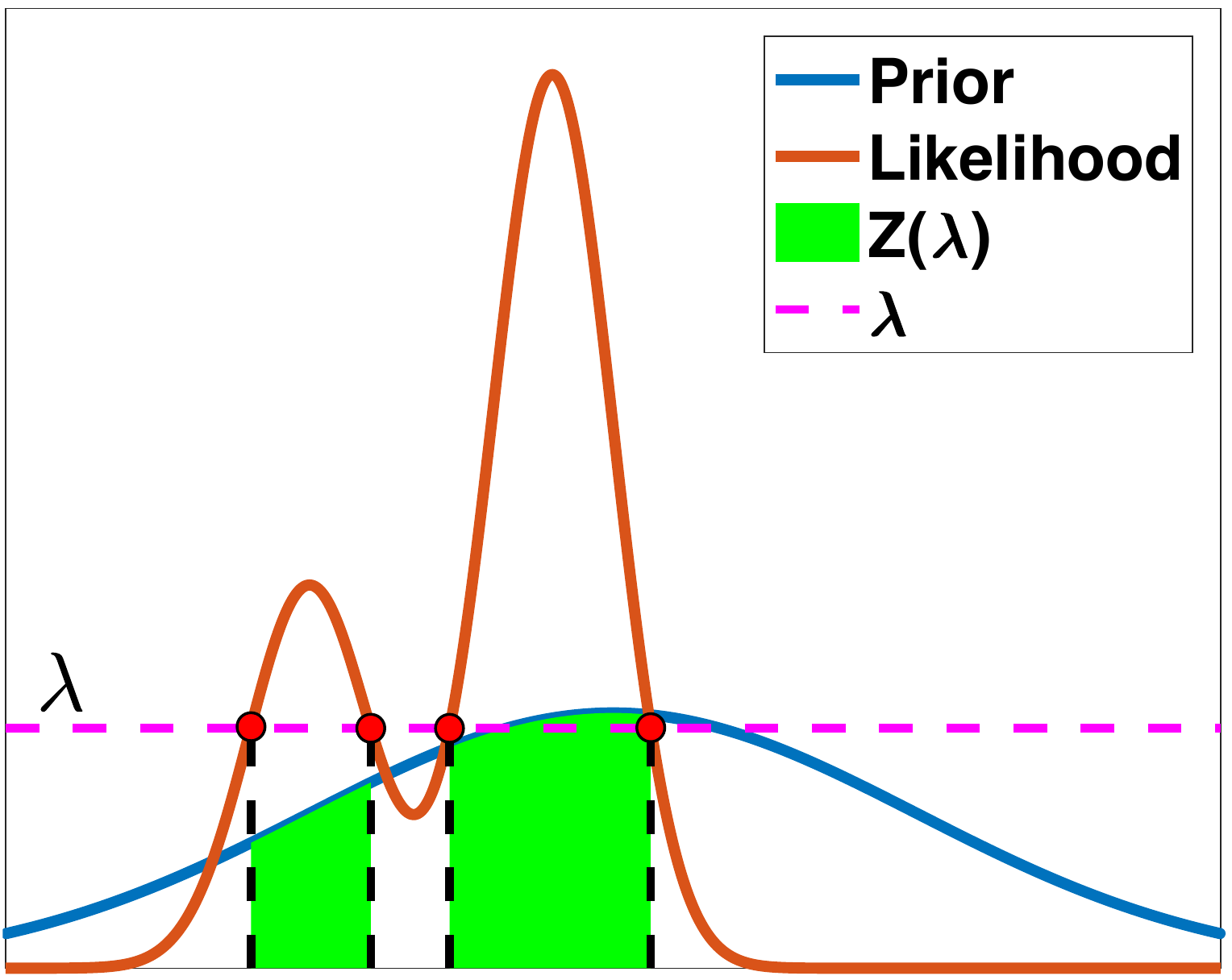}}
\hspace{0.5cm}
\subfigure[]{\includegraphics[width=5cm]{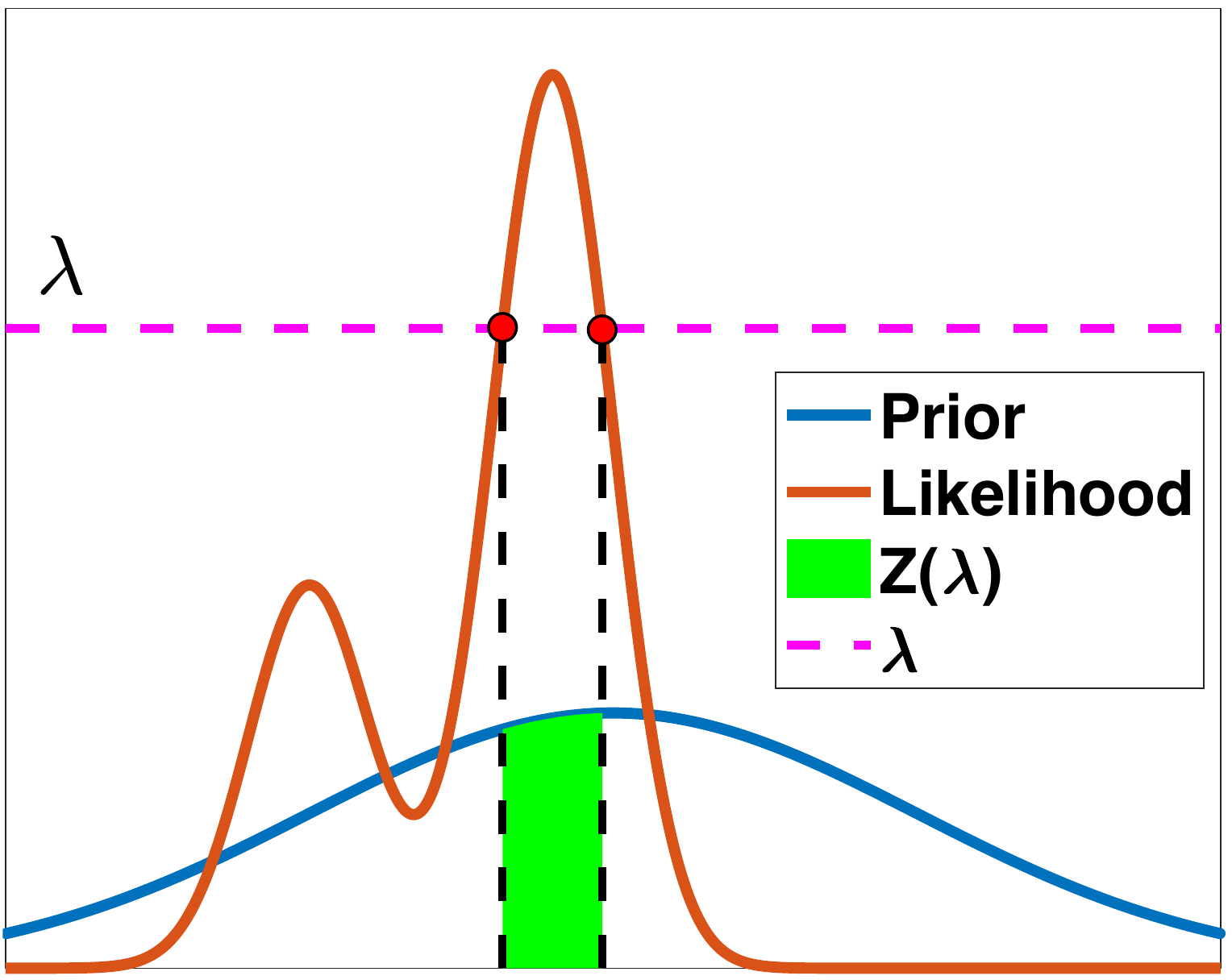}}
\vspace{-0.4cm}
	\caption{Two examples of the area below the truncated prior $g(\x|\lambda)$, i.e., the function $Z(\lambda)$.  Note that in figure (b) the value of $\lambda$ is greater than in figure (a), so that the area $Z(\lambda)$  decreases. If $\lambda$ is bigger than the maximum of the likelihood function then $Z(\lambda)=0$. 
	}
	\label{figNested}
\end{figure}

Note that $Z(\lambda)$ is also the normalizing constant of the following truncated prior pdf 
\begin{eqnarray}
g(\x|\lambda)=\frac{1}{Z(\lambda)}\mathbb{I}\{\ell(\y|\x)>\lambda\} g(\x),
\end{eqnarray}
where $g(\x|0)=g(\x)$ and $g(\x|\lambda)$ for $\lambda>0$. Two graphical examples of $g(\x|\lambda)$ and $Z(\lambda)$ are given in Figure \ref{figNested}.

\noindent
{\bf Sampling  from $\bm{g(\x|\lambda)}$ and $\bm{F(\lambda|\lambda_0)}$.}  Given a fixed value $\lambda_0\geq 0$, in order to  generate samples from $g(\x|\lambda_0)$  one alternative is to use an MCMC procedure.
However, in this case, the following acceptance-rejection procedure can be also employed \cite{martino2018independent}:
\begin{enumerate}
\item For $n=1,...,N$:
\begin{enumerate}
\item Draw $\x'\sim g(\x)$.   
\item if $\ell(\y|\x')> \lambda_0$ then set $\x_n=\x'$ and $\lambda_n=\ell(\y|\x')$.
\item if $\ell(\y|\x')\leq \lambda_0$, then reject $\x'$ and repeat from step 1(a).
\end{enumerate}
\item Return $\{\x_n\}_{n=1}^N$ and $\{\lambda_n\}_{n=1}^N$.
\end{enumerate}
Observe that $\x_n \sim g(\x|\lambda_0)$, for all $n=1,...,N$, and the probability of accepting a generated sample $\x'$ is exactly $Z(\lambda)$.  The values $\lambda_n=\ell(\y|\x_n)$ where $\x_n \sim g(\x|\lambda_0)$, have the following {\it truncated} cumulative distribution
\begin{equation}
F(\lambda|\lambda_0)=\frac{F(\lambda)-F(\lambda_0)}{1-F(\lambda_0)}, \quad \mbox{ with } \lambda\geq \lambda_0,
\end{equation}
i.e., we can write $\lambda_n\sim F(\lambda|\lambda_0)$.
\subsubsection{Distribution of $\bm{a_n=Z(\lambda_n)}$ and $\widetilde{a}_n=\frac{a_n}{a_0}$ if $\bm{\lambda_n\sim F(\lambda|\lambda_0)}$}
\label{ZandSampling3}
 Considering the values $\lambda_n=\ell(\y|\x_n)$ where $\x_n \sim g(\x|\lambda_0)$, then $\lambda_n\sim F(\lambda|\lambda_0)$. Therefore, considering the values  $a_0=Z(\lambda_0)\leq 1$ and $a_n=Z(\lambda_n)$, with a similar argument used above in Eqs. \eqref{AquiAn}-\eqref{AquiAn2} we can write
\begin{align*}
a_n &\sim \mathcal{U}([0,a_0]), \\
\widetilde{a}_n&=\frac{a_n}{a_0} \sim \mathcal{U}([0,1]), \quad \forall n=1,...,N. 
\end{align*}
In summary, with $a_0=Z(\lambda_0)$, we have that  
\begin{align}
 \mbox{ if } \x_n \sim g(\x|\lambda_0)    \mbox{ and }  \lambda_n=\ell(\y|\x_n) \sim F(\lambda|\lambda_0), \quad \mbox{then} \quad	Z(\lambda_n) \sim \mathcal{U}([0,a_0]),
\end{align}
and the ratio $\widetilde{a}_n=\frac{a_n}{a_0}\sim \mathcal{U}([0,1])$.
\newline
\newline
\subsubsection{Distributions $\widetilde{a}_{\texttt{max}}$}
\label{ZandSampling4}
 Let us consider $\lambda_1,....,\lambda_n \sim F(\lambda|\lambda_0)$ and  the minimum and maximum values   
\begin{equation}\label{IMP_amax0}
\lambda_{\texttt{min}}=\min_n \lambda_n, \quad a_{\texttt{max}}=Z(\lambda_{\texttt{min}}), \quad  \mbox{ and } \quad \widetilde{a}_{\texttt{max}}=\frac{a_{\texttt{max}}}{a_0}=\frac{Z(\lambda_{\texttt{min}})}{Z(\lambda_0)}.
\end{equation}
Let us recall $\widetilde{a}_n=\frac{a_n}{a_0} \sim \mathcal{U}([0,1])$.
Then, note that $\widetilde{a}_{\texttt{max}}$ is  maximum of $N$ uniform random variables 
$$
\widetilde{a}_1, ...,\widetilde{a}_N  \sim \mathcal{U}([0,1]).
$$
Then  it is well-known that the cumulative distribution of the maximum value  
$$
\widetilde{a}_{\texttt{max}}= \max_n \widetilde{a}_n \sim \mathcal{B}(N,1),
$$
is distributed according to a Beta distribution $\mathcal{B}(N,1)$, i.e.,
$F_{\texttt{max}}(\widetilde{a})= \widetilde{a}^N$ and density $f_{\texttt{max}}(\widetilde{a})=\frac{dF_\text{max}(\widetilde{a})}{d\widetilde{a}}= N\widetilde{a}^{N-1}$ \cite[Section 2.3.6]{martino2018independent}. In summary, we have
\begin{align}\label{IMP_amax}
	\widetilde{a}_{\texttt{max}}= \frac{Z(\lambda_{\texttt{min}})}{Z(\lambda_0)} \sim \mathcal{B}(N,1), \mbox{ where } \lambda_{\texttt{min}}=\min_n \lambda_n, \enskip \text{ and } \enskip \lambda_n \sim F(\lambda|\lambda_0).
\end{align}
 This result is important for deriving the standard version of the  nested sampling method, described in the next section. A summary of the relationships presented above is provided in Table \ref{SummaryVariables}.

\begin{table}[!h]
	\caption{Summary of the relationships among the random variables introduced above.}
	\label{SummaryVariables}
	\vspace{-0.5cm}
	\begin{center}
	\footnotesize
		\begin{tabular}{|c||c|}  
		   	\hline 
		     {\bf Sections} & {\bf Relationships} \\
			\hline 
			\hline 
\multirow{3}{*}{\ref{ZandSampling}} 	&		\\ 
                   &  $Z(\lambda)=\mathbb{P}\left(\lambda<\ell(\y|\x)\right), \quad \mbox{ and } \quad F(\lambda)=1- Z(\lambda)=\mathbb{P}\left( \ell(\y|\x)\leq\lambda\right), \quad \mbox{ where } \quad \x \sim g(\x).$ \\
	&		\\ 
			\hline
\multirow{3}{*}{\ref{ZandSampling}} 	&		\\
	& $\mbox{ If }  \x_n \sim g(\x),   \mbox{ we have  }  \lambda_n=\ell(\y|\x_n) \sim F(\lambda) \ \mbox{ and }    a_n=Z(\lambda_n) \sim \mathcal{U}([0,1]).$  \\
	&    \\
\hline
\multirow{4}{*}{\ref{ZandSampling2}}  & \\		      
\multirow{4}{*}{\ref{ZandSampling3}}&$\mbox{ If } \x_n \sim g(\x|\lambda_0), \mbox{ we have  }  \lambda_n=\ell(\y|\x_n) \sim F(\lambda|\lambda_0) \quad \mbox{ and } \mbox{ }   a_n=Z(\lambda_n) \sim \mathcal{U}([0,a_0]), \mbox{ with } a_0=Z(\lambda_0).$  \\
& \\
&Moreover, $\widetilde{a}_n=\frac{a_n}{a_0} \sim \mathcal{U}([0,1])$. \\
 &\\
 \hline
\multirow{5}{*}{\ref{ZandSampling4}}  &\\
 &$\mbox{ If } \x_n \sim g(\x|\lambda_0), \mbox{ we have  }  \lambda_n=\ell(\y|\x_n) \sim F(\lambda|\lambda_0)  \mbox{ }$	and $ \mbox{ } \widetilde{a}_{\texttt{max}}=  \frac{Z(\lambda_{\texttt{min}})}{Z(\lambda_0)}  \sim \mathcal{B}(N,1), $
  $\mbox{ where } \lambda_{\texttt{min}}=\min \lambda_n.$	\\ 
&	\\
&Note also that $\widetilde{a}_{\texttt{max}}=\max \widetilde{a}_n$.	\\
& \\
			\hline
		\end{tabular}
	\end{center}
	
\end{table}

\subsubsection{Second one-dimensional representation}
Now let consider a specific {\it area} value $a=Z(\lambda)$. The inverse function 
\begin{align}
\Psi(a) = Z^{-1}(a) =\sup\{\lambda: Z(\lambda)>a  \},
\end{align} 
is also non-increasing. Note that $Z(\lambda)>a$ if and only if $\lambda < \Psi(a)$. Then, we can write 
\begin{align}
Z &= \int_0^\infty Z(\lambda)d\lambda \nonumber \\
&= \int_0^\infty d\lambda \int_0^1 \mathbb{I}\{a<Z(\lambda)\}da \qquad \text{(again the extended space ``trick'')}  \nonumber\\
&= \int_0^1da \int_0^\infty\mathbb{I}\{u<Z(\lambda)\} d\lambda  \qquad \text{(switching  the integration order)}  \nonumber \\
&= \int_0^1da \int_0^\infty \mathbb{I}\{\lambda<\Psi(a)\}d\lambda \qquad \text{(using } Z(\lambda)>a \iff \lambda<\Psi(a) \text{)}   \nonumber\\
&=\int_0^1\Psi(a)da. \label{SecondVertical}
\end{align}
\subsubsection{Summary of the one-dimensional representations}
Thus, finally we have obtained two one-dimensional integrals for expressing the Bayesian evidence $Z$,
\begin{align}\label{ZasOneDimIntegral}
Z = \int_0^{{\sup \ell(\y|\x)}} Z(\lambda)d\lambda =\int_0^1 \Psi(a)da.
\end{align}
Now that we have expressed the quantity $Z$ as an integral of a function over $\mathbb{R}$, we could think of applying simple quadrature: choose a grid of points in $[0,\sup\ell(\y|\x)]$ ($\lambda_i> \lambda_{i-1}$) or in $[0,1]$ ($a_i> a_{i-1}$), evaluate $Z(\lambda)$ or $\Psi(a)$ and use the quadrature formulas
\begin{align}
	\widehat{Z} &= \sum_{i=1}^I (\lambda_i - \lambda_{i-1})Z(\lambda_i), \enskip \text{or} \\
	\widehat{Z} &= \sum_{i=1}^I (a_i - a_{i-1})\Psi(a_i).
\end{align}
However, this simple approach is not desirable since (i) the functions $Z(\lambda)$ and $\Psi(a)$ are intractable in most cases and (ii) they change much more rapidly over their domains than does $\pi(\x|\y)=\ell(\y|\x)g(\x)$, hence the quadrature approximation can have very bad performance, unless the grid of points is chosen with extreme care. Table \ref{OnedimRepresentationZ}  summarizes the one-dimensional expression for $\log Z$ and $Z$ contained in this work.
Clearly, in all of them,  the integrand function depends, explicitly or implicitly, on the variable $\x$.

\begin{table}[!h]
	\caption{One-dimensional integrals for $\log Z$ and $Z$. Note that, in all cases, the integrand function contains the dependence on $\x$.}
	\label{OnedimRepresentationZ}
	\begin{center}
	\footnotesize
		\begin{tabular}{|c|c|c|}  
		   	\hline
			Method & Expression & Equations \\
			\hline
			\hline
			& &Ê\\ 
			path sampling & $\log Z= \int_{0}^1\frac{1}{Z(\beta)}\frac{\partial}{\partial\beta}\left(\int_{\Theta}\pi(\x|\y,\beta) d\x\right)d\beta$ & \eqref{PathSamplingIdentity2} \\ 
			&  &\\
		    power-posteriors  & $\log Z= \int_0^1 \E_{\post(\x|\y,\beta)}\left[ \log \ell(\y|\x)\right]d\beta$ &  \eqref{PowerOneDim} \\
			 &  &\\
			 vertical representation-1 & $Z = \int_0^{{\sup \ell(\y|\x)}} Z(\lambda)d\lambda$  &  \eqref{aquiHlam}-\eqref{aquiHlam2}  \\
			 &  &\\
			 vertical representation-2 & $Z = \int_0^1 \Psi(a)da$ &  \eqref{SecondVertical} \\ 
			 \hline
		\end{tabular}
	\end{center}
\end{table}

\subsection{Nested Sampling}\label{NestedSampling}
Nested sampling is a technique for estimating the marginal likelihood that exploits the second identity in \eqref{ZasOneDimIntegral} \cite{skilling2006nested, chopin2010properties, polson2014vertical}. Nested Sampling estimates $Z$ by a quadrature using nodes (in {\it decreasing} order),
$$
0<a_{\texttt{max}}^{(I)}<\dots<a_{\texttt{max}}^{(1)}<1
$$
and the quadrature formula
\begin{align}
\label{EqNestedZ}
	\widehat{Z} = \sum_{i=1}^{I} (a_{\texttt{max}}^{(i-1)} -a_{\texttt{max}}^{(i)} )\Psi(a_{\texttt{max}}^{(i)})=  \sum_{i=1}^{I}  (a_{\texttt{max}}^{(i-1)} -a_{\texttt{max}}^{(i)} ) \lambda_{\texttt{min}}^{(i)},
\end{align}
with $a_{\texttt{max}}^{(0)}=1$. We have to specify the  grid points  $a_{\texttt{max}}^{(i)}$'s (possibly well-located, with a suitable strategy) and the corresponding values $\lambda_{\texttt{min}}^{(i)}= \Psi(a_{\texttt{max}}^{(i)})$. Recall that the function $\Psi(a)$, and its inverse $a=\Psi^{-1}(\lambda)=Z(\lambda)$, are generally intractable, so that  it is not even possible to evaluate $\Psi(a)$ at a grid of chosen $a_{\texttt{max}}^{(i)}$'s. 

{\rem The nested sampling algorithm works in the other way around: it suitably selects the ordinates $\lambda_{\texttt{min}}^{(i)}$'s and find some approximations $\widehat{a}_i$'s of the corresponding values $a_{\texttt{max}}^{(i)}=Z(\lambda_{\texttt{min}}^{(i)})$. This is possible since the distribution of $a_{\texttt{max}}^{(i)}$ is known (see Section \ref{ZandSampling4}).}


\subsubsection{Choice of $\lambda_{\texttt{min}}^{(i)}$ and $a_{\texttt{max}}^{(i)}$ in nested sampling}
\label{superIMPNested}

Nested sampling employs an iterative procedure  in order to  generate an {\it increasing} sequence of likelihood ordinates $\lambda_{\texttt{min}}^{(i)}$, $i=1,...,I$, such that
\begin{equation}
\lambda_{\texttt{min}}^{(1)}<\lambda_{\texttt{min}}^{(2)}<\lambda_{\texttt{min}}^{(3)}....<\lambda_{\texttt{min}}^{(I)}.
\end{equation}
The details of the algorithm is given in Table \ref{tableNestedSampling} and it is based on the sampling of the truncated prior pdf $g(\x|\lambda_{\texttt{min}}^{(i-1)})$ (see Sections from \ref{ZandSampling} to \ref{ZandSampling4}), where $i$ denotes the iteration index. The nested sampling procedure is explained below:
\begin{itemize}
\item At the first iteration ($i=1$), we set $\lambda_{\texttt{min}}^{(0)}=0$ and $a_{\texttt{max}}^{(0)}=Z(\lambda_{\texttt{min}}^{(0)})=1$. Then, $N$ samples are drawn from the prior $\x_n\sim g(\x|\lambda_{\texttt{min}}^{(0)})=g(\x)$ obtaining a cloud $\mathcal{P}=\{\x_n\}_{n=1}^N$ and  then set $\lambda_n=\ell(\y|\x_n)$, i.e.,
$\{\lambda_n\}_{n=1}^N\sim F(\lambda)$ as shown in Section \ref{ZandSampling}. Thus, the first ordinate is chosen as
$$
\lambda_{\texttt{min}}^{(1)}= \min_{n}  \lambda_n=  \min_{n} \ell(\y|\x_n)= \min\limits_{\x\in \mathcal{P}}\ell(\y|\mathcal{P}). 
$$
 Since $\{\lambda_n\}_{n=1}^N\sim F(\lambda)$,  using the result in  Eq. \eqref{IMP_amax}, we have that 
$$
\widetilde{a}_{\texttt{max}}^{(1)}=\frac{a_{\texttt{max}}^{(1)}}{a_{\texttt{max}}^{(0)}}=\frac{Z(\lambda_{\texttt{min}}^{(1)})}{Z(\lambda_{\texttt{min}}^{(0)})}\sim \mathcal{B}(N,1).
$$
Since $a_{\texttt{max}}^{(0)}=Z(\lambda_{\texttt{min}}^{(0)})=1$, then $\widetilde{a}_{\texttt{max}}^{(1)}=a_{\texttt{max}}^{(1)} \sim \mathcal{B}(N,1)$. The corresponding $\x^*=\arg\min\limits_{\x\in \mathcal{P}}\ell(\y|\mathcal{P})$ is also removed from $\mathcal{P}$, i.e., $\mathcal{P}=\mathcal{P}\backslash \{\x^*\}$ (now $|\mathcal{P}|=N-1$).
\item At a generic $i$-th iteration ($i\geq 2$), a unique additional sample $\x'$ is drawn from the truncated prior $g(\x|\lambda_{\texttt{min}}^{(i-1)})$ and  added to the current cloud of samples, i.e., $\mathcal{P}=\mathcal{P}\cup \x'$ (now again $|\mathcal{P}|=N$). First of all, note that the value $\lambda' =\lambda_n=\ell(\y|\x')$ is distributed as $F(\lambda|\lambda_{\texttt{min}}^{(i-1)})$ (see Section \ref{ZandSampling2}). More precisely, note that all the $N$ ordinate values 
$$
\{\lambda_n\}_{n=1}^N=\ell(\y|\mathcal{P})=\{\lambda_n=\ell(\y|\x_n)  \mbox{ }  \mbox{ for all }  \mbox{ }  \x_n\in \mathcal{P}\}
$$
are distributed as $F(\lambda|\lambda_{\texttt{min}}^{(i-1)})$, i.e., $\{\lambda_n\}_{n=1}^N \sim F(\lambda|\lambda_{\texttt{min}}^{(i-1)})$. This is due to how the population $\mathcal{P}$ has been built in the previous iterations. Then, we choose the new minimum value as
$$
\lambda_{\texttt{min}}^{(i)}= \min_{n}  \lambda_n=  \min_{\x\in \mathcal{P}} \ell(\y|\mathcal{P}).
$$
 Moreover, since $\lambda_{\texttt{min}}^{(i)}$ is the minimum value of $\{\lambda_1,...,\lambda_N\}\sim  F(\lambda|\lambda_{\texttt{min}}^{(i-1)})$, in Section \ref{ZandSampling4} we have seen that  
\begin{equation}
\widetilde{a}_{\texttt{max}}^{(i)}=\frac{a_{\texttt{max}}^{(i)}}{a_{\texttt{max}}^{(i-1)}}=\frac{Z(\lambda_{\texttt{min}}^{(i)})}{Z(\lambda_{\texttt{min}}^{(i-1)})} \sim \mathcal{B}(N,1),
\end{equation}
where we have used Eq. \eqref{IMP_amax}. We remove again the corresponding sample $\x^*=\arg\min\limits_{\x\in \mathcal{P}}\ell(\y|\mathcal{P})$, i.e., we set $\mathcal{P}=\mathcal{P} \backslash \{\x^*\}$ and the procedure is repeated. Note that we have also found the recursion among the following random variables,
\begin{equation}\label{RecursionAi}
a_{\texttt{max}}^{(i)}= \widetilde{a}_{\texttt{max}}^{(i)} a_{\texttt{max}}^{(i-1)}, 
\end{equation}
for  $i=1,...,I$ and $a_{\texttt{max}}^{(0)}=1$. 
\item The random value $\widetilde{a}_{\texttt{max}}^{(i)}$ could be estimated and replaced with the expected value of the Beta distribution $\mathcal{B}(N,1)$, i.e.,
\begin{equation} \label{approxMeanBeta}
 \widetilde{a}_{\texttt{max}}^{(i)} \approx \widehat{a}_1=\frac{N}{N+1} \approx \exp\left(-\frac{1}{N}\right).
\end{equation} 
where $\mathbb{E}[\mathcal{B}(N,1)]=\frac{N}{N+1}$, and $ \exp\left(-\frac{1}{N}\right)$ becomes a very good approximation as $N$ grows.  
In that case, the recursion above becomes
\begin{equation}\label{approxMeanBeta2}
a_{\texttt{max}}^{(i)}\approx \exp\left(-\frac{1}{N}\right) a_{\texttt{max}}^{(i-1)}=\exp\left(-\frac{i}{N}\right).
\end{equation} 
Then, denoting $\widehat{a}_i=\exp\left(-\frac{i}{N}\right)$, we can use $\widehat{a}_i$ as an approximation of $a_{\texttt{max}}^{(i)}$. 
\end{itemize}
{\rem The intuition behind the iterative approach above is to accumulate more ordinates $\lambda_i$ close to the $\sup \ell(\y|\x)$. They are also more dense around $\sup \ell(\y|\x)$. Moreover, using this scheme, we can employ  $\widehat{a}_i=\exp\left(-\frac{i}{N}\right)$ as an approximation of $a_{\texttt{max}}^{(i)}$.}

{\rem An implicit optimization of the likelihood function is performed in the nested sampling algorithm. All population  of $\lambda_i\in \mathcal{P}$  approaches the value $\sup \ell(\y|\x)$.
}

 
 \begin{table}[!h]
\caption{\normalsize The standard Nested Sampling procedure. }
\label{tableNestedSampling}
\begin{tabular}{|p{0.95\columnwidth}|}
   \hline
   \vspace{-0.7cm}
\begin{enumerate}
\item Choose $N$ and set $\widehat{a}_0=1$.
\item Draw $\{\x_{n}\}_{n=1}^N \sim g(\x)$ and define the set $\mathcal{P}=\{\x_{n}\}_{n=1}^N$. Let us also define the notation 
\begin{equation}
\ell(\y|\mathcal{P})=\{\lambda_n=\ell(\y|\x_n)  \mbox{ }  \mbox{ for all }  \mbox{ }  \x_n\in \mathcal{P}\},
\end{equation}
\item Set $\lambda_{\texttt{min}}^{(1)} =\min\limits_{\x\in \mathcal{P}}\ell(\y|\mathcal{P})$ and $\x^*=\arg\min\limits_{\x\in \mathcal{P}}\ell(\y|\mathcal{P})$.
\item Set $\mathcal{P}=\mathcal{P}\backslash \{\x^*\}$, i.e., eliminate $\x^*$ from $\mathcal{P}$.
\item Find an approximation $\widehat{a}_1$ of $a_{\texttt{max}}^{(1)}=Z(\lambda_{\texttt{min}}^{(1)})$. One usual choice is $\widehat{a}_1=\exp\left(-\frac{1}{N}\right)$.
\item For $i=2,..,I:$
\begin{enumerate}
\item Draw $\x' \sim g(\x|\lambda_{\texttt{min}}^{(i-1)} )$ and add to the current cloud of samples, i.e., $\mathcal{P}=\mathcal{P}\cup \x'$. 
\item Set  $\lambda_{\texttt{min}}^{(i)} =\min\limits_{\x\in \mathcal{P}}\ell(\y|\mathcal{P})$ and  $\x^*=\arg\min\limits_{\x\in \mathcal{P}}\ell(\y|\mathcal{P})$.
\item Set $\mathcal{P}=\mathcal{P} \backslash \{\x^*\}$.
\item Find an approximation $\widehat{a}_i$ of $a_{\texttt{max}}^{(i)}=Z(\lambda_{\texttt{min}}^{(i)})$. One usual choice is 
\begin{equation}
\widehat{a}_i=\exp\left(-\frac{i}{N}\right),
\end{equation}
The rationale behind this choice is explained in the section above.
\end{enumerate}
\item Return 
\begin{align}
	\widehat{Z} = \sum_{i=1}^I(\widehat{a}_{i-1}-\widehat{a}_i)\lambda_{\texttt{min}}^{(i)}= \sum_{i=1}^I(e^{-\frac{i-1}{N}}-e^{-\frac{i}{N}})\lambda_{\texttt{min}}^{(i)}.
\end{align}
\end{enumerate} \\ \\
\hline 
\end{tabular}
\end{table}

\subsubsection{Further considerations} \label{NestedApprox}

Perhaps, the most critical task of  the nested sampling implementation consists in drawing from the truncated priors. 
For this purpose, one can use a rejection sampling or an MCMC scheme. In the first case, we sample from the prior and then accept only the samples $\x'$ such that $\ell(\y|\x')>\lambda$. However, as $\lambda$ grows, its performance deteriorates  since the acceptance probability gets smaller and smaller. The MCMC algorithms could also have poor performance due to the sample correlation, specially when the support of the constrained prior is formed by disjoint regions or distant modes \cite{chopin2010properties}. Moreover, in the derivation of the standard nested sampling method  we have considered different approximations. First of all, for each likelihood value $\lambda_i$, its corresponding $a_i = \Psi^{-1}(\lambda_i)$ is approximated by replacing the expected value of a Beta random variable within a recursion involving $a_i$ (Eq. \eqref{RecursionAi}). Then this expected value is again approximated with an exponential function in Eq. \eqref{approxMeanBeta}. This step could be avoided, keeping directly $\frac{N}{N+1}$. The simplicity of the final formula $\widehat{a}_i=\exp\left(-\frac{i}{N}\right)$ is perhaps the reason of using the approximation $\frac{N}{N+1} \approx \exp\left(-\frac{1}{N}\right)$. 
A further approximation $\E[a_{\texttt{max}}^{(i)}]\approx \E[\widetilde{a}_{\texttt{max}}^{(i)}] \E[a_{\texttt{max}}^{(i-1)}]$ is also implicitly applied in \eqref{approxMeanBeta2}.
Additionally, if an MCMC method is run for sampling from the constrained prior, also the likelihood values $\lambda_i$ are in some sense approximated due to the possible burn-in period of the chain.

\subsubsection{Generalized Importance Sampling based on vertical representations}
Let us recall the estimator IS vers-2  with proposal density $\bar{q}(\x)\propto q(\x)$,
\begin{align}\label{aquiZdos}
 \widehat{Z} =\sum_{n=1}^N \bar{\rho}_n \ell(\y|\x_n), \qquad \{\x_n \}_{n=1}^N \sim \bar{q}(\x),
\end{align}
where $\rho_n=\frac{g(\x_n)}{q(\x_n)}$ and $ \bar{\rho}_n=\frac{\rho_n}{\sum_{n=1}^{N} \rho_n}$. In \cite{polson2014vertical}, the authors consider the use of the following proposal pdf
\begin{align}\label{VerticalProposalDensity}
\bar{q}_w(\x) = \frac{g(\x) W(\ell(\y|\x))}{Z_w} \propto  q_w(\x)=g(\x) W(\ell(\y|\x)),
\end{align}
where the function $W(\lambda): \mathbb{R}^{+}\rightarrow \mathbb{R}^{+}$ is defined by the user.   Using $\bar{q}_w(\x) $ leads to the weights of the form
\begin{align}
\rho_n = \frac{g(\x_n)}{q_w(\x_n)}= \frac{1}{W(\ell(\y|\x_n))}, \quad\x_n \sim \bar{q}_w(\x).
\end{align}
Note that choosing $W(\lambda)=\lambda$ we have $W(\ell(\y|\x))=\ell(\y|\x)$, and $\bar{q}_w(\x) = \post(\x|\y)$, recovering the harmonic mean estimator.  With $W(\lambda)=\lambda^{\beta}$, we have $W(\ell(\y|\x))=\ell(\y|\x)^{\beta}$ and $\bar{q}_w(\x) =  \frac{g(\x)\ell(\y|\x)^{\beta}}{Z(\beta)}$, recovering the method in Section \ref{TemperedLuca} that uses a power posterior as a proposal pdf. Nested sampling seems that can be also included in this framework \cite{polson2014vertical}. 

\section{On the marginal likelihood approach and other strategies }\label{SectNuovaBella}
In this section, we examine the marginal likelihood approach to Bayesian model selection and compare it to other strategies such as the well-known {\it posterior predictive check} approach.

\subsection{Dependence on the prior and related discussion}\label{SuperSuperIMPSect}

The marginal likelihood approach for model selection and hypothesis testing naturally appears as a consequence of the application of Bayes' theorem to derive posterior model probabilities $p(\mathcal{M}_m|\y) \propto p_mZ_m$. Under the assumption that one of $\mathcal{M}_m$ is the true generating model, the Bayes factor will choose the correct model as the number of data grows, $D_y \to \infty$ \cite{kass1995bayes}.  We can also apply the posterior model probabilities $p(\mathcal{M}_m|\y)$ to combine inferences across models, a setting called Bayesian model averaging \cite{BMA99,Martino15PF}.

\subsubsection{Dependence on the prior}  

In Section \ref{ModelFitSect}, we have seen the marginal likelihood $Z$ contains intrinsically a penalization for the model complexity. This penalization is related to the choice of the prior and its ``overlap'' with likelihood function. Indeed, $Z=\int_{\Theta} \ell(\y|{\bm \theta}) g({\bm \theta}) d{\bm \theta}$ is by definition a continuous mixture of the likelihood values weighted according to the prior. In this sense, depending on  the choice of the prior, the evidence $Z$ can take any possible value in the interval $[\ell(\y|\x_\text{min}),  \ell(\y|\x_\text{max})]$ (see Section \ref{ModelFitSect}, for more details). Hence, the marginal likelihood even with strong data (unlike the posterior density) is highly sensitivity to the choice of prior density. See also the examples in the Supplementary Material.
\newline
\newline
{\bf Improper priors.} The use of improper priors, $\int_{\Theta} g(\x) d\x=\infty$, is allowed when $\int_{\Theta} \ell(\y|\x) g(\x) d\x  <Ê\infty$, since the corresponding  posteriors are proper. 
However,  this is an issue for the model selection with $Z$. Indeed,
the prior $g(\x)=ch(\x)$ is not completely specified, since $c>0$ is arbitrary. Some possible solutions are given in Section \ref{AppBayeFactImproperPrior}. 
\newline
\newline
 Generally, the use of more diffuse (proper) priors provides smaller values of $Z$. Therefore, different choices of the priors can yield different selected models. For this fact, some authors criticize the use of evidence $Z$ for model comparison.

\subsubsection{Safe scenarios for fair comparisons}  

In a Bayesian framework, the best scenario is clearly when the practitioners and/or researchers have strong beliefs that can be translated into informative priors. Hence, in this setting, the priors truly encode some relevant information about the inference problem.
  When this additional information is not available, different strategies could be considered. We consider as a safe scenario for comparing different models,  a scenario where the choice of the priors is {\it virtually} not favoring any of the models.
Below and in Sections \ref{AppBayeFactImproperPrior} and \ref{SuperSect_PredPost}, we describe some interesting scenarios and some possible solutions for reducing, in some way, the dependence of the model comparison on the choice of the priors.  
\newline
\newline
  {\bf Same priors.} Generally, we are interested in comparing two or more models. The use of the same (even improper) priors is possible when the models have the same parameters  (and hence also share the same support space). 
With this choice,  the resulting comparison seems fair and reasonable. However, this scenario is very restricted in practice.
 An example is when we have nested models. As noted in \cite[Sect. 5.3]{kass1995bayes}, in the context of testing hypothesis, some authors have considered improper priors on nuisance parameters that appear on both null and alternative hypothesis. Since the nuisance parameters appear on both models, the multiplicative constants cancel out in the Bayes factor.  
\newline
\newline
   {\bf Likelihood-based priors.} When $\int_{\Theta} \ell(\y|\x)d\x <\infty$, we can build a prior based on the data and the observation model. For instance, we can choose $g_\text{like}(\x) =\frac{\ell(\y|\x)}{\int_{\Theta} \ell(\y|\x)d\x}$, then the marginal likelihood is 
	\begin{align}
	Z = \int_{\Theta} \ell(\y|\x) g_\text{like}(\x)d\x = \frac{\int_{\Theta}  \ell^2(\y|\x)d\x}{\int_{\Theta}  \ell(\y|\x)d\x}. 
	\end{align}
	This idea is connected to {\it posterior predictive approach}, described in Section \ref{SuperSect_PredPost}. Indeed, the marginal likelihood above can be written as $Z=E_{P(\x|\y)}[\ell(\y|\x)]=\int_{\Theta} \ell(\y|\x) P(\x|\y)d\x$ when $g(\x)=1$. 
	Less informative likelihood-based priors can be constructed using a tempering effect with a parameter $0<\beta\leq 1$ or considering only a subset of data $\y_\text{sub}$. For instance, when $\int_{\Theta} \ell(\y|\x)^{\beta}d\x <\infty$ or $\int_{\Theta} \ell(\y_\text{sub}|\x)d\x <\infty$, then we can choose $g_\text{like}(\x) \propto \ell(\y|\x)^{\beta}$ or  $g_\text{like}(\x) \propto \ell(\y_\text{sub}|\x)$, the marginal likelihood is 
	\begin{align}
	Z= \frac{\int_{\Theta} \ell(\y|\x)^{\beta+1}d\x}{\int_{\Theta} \ell(\y|\x)^{\beta}d\x}, \quad \mbox{ or }  \quad Z=\frac{\int_{\Theta} \ell(\y|\x) \ell(\y_\text{sub}|\x) d\x}{\int_{\Theta} \ell(\y_\text{sub}|\x) d\x}.
	\end{align}
This is also the key idea underlying the partial and intrinsic Bayes factors described in the next section. 
\subsection{Bayes factors with improper priors}
\label{AppBayeFactImproperPrior}

So far we have considered proper priors, i.e., $\int_\Theta g(\x) d\x=1$.
The use of improper priors is common in Bayesian inference to represent weak prior information. Consider $g(\x)\propto h(\x)$ where $h(\x)$ is a non-negative function whose integral over the state space does not converge,  $\int_\Theta g(\x)d\x=\int_\Theta h(\x)d\x= \infty$. In that case, $g(\x)$ is not completely specified. Indeed, we can have different  definitions $g(\x)=ch(\x)$ where $c>0$ is (the inverse of) the  ``normalizing'' constant,  not uniquely determinate since $c$ formally does not exist. Regarding the parameter inference and posterior definition, the use of improper priors poses no problems as long as $\int_{\Theta} \ell(\y|\x)h(\x)d\x <\infty$, indeed
\begin{align}
	\post(\x|\y) &= \frac{1}{Z} \pi(\x|\y)=\frac{\ell(\y|\x)ch(\x)}{\int_{\Theta} \ell(\y|\x)ch(\x)d\x} = \frac{\ell(\y|\x)h(\x)}{\int_{\Theta} \ell(\y|\x)h(\x)d\x}, \nonumber \\
	 &=\frac{1}{Z_h}\ell(\y|\x)h(\x)
\end{align}
where $Z=\int_{\Theta} \ell(\y|\x)g(\x)d\x$,   $Z_h=\int_{\Theta} \ell(\y|\x)h(\x)d\x$ and $Z=cZ_h$. Note that the unspecified constant $c>0$ is canceled out, so that the posterior $\post(\x|\y)$ is well-defined even with an improper prior if $\int_{\Theta} \ell(\y|\x)h(\x)d\x <\infty$. However, the issue is not solved when we compare different models, since $Z=cZ_h$ depends on $c$.  For instance, the Bayes factors depend on the undetermined constants $c_1, c_2>0$ \cite{spiegelhalter1982bayes}, 
\begin{align}
	\text{BF}(\y) =\frac{c_1}{c_2} \frac{\int_{\Theta_1} \ell_1(\y|\x_1)h_1(\x_1)d\x_1}{\int_{\Theta_2} \ell_2(\y|\x_2)h_2(\x_2)d\x_2}
	=\frac{Z_{1}}{Z_{2}}=\frac{c_1Z_{h_1}}{c_2Z_{h_2}},
\end{align}
so that different choices of $c_1, c_2$ provide different preferable models.
There exists various approaches for dealing with this issue. Below we describe some relevant ones.
\newline
\newline
{\bf Partial Bayes Factors.} The idea behind the partial Bayes factors consists of using a subset of  data to build proper priors and, jointly with the remaining data, they are used to calculate the Bayes factors. This is related to the likelihood-based prior approach, described above.
The method starts by dividing the data in two subsets, $\y=(\y_\text{train},\y_\text{test})$. The first subset $\y_\text{train}$ is used to obtain partial posterior distributions, 
\begin{align}\label{AquiPriorGm}
\bar{g}_m(\x_m|\y_\text{train})=\frac{c_m}{Z_\text{train}^{(m)}} \ell_m(\y_\text{train}|\x_m)h_m(\x_m),
\end{align}
using the improper priors. The partial posterior $\bar{g}_m(\x_m|\y_\text{train})$ is then employed as prior.
 Note that 
$$ 
Z_\text{train}^{(m)}= c_m \int_{\Theta_m}\ell_m(\y_\text{train}|\x_m)h_m(\x_m)d\x_m.
$$
 Recall that the complete posterior of $m$-th model is
\begin{align}\label{FernandoFollador}
\post_m(\x|\y)=\post_m(\x|\y_\text{test},\y_\text{train})=\frac{c_m}{Z_m}\ell_m(\y|\x_m) h_m(\x_m),
\end{align} 
 where 
$$
Z_m=c_m\int_{\Theta_m} \ell_m(\y|\x_m) h_m(\x_m) d\x_m.
$$
Note that $Z_\text{train}^{(m)}$ and $Z_m$ both depend on the unspecified constant $c_m$.
Considering the conditional likelihood  $\ell_m(\y_\text{test}|\x_m,\y_\text{train})$ of the remaining data $\y_\text{test}$,\footnote{In case of conditional independence of the data given $\x$, we have $\ell_m(\y_\text{test}|\x_m,\y_\text{train})=\ell_m(\y_\text{test}|\x_m)$.  } we can study another posterior of  $\y_\text{test}$, 
\begin{align}\label{FernandoFollador2} 
P_\text{test}^{(m)}(\x|\y_\text{test})=\frac{1}{Z_{\text{test}|\text{train}}^{(m)}} \ell_m(\y_\text{test}|\x_m,\y_\text{train})\bar{g}_m(\x_m|\y_\text{train}),  
\end{align}
where $\bar{g}_m(\x_m|\y_\text{train})$ in \eqref{AquiPriorGm} plays the role of a prior pdf, and  
\begin{align*}
Z_{\text{test}|\text{train}}^{(m)}&=\int_{\Theta_m} \ell_m(\y_\text{test}|\x_m,\y_\text{train})\bar{g}_m(\x_m|\y_\text{train}) d\x_m, \\
&=\int_{\Theta_m} \ell_m(\y_\text{test}|\x_m,\y_\text{train}) \frac{c_m}{Z_\text{train}^{(m)}} \ell_m(\y_\text{train}|\x_m)h_m(\x_m) d\x_m, \\
&=\frac{c_m}{Z_\text{train}^{(m)}} \int_{\Theta_m} \ell_m(\y_\text{test}|\x_m,\y_\text{train})  \ell_m(\y_\text{train}|\x_m)h_m(\x_m) d\x_m, \\
&=\frac{c_m}{Z_\text{train}^{(m)}} \int_{\Theta_m}  \ell_m(\y|\x_m)h_m(\x_m) d\x_m, \\
&= \frac{Z_m}{Z_\text{train}^{(m)}}. 
\end{align*}
Thus, $Z_{\text{test}|\text{train}}^{(m)}$ does not depend on $c_m$. 
Therefore, considering the partial posteriors $\bar{g}_m(\x_m|\y_\text{train})$ as proper priors, we can define the following {\it partial} Bayes factor
\begin{align}\label{PartialBayesFactor}
	\text{BF}(\y_\text{test}|\y_\text{train}) &= \frac{Z_{\text{test}|\text{train}}^{(1)}}{Z_{\text{test}|\text{train}}^{(2)}}= \frac{\frac{Z_1}{Z_\text{train}^{(1)}}}{\frac{Z_2}{Z_\text{train}^{(2)}}}, \nonumber \\
	&=  \frac{\frac{Z_1}{Z_2}}{\frac{Z_\text{train}^{(1)}}{Z_\text{train}^{(2)}}}=\frac{\text{BF}(\y)}{\text{BF}(\y_\text{train})}. \quad \text{{\it(``Bayes law for Bayes Factors'')}}.
\end{align}
Therefore, one can approximate firstly $\text{BF}(\y_\text{train})$, secondly $\text{BF}(\y)$ and then compare the model using the partial Bayes factor $\text{BF}(\y_\text{test}|\y_\text{train})$.
\begin{rem}
	The trick here consists in computing {\it two normalizing constants} for each model, instead of only one.  The first normalizing constant is  used for building an auxiliary proper prior, depending on $\y_\text{train}$.  The difference with the likelihood-based prior approach in previous section is that $\y_{\text{train}}$ is used only once (in the auxiliary proper prior).
\end{rem}
\noindent A training dataset $\y_\text{train}$ is  proper if $\int_{\Theta_m}\ell_m(\y_\text{train}|\x_m)h_m(\x_i)d\x_m < \infty$  for all models, and it is called {\it minimal} if  is proper and no subset of  $\y_\text{train}$ is proper. If we use actually proper prior densities, the minimal training dataset is the empty set and the fractional Bayes factor reduces to the classical Bayes factor. However, the main drawback of the partial Bayes factor approach is the dependence on the choice of $\y_\text{train}$ (which could affect the selection of the model). The authors suggest finding the {\it minimal} suitable training set $\y_\text{train}$, but this task is not straightforward. Two alternatives in the literature have been proposed, the fractional Bayes factors and the intrinsic Bayes factors.
\newline
\newline
{\bf Fractional Bayes Factors \cite{o1995fractional}.} Instead of using a training data, it is possible to use power posteriors,  i.e.,
\begin{align}\label{FractionalBayesFactor}
	\text{FBF}(\y) =\frac{\text{BF}(\y)}{\text{BF}(\y|\beta)},
\end{align}
where the denominator is 
\begin{align}\label{FractionalBayesFactor2}
	\text{BF}(\y|\beta)=\frac{\int_{\Theta_1} \ell_1(\y|\x_1)^{\beta} g_1(\x_1)d\x_1}{\int_{\Theta_2} \ell_2(\y|\x_2)^{\beta} g_2(\x_2)d\x_2}=
	\frac{c_1\int_{\Theta_1} \ell_1(\y|\x_1)^{\beta} h_1(\x_1)d\x_1}{c_2\int_{\Theta_2} \ell_2(\y|\x_2)^{\beta} h_2(\x_2)d\x_2}.
\end{align}
with $0< \beta < 1$, and $\text{BF}(\y|1)=\text{BF}(\y)$. Note that the value $\beta=0$ is not admissible since $\int_{\Theta_m}h_m(\x_m)d\x_m =\infty$ for $m=1,2$.
Again, since both $\text{BF}(\y)$ and $\text{BF}(\y|\beta)$ depend on the ratio $\frac{c_1}{c_2}$, the fractional Bayes factor $\text{FBF}(\y)$ is independent on $c_1$ and $c_2$ by definition.
 \newline
\newline
{\bf Intrinsic Bayes factors \cite{berger1996intrinsic}.} The partial Bayes factor \eqref{PartialBayesFactor} will depend on the choice of (minimal) training set $\y_\text{train}$. These authors solve the problem of choosing the training sample by averaging the partial Bayes factor over all possible minimal training sets. They suggest using the arithmetic mean,  leading to the {\it arithmetic} intrinsic Bayes factor, or the geometric mean, leading to the {\it geometric} intrinsic Bayes factor.


\subsection{Marginal likelihood as a prior predictive approach} \label{PriorPredictive}

Due to the definition of the marginal likelihood $Z=E_g[\ell(\y|\x)]=\int_\Theta \ell(\y|\x)g(\x)d\x$ is also called or related to the so-called {\it prior predictive approach}. As in the Approximate Bayesian Computation (ABC) \cite{LuengoMartino2020}, the idea is that we can generate artificial data $\widetilde{\y}_{i,m}$, $i=1,...,L$ from each $m$-th model with the following procedure: (a) draw $\x_{i,m}$ from the $m$-th prior, $g_m(\x)$ and $\widetilde{\y}_{i,m}$ from the $m$-th likelihood $\ell_m(\y|\x_{i,m})$. Given each set of fake data $\mathcal{S}_m=\{\widetilde{\y}_{i,m}\}_{i=1}^L$, we can use different classical hypothesis testing techniques for  finding the set $\mathcal{S}_m$ closest to the true data $\y$ (for instance, based on $p$-values). Another possibility, we could approximate the value $Z_m=p_m(\y)$ applying kernel density estimation $\widehat{p}_m$ to each set $\mathcal{S}_m$.
\newline
In the next section, we describe the posterior predictive approach, which consider the expected value of likelihood evaluated in a generic $\widetilde{\y}$ with respect to (w.r.t.) the posterior $P(\x|\y)$, instead of w.r.t. the prior $g(\x)$.  The  posterior predictive idea can be considered an alternative model selection approach w.r.t. the marginal likelihood approach, which includes several well-known model selection schemes.

\subsection{Other ways of model selection: the posterior predictive approach}\label{SuperSect_PredPost}

The marginal likelihood approach is not the unique approach for model selection in Bayesian statistics. Here, we discuss some alternatives which are based on the concept of prediction.
\newline
After fitting a Bayesian model, a popular approach for model checking (i.e. assessing the adequacy of the model fit to the data) consists in measuring its predictive accuracy \cite[Chapter 6]{gelman2013bayesian}\cite{piironen2017comparison}.
 Hence, a key quantity in these  approaches is the posterior predictive distribution of generic different data $\widetilde{\y}$ given $\y$,
\begin{align}\label{eq:PPD}
	p(\widetilde{\y}|\y) = E_{\post(\x|\y)}[\ell(\widetilde{\y}|\x)] =\int_\Theta \ell(\widetilde{\y}|\x)\post(\x|\y)d\x.
\end{align}
 Considering $\widetilde{\y}=\y$, note that exists a clear connection with likelihood-based priors described in Section \ref{SuperSuperIMPSect}. 
 
{\rem The posterior predictive distribution in \eqref{eq:PPD} is an expectation w.r.t. the posterior, which is robust to the prior selection with informative data, unlike the marginal likelihood. Therefore, this approach is less affected by the prior choice.}
\newline
\newline
 Note that we can consider  posterior predictive distributions $p(\widetilde{\y}|\y)$ for vectors $\widetilde{\y}$ smaller than $\y$ (i.e., with less components).
  The {\it posterior predictive checking} is based on the main idea of considering some simulated data $\widetilde{\y}_i \sim p(\widetilde{\y}|\y)$, with $i=1,...,L$, and comparing with the observed data $\y$.  After obtaining a set of fake data $\{\widetilde{\y}_i\}_{i=1}^L$, we have to measure the discrepancy between the true observed data $\y$ and the set $\{\widetilde{\y}_i\}_{i=1}^L$. This comparison can be made with test quantities and graphical checks (e.g., posterior predictive p-values). 
\newline
\newline
Alternatively, different measures of predictive accuracy can be employed. An example, is the {\it expected log pointwise predictive density} (ELPD) \cite{vehtari2017practical}. Let recall that ${\bf y}=[y_1,\ldots,y_{D_y}]\in \mathbb{R}^{D_y}$, and define as $\bar{y} \in \mathbb{R}$ any alternative scalar data. Considering $M$ alternative scalar data $\bar{y}_i$ with density $p_{\text{true}}(\bar{y}_i)$, the ELPD is defined as
\begin{align}\label{expectedLPD}
\text{ELPD} &= \sum_{i=1}^{M}\int_{\mathbb{R}} \log p(\bar{y}_i|\y)p_{\text{true}}(\bar{y}_i) d\bar{y}_i \nonumber\\
&=\sum_{i=1}^{M}\int_{\mathbb{R}}  \log\left[\int_{\Theta}  \ell(\bar{y}_i|\x)\post(\x|\y)d\x \right] p_{\text{true}}(\bar{y}_i)d\bar{y}_i.
\end{align}
Note that $p_{\text{true}}(\bar{y}_i)$ is the density representing the true data generating process for $\bar{y}_i$, which is clearly unknown. 
Therefore, some approximations are required. First all, we define an over-estimation of the ELPD, considering the observed data in ${\bf y}=[y_1,\ldots,y_{D_y}]$ instead new alternative data $\bar{y}_i$, so that $M=D_y$ and $\int_{\mathbb{R}} \log p(\bar{y}_i|\y)p_{\text{true}}(\bar{y}_i) d\bar{y}_i \approx  \log p(y_i|\y)$, i.e.,
\begin{equation}\label{expectedLPD2}
\widehat{\text{ELPD}}=\sum_{i=1}^{D_y}\log p(y_i|\y)= \sum_{i=1}^{D_y}\ \log\left[\int_{\Theta}  \ell(y_i|\x)\post(\x|\y)d\x \right].
\end{equation}
In practice, we need an additional approximation for computing $p(y_i|\y)=\int_{\Theta}  \ell(\bar{y}_i|\x)\post(\x|\y)d\x$. We can use MCMC samples from $\post(\x|\y)$, i.e., 
\begin{equation}\label{expectedLPD3}
\widehat{\text{ELPD}}=\sum_{i=1}^{D_y}\log \widehat{p}(y_i|\y)=\sum_{i=1}^{D_y}\log\left[\frac{1}{N}\sum_{n=1}^{N} \ell(y_i|\x_n) \right], \quad \text{ with }  \quad \x_n\sim \post(\x|\y).
\end{equation}
{\bf LOO-CV.} However, we know that the approximation above overestimates ELPD.  One possibility is to use cross-validation (CV), such as the leave-one-out cross-validation (LOO-CV). In LOO-CV, we consider   $p(y_i|\y_{-i})$ instead of $p(y_i|\y)$ in Eq. \eqref{expectedLPD2}, where $\y_{-i}$ is  vector $\y$ leaving out the $i$-th data, $y_i$.  Hence, 
\begin{equation}\label{expectedLPD4}
\widehat{\text{ELPD}}_{\text{LOO-CV}}=\sum_{i=1}^{D_y}\log p(y_i|\y_{-i})= \sum_{i=1}^{D_y}\ \log\left[\int_{\Theta}  \ell(y_i|\x)\post(\x|\y_{-i})d\x \right].
\end{equation}
 For approximating $p(y_i|\y_{-i})=\int_{\Theta}  \ell(y_i|\x)\post(\x|\y_{-i})d\x$, we draw again from the full posterior by means of an MCMC technique,  $\x_n\sim \post(\x|\y)$, and apply importance sampling \cite{vehtari2017practical}, 
\begin{equation}\label{expectedLPD5}
p(y_i|\y_{-i})\approx \widehat{p}(y_i|\y_{-i}) =\sum_{n=1}^N {\bar w}_{i,n}  \ell(y_i|\x_n), \qquad  \x_n\sim \post(\x|\y),
\end{equation}  
where ${\bar w}_{i,n} =\frac{w_{i,n}}{\sum_{k=1}^N w_{i,k}}$ and, in the case the data are conditionally independent,
$$
w_{i,n}=\frac{1}{\ell(y_i|\x_n)}\propto \frac{\post(\x_n|\y_{-i})}{\post(\x_n|\y)}.
$$ 
Thus, replacing in \eqref{expectedLPD5}, we obtain 
\begin{equation}\label{expectedLPD6}
p(y_i|\y_{-i})\approx \widehat{p}(y_i|\y_{-i}) =\frac{1}{\sum_{n=1}^N \frac{1}{\ell(y_i|\x_n)}}, \qquad  \x_n\sim \post(\x|\y),
\end{equation}  
which resembles the harmonic mean estimator but with just one data point. However, since the full posterior $\post(\x_n|\y)$ has smaller variance of $\post(\x_n|\y_{-i})$, the direct use of \eqref{expectedLPD6} is quite unstable, since the IS weights can have high or infinite variance. 
See \cite{vehtari2017practical} for stable computations of LOO-CV and using posterior simulations.
Moreover, see also \cite{piironen2017comparison} for a quantitative comparison of methods for estimating the predictive ability of a model.  The marginal likelihood can also be interpreted as a measure of predictive performance \cite[Sect. 3.2]{kass1995bayes}. In \cite{fong2020marginal}, the authors show that the marginal likelihood is equivalent, in some sense, to a leave-p-out cross-validation procedure. For further discussions about model selection strategies, see \cite{alston2005bayesian,o2009review}.


\section{Numerical comparisons}
\label{NumSimu}
In this section, we compare the performance of different marginal likelihood estimators in different experiments.  
First of all, we consider 3 different illustrative scenarios in Section \ref{Comparison IS vs RIS}, \ref{Example sspp} and  \ref{Example mixture} each one considering different challenges: different overlap between prior and likelihood (changing the number of data, or the variance and mean of the prior), multi-modality and different dimensions of the inference problem. The first experiment also considers two different sub-scenarios.
Additional theoretical results related to the experiments in Sect. \ref{Comparison IS vs RIS} are provided in the Supplementary Material.

\noindent The last two experiments involves a real data analysis. In Section \ref{Example Diccicio}, we test several estimators in a nonlinear regression problem with real data (studied also in \cite{diciccio1997computing}), where the likelihood function has non-elliptical contours. 
Finally, in Section \ref{Example covid} we consider another   regression problem employing non-linear localized bases with real data of the COVID-19 outbreak. 
  

\subsection{First experiment}\label{Comparison IS vs RIS}

\subsubsection{First setting: Gaussians with same mean and different variances}\label{Example aum disp}

In this example, our goal is to compare by numerical simulations different schemes for estimating the normalizing constant of a Gaussian target $\pi(\theta) = \exp( -\frac{1}{2}\theta^2)$. We know the ground-truth  $Z =\int_{-\infty}^{\infty}\pi(\theta)d\theta = \sqrt{2\pi}$, so $\post(\theta) = \frac{\pi(\theta)}{Z} = \mathcal{N}(\theta|0,1)$. 
{Since this is a data-independent example, $\pi(\theta)$ and $\post(\theta)$ have no dependence on $\y$. We compare several estimators enumerated below, considering one or two proposals.}
\newline
\newline
{\bf One proposal estimators (IS and RIS)}. First of all, we recall that the IS vers-1 estimator with importance density $\bar{q}(\theta)$ and the RIS estimator with auxiliary density $f(\theta)$ are
\begin{align*}
\widehat{Z}_\text{IS} = \frac{1}{N}\sum_{i=1}^{N} \frac{\pi(z_i)}{\bar{q}(z_i)}, \quad z_i \sim \bar{q}(\theta), \quad \widehat{Z}_\text{RIS} = \frac{1}{\frac{1}{N}\sum_{i=1}^{N}\frac{f(\theta_i)}{\pi(\theta_i)}}, \quad \theta_i \sim \post(\theta).
\end{align*}
For a fair comparison, we consider
\begin{align*}
	\bar{q}(\theta)=f(\theta) = \mathcal{N}(\theta|0,h^2) = \frac{1}{\sqrt{2\pi h^2}}\exp\left(- \frac{1}{2h^2}\theta^2\right).
\end{align*}
where $h>0$ is the standard deviation.  We desire to study the performance of the two estimators as $h$ varies. {Moreover, a  theoretical comparison of IS and RIS estimators is given in the Supplementary Material.}
{
\newline
\newline	
{\bf Estimators using with two proposals.}	The IS and RIS estimators use a single set of samples from $\bar{q}(\theta)$ or $\post(\theta)$, respectively. Now, we consider the comparison, in terms of MSE, against several estimators that use sets of samples from both densities, $\bar{q}(\theta)$ and $\post(\theta)$, at the same time. Let $\{z_i\}_{i=1}^M$ and $\{\theta_j\}_{j=1}^N$ denote sets of iid samples from $\bar{q}(\theta)$ and $\post(\theta)$, respectively. When $M=N=500$, the set $\{\{z_i\}_{i=1}^M, \{\theta_j\}_{j=1}^N\}$ can be considered as a unique set of samples drawn from the mixture $\frac{1}{2}\post(\theta) + \frac{1}{2}\bar{q}(\theta)$ \cite{ElviraMIS15}. For a fair comparison, these estimators use $\frac{M}{2}$ samples from $\bar{q}(\theta)$  and $\frac{N}{2}$ samples from $\post(\theta)$.
\newline
\newline	
{\bf Ideal and realistic scenarios.} Furthermore, we consider two scenarios, corresponding to whether we can evaluate $\post(\theta)$ (ideal and impossible scenario) or we evaluate $\pi(\theta) \propto \post(\theta)$ (realistic scenario). Note that the first scenario is simply for illustration purposes.
\newline
\newline	
Jointly with IS and RIS estimator, we test several other estimators of $Z$, introduced in Section \ref{TwoDensitiesIS}, that use two sets of samples simultaneously. 
	\begin{itemize}
		\item {\bf Opt-BS:} The optimal bridge sampling estimator  with $\alpha(\theta) = (\frac{1}{2}\post(\theta) + \frac{1}{2}\bar{q}(\theta))^{-1}$.
		\item {\bf Mix-IS:}  IS vers-1 with the mixture $\frac{1}{2}\post(\theta) + \frac{1}{2}\bar{q}(\theta)$, instead of $\bar{q}(\theta)$, as proposal.
		\item {\bf Mix-self IS:} The self-IS estimator, with $f(\theta)=\bar{q}(\theta)$, and the mixture $\frac{1}{2}\post(\theta) + \frac{1}{2}\bar{q}(\theta)$ as the proposal.
\end{itemize}
Moreover, we consider another one proposal estimator, described in Section \ref{UmbrellaForZsect}:
\begin{itemize}
		\item {\bf Opt-self IS:}  The optimal self-IS estimator, with $f(\theta)=\bar{q}(\theta)$. Note that this estimator use samples from a density  to $\bar{q}^{\text{opt}}(\theta)\propto |\post(\theta)-\bar{q}(\theta)|$.  We include it as a reference, for its optimality, and since $\bar{q}^{\text{opt}}(\theta)$ involves both, $\post(\theta)$ and $\bar{q}(\theta)$.
\end{itemize}

{\rem Clearly, in the realistic scenario, all of the schemes above must replaced for their iterative versions, since we cannot evaluate $\post(\theta)$  but only $\pi(\theta) \propto \post(\theta)$.}
\newline
\newline
{\bf Results in ideal scenario.} Figures \ref{comparison2}(a)-(b) show the MSE of the estimators versus $h$ (which is the standard deviation of $\bar{q}(\theta)$) in the ideal scenario.
IS vers-1 can have very high MSE when $h<1$, i.e., $\bar{q}(\theta)$ has smaller variance then the $\post(\theta)$. 
Whereas, IS vers-1 is quite robust when $h> 1$. 
The MSE of RIS has the opposite behavior of IS vers-1. This is because RIS needs that $\bar{q}(\theta)$ has lighter tails  than $\post(\theta)$. 
In this example, optimal bridge sampling seems to provide performance in-between the IS and RIS estimators. The MSE of Opt-BS is closer to RIS for $h<1$, whereas Opt-BS becomes closer to IS for $h>1$. Conversely,  the MSE of Opt BS is not smaller than that of IS or RIS for any $h$ in this example. Finally, Mix-IS and Mix-self-IS provide the best performance, even  better than the optimal self-IS estimator. But this is due to we are in an ideal, unrealistic scenario.
\newline
\newline
{\bf Results in the realistic scenario.}	  Since $Z$ is unknown we cannot evaluate $\post(\theta)$ but only $\pi(\theta) \propto \post(\theta)$. Only IS and RIS can be truly applied. The rest of above estimators must employ an iterative procedure (see  Section \ref{UmbrellaForZsect} and Section \ref{TwoDensitiesIS}). 
	The iterative versions of these estimators evaluate $\frac{1}{2}\pi(\theta)/\widehat{Z}^{(t)} + \frac{1}{2}\bar{q}(\theta)$, where $\widehat{Z}^{(t)}$ is the current approximation.
	In Figure \ref{iteratives}, we show these three estimators after $T=5$ and $T=15$ iterations. 
	Interestingly, note that they all converge to the results of Opt-BS estimator. This means that the iterative versions of Mix-IS and Mix-self-IS are two alternative of Opt-BS in practice, and the performance obtained in ideal scenario are unachievable. However, the iterative version of Opt-BS seems to have the fastest convergence (to the results of the ideal Opt-BS), w.r.t. the iterative versions of Mix-IS and Mix-self-IS.
	\newline
	We also include a two-stage version of the Opt-selfIS estimator (see Section \ref{UmbrellaSect}). This estimator employs $\frac{N}{4}$ to obtain an approximation $\widehat{Z}$ via standard IS, and then draws $\frac{3}{4}N$ samples from a density proportional to $|\pi(\theta)/\widehat{Z} - \bar{q}(\theta)|$. This two-stage Opt-selfIS depends on the quality of the initial approximation of $\widehat{Z}$. Since this initial approximation is provided by IS, and since IS is problematic when $h<1$, the two-stage self-IS does not perform better than Opt-BS  for $h<1$.


	\begin{figure}[!h]
		\centering
		\subfigure[]{\includegraphics[width=8cm]{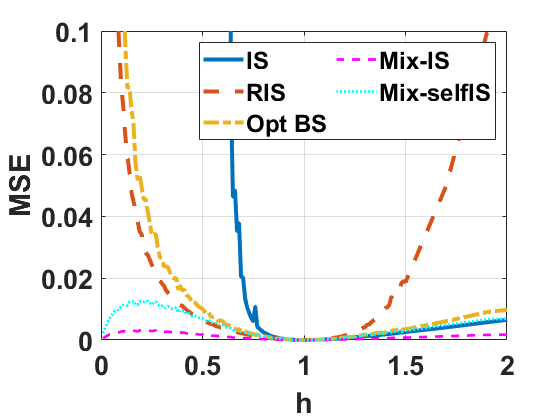}}
		\subfigure[]{\includegraphics[width=8cm]{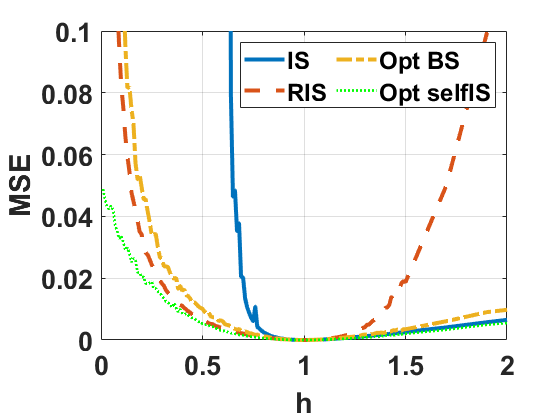}}
		
		\caption{{ Numerical comparison with estimators using samples from $\bar{q}(\theta)$ and $\post(\theta)$, and optimal self-IS. The figure shows the MSE of each method (averaged over 2000 simulations) as a function of $h$. }}
		\label{comparison2}
	\end{figure}

	\begin{figure}[!h]
		\centering
		\subfigure[T=5]{\includegraphics[width=8cm]{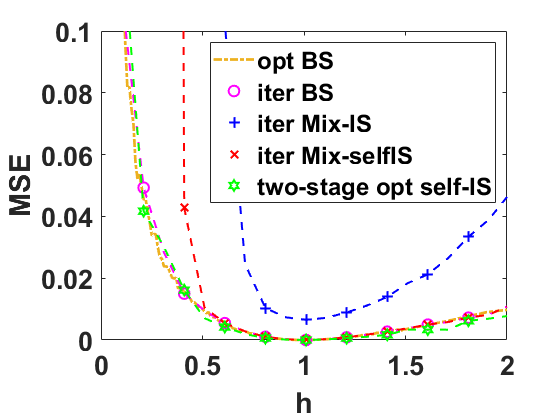}}
		\subfigure[T=15]{\includegraphics[width=8cm]{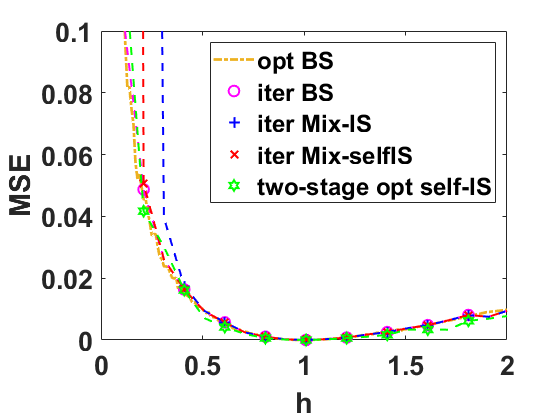}}
		
		\caption{{ Comparison of iterative version of estimators with a very far starting value, $\widehat{Z}^{(0)}=5000$, with $T=5$  and $T=15$. Note that the two-stage self-IS is not iterative (see Section \ref{UmbrellaSect}).} }
		\label{iteratives}
	\end{figure}

	\subsubsection{Second setting: Gaussians with same variance and different means}\label{Example alejandose}

	In this setting, we consider  again $\post(\theta) \propto \pi(\theta) = \exp\left(-\frac{1}{2}\theta^2\right)$,i.e., $\post(\theta) = \frac{\pi(\theta)}{Z}=\mathcal{N}(\theta|0,1)$, but the proposal is $\bar{q}(\theta)=\mathcal{N}(\theta|\mu,1)$ for $\mu\geq 0$. Namely, as $\mu$ grows, $\bar{q}(\theta)$ and  $\post(\theta)$ are more distant. A theoretical comparison of IS and RIS estimators is given in the Supplementary Material, also for this setting.
\newline	
\newline		
	Similarly, we compare the MSE as function of  $\mu$ of  different estimators of $Z$: (a) IS vers-1, (b) RIS, (c) optimal BS (Opt-BS), (d) a suboptimal self-IS estimator with $f(\theta)=\bar{q}(\theta)$ and using $\bar{q}(\theta)=\mathcal{N}(\frac{\mu}{2},1)$ as proposal, 
	and (e) the Opt-self IS estimator with $f(\theta)=\bar{q}(\theta)$ and proposal $\bar{q}^{\text{opt}}(\theta)\propto |\post(\theta)-\bar{q}(\theta)|$.   Each estimator is computed using $500$ samples in total and the results are  averaged over 2000 independent simulations.
\newline	
\newline	
{\bf Results of the second setting.} Unlike in the first setting, here we consider only the ideal scenario (i.e., without iterative procedures).	
However, note that the  suboptimal self-IS  scheme would not require an iterative version.
The results are shown in Figure \ref{comparison_alej}. The MSE of both IS and RIS diverge as $e^{\mu^2}$. 
	Opt-BS shows better performance than IS vers-1 and RIS. 
	The suboptimal self-IS estimator performs similarly to the Opt-BS, but both are worse than the Opt-self IS estimator. In this example, the estimators that use a middle density (as Opt-BS and the self-IS estimators) are less affected by the problem of $\post(\theta)$ and $\bar{q}(\theta)$ becoming further apart. 
	As in the previous setting, we expect that the iterative versions of Opt-BS converges to the results of the ideal Opt-BS, provided in Figure \ref{comparison_alej}. 
	Recall that, for approximating the Opt-self IS, we require a two-stage procedure. However, a procedure with just two stages could be not enough, as we showed in the previous setting. Hence, an iterative application of the two-stage procedure could be employed (becoming actually an adaptive importance sampler).

	\begin{figure}[!h]
		\centering
		{\includegraphics[width=8cm]{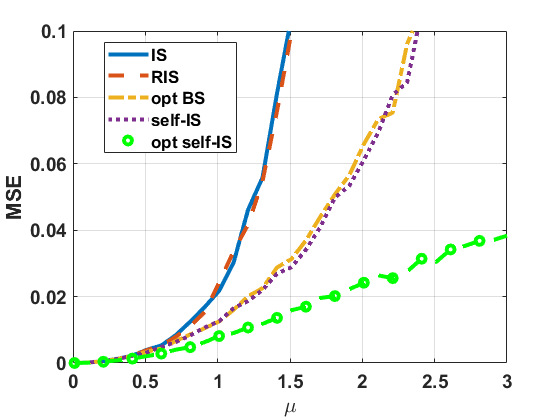}}
		\caption{{Numerical comparison of IS, RIS, Opt-BS, suboptimal self-IS and Opt self-IS. The figure shows the MSE of each method (averaged over 2000 simulations) as a function of $\mu$. Greater $\mu$ means $\post(\theta)$ and $\bar{q}(\theta)$ are further apart.}}
		\label{comparison_alej}
	\end{figure}

	\subsection{Second experiment: Gaussian likelihood and uniform prior} \label{Example sspp} 
	Let us consider the following one-dimensional  example. More specifically,  we consider independent data $\y = [y_1,\dots,y_{D_y}]$ generated according to a Gaussian observation model,
	$$
	\ell(\y|\theta) = \prod_{i=1}^{D_y}\ell(y_i|\theta) = \frac{1}{(\sqrt{2\pi}\sigma)^{D_y}}\exp\left\{-\frac{D_y}{2\sigma^2}[(\theta - \bar{y})+s_{y}]\right\},
	$$
	where $\sigma=3$, $\bar{y}$ and $s_{y}$ denote the sample mean and sample variance of $\y$, respectively. We consider a uniform prior $ g(\theta) = \frac{1}{2\Delta},\ \theta \in [-\Delta,\Delta]$ with $\Delta>0$ being the prior width. 
	In this setting, the marginal likelihood $Z$ can be obtained in closed-form as a function of $\Delta$ and $n$ (considering the evaluation of the error function $\mbox{erf}(x)$).
	 The posterior is a truncated Gaussian $\post(\theta|\y) \propto \mathcal{N}(\theta|\bar{y},\frac{\sigma^2}{D_y}),\ \theta\in[-\Delta,\Delta]$.
	Let $\beta\in[0,1]$ denote an inverse temperature, the power posterior is 
	\begin{align}
	\post(\theta|\y,\beta) \propto \mathcal{N}\left(\theta\Big|\bar{y}, \frac{\sigma^2}{D_y\beta}\right), \quad \mbox{ restricted to } \theta\in[-\Delta,\Delta].
	\end{align}
	For any $\beta$, we can sample $\post(\theta|\y,\beta) \propto \ell(\y|\theta)^{\beta}g(\theta)$
	with rejection sampling by drawing from $\mathcal{N}(\theta|\bar{y},\frac{\sigma^2}{D_y\beta})$ and discarding the samples that fall outside $[-\Delta, \Delta]$. 
\newline
\newline
{\bf  Scenario 1: $\Delta=10$ and $D_y=10$.} We start by setting $\Delta=10$ and generating $D_y=10$ data points from $\mathcal{N}(0,\sigma^2)$ with $\sigma=3$. 
	The value of the marginal likelihood is $\log Z = -25.2853$.
	We aim to compare the performances of several methods in estimating $\log Z$: 
	(a) Naive Monte Carlo (NMC), (b) Harmonic mean (HM), (c)  IS with a tempered posterior as proposal (IS-P),
      (d)   stepping stone sampling (SS), (e)  power posterior method (PP), and (f)  path sampling (PS). 
      
      {\rem Estimating $\log Z$, instead of directly $Z$, helps  the methods of PP and PS, with respect to  NMC, HM, IS-P and SS (making their results worse).   }
	\newline
	\newline
	We establish a total budget of $N=10^3$ likelihood evaluations. 
	For SS and PP, we set $K+1$ values of $\beta$, from $\beta_0=0$ to $\beta_K=1$, chosen (i) uniformly, i.e., $\beta_k = \frac{k}{K}$ for $k=1,\dots,K$, or (ii) concentrated around $\beta=0$, i.e., $\beta_k = \left(\frac{k}{K}\right)^{1/\alpha}$ with $\alpha=0.25$.
	Hence the uniform case is obtained when $\alpha=1$.
	Note that SS draws samples from $K$ distributions, while PP draw samples from $K+1$ distributions.
	For fair comparison, we sample $\floor*{\frac{N}{K}}$ times from each $\post(\theta|\y,\beta_k)$, for $k=0,\dots,K-1$, in SS, and $\floor*{\frac{N}{K+1}}$ times from of each $\post(\theta|\y,\beta_k)$, for $k=0,\dots,K$, in PP. For IS-P we test $\beta_1=0.5$ and $\beta_2=0.5^4$ and draw $N$ samples from each of the $\post(\theta|\y,\beta_1)$ and $\post(\theta|\y,\beta_2)$. For PS, we sample $N$ pairs $(\beta',\theta')$ as follows: we first sample $\beta'$ from a $\mathcal{U}(0,1)$ and then sample $\theta'$ from the corresponding power posterior $\post(\theta|\y,\beta')$. Naive Monte Carlo uses $N$ independent samples from prior and HM uses $N$ independent samples from the posterior.
\newline
\newline
{\bf Results scenario 1.} In Figure \ref{10_datos}(a), we show 500 independent estimations from each method. We observe that NMC works very well in this scenario since the prior acts as a good proposal. SS with $K=2$ provides also good performance, since half the samples come from the prior with this choice of $K$.  The value of $\alpha$ seems to be not important for SS in this case. 
	PS  performs as well as NMC and SS, but shows a slightly bigger dispersion.  HM tends to overestimate the marginal likelihood, which is a well-known issue. 
	The estimation provided by IS-P depends on the choice of $\beta$. For $\beta_1=0.5$, the power posterior is closer to the posterior so its behavior is similar to HM. For $\beta_2=0.0625$ the power posterior is close to the prior, and IS-P tends to underestimate $Z$. Recall that IS-P has a bias since it is a special case of IS vers-2.  PP performs poorly with $K=2$, due to the discretization error in \eqref{PowerOneDim}, which improves when considering the value $K=35$. The choice $\alpha=0.25$, w.r.t. $\alpha=1$, improves the performance in PP. 
\newline	
In Figure \ref{10_datos}(b), we show the mean absolute error (MAE) in estimating $\log Z$ of SS and PP as a function of $K$. We depict two curves for each method, corresponding to the choices $\alpha=1$ and $\alpha=0.25$. We can observe that the errors obtained in SS and PP when $\alpha=0.25$ are smaller than when $\alpha=1$ for any $K$. This is in line with the recommendations provided in their original works. We note that the error of SS slightly deteriorates as $K$ grows: for $K>2$, less and less samples are drawn from the prior, which is a good proposal in  this scenario (with $\Delta=10$ and $D_y=10$). The performance of PP improves drastically as $K$ grows, since larger $K$ means that the trapezoidal rule is more accurate in approximating \eqref{PowerOneDim}. SS and PP, for $\alpha=1$ and $\alpha=0.25$, approach the same limit when $K$ grows, achieving an error which is always greater than the one obtained by NMC, in this scenario. 
\newline
\newline	
{\bf  Scenario 2: $\Delta=1000$ and $D_y=100$.} Now, we replicate the previous experiment  increasing the number of data, $D_y=100$, and the width of the prior, $\Delta=1000$. The joint effect of increasing $D_y$ and $\Delta$ makes the likelihood become extremely concentrated w.r.t. the prior, hence decreasing the value of the marginal likelihood, being $\log Z = -267.6471$.
	Moreover, this high discrepancy between prior and posterior is reflected in the power posteriors $\post(\theta|\y,\beta)$, which will be very similar to the posterior except for very small values of $\beta$. 
	We compare all the methods described before with a total budget of $N=10^3$ likelihood evaluations.  Additionally, we also test a PS where $\beta'\sim \mathcal{B}(0.25,1)$, i.e., from a beta distribution which provides more $\beta'$ values closer to $0$.
\newline
\newline
{\bf Results scenario 2.}	In Figure \ref{100_datos}(a), we can see that, unlike in the previous scenario, the NMC tends to underestimate the marginal likelihood, since the likelihood is much more concentrated than the prior. 
	The HM and the two implementations of IS-P provide similar results, overestimating $Z$: in this case, the posterior is so different from the prior that $\post(\theta|\y,0.5)$ and $\post(\theta|\y,0.06)$ are very similar to the posterior.
	PS with  $\beta' \sim [0,1]$ tends to overestimate $Z$: since the $\beta'$'s are drawn uniformly in $[0,1]$, many samples $(\beta',\theta')$ are drawn in high-valued likelihood zones. Indeed, at least the bias is reduced when we test PS with  $\beta'\sim \mathcal{B}(0.25,1)$. We also show the results of one implementation of SS (with $K=10$ and $\alpha=0.25$) and PP (with $K=70$ and $\alpha=0.25$).  Both greatly outperform the rest of estimators in this scenario, providing accurate estimations. In Figure \ref{100_datos}(b), we show again the MAE of SS and PP as a function of $K$ for two values $\alpha=1$ and $\alpha=0.25$. The error of PP, with either $\alpha=1$ or $\alpha=0.25$, decreases as $K$ grows, although it decreases more rapidly when considering $\alpha=0.25$. The error of SS with $\alpha=0.25$ decreases as $K$ grows, but increases with $K$ when $\alpha=1$. 	
Again, PP requires the use bigger values of $K$ with respect to SS. In both methods, the choice of $\alpha<1$, i.e., concentrating $\beta$'s near $\beta=0$ where $\post(\theta|\y,\beta)$ is usually changing rapidly, shows to improve the overall performance.

	\begin{figure}[!h]
		\centering
		\subfigure[]{\includegraphics[width=8cm]{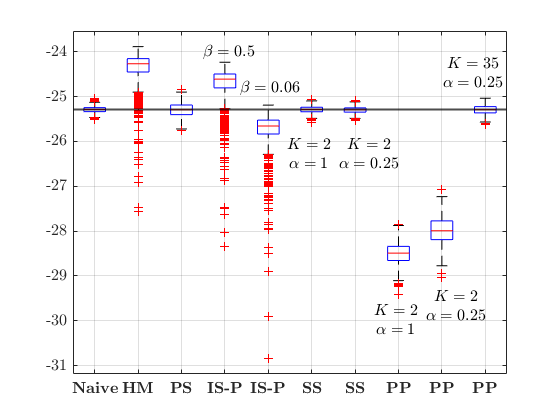}}
		\subfigure[]{\includegraphics[width=8cm]{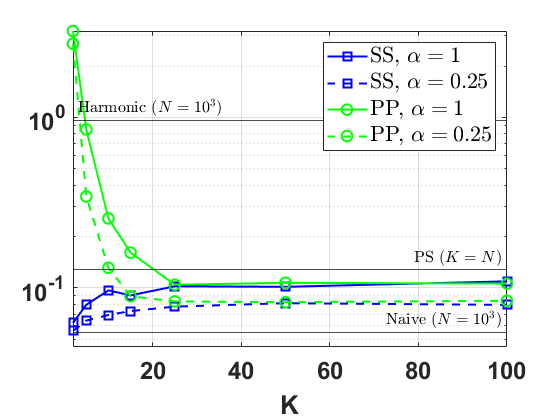}}
		
		\caption{{Simulations when $D_y=10$ and $\Delta=10$: (a) Estimates of $\log Z$ in 500 independent simulations, (b) MAEs of SS and PP as a function of $K$ for two values of $\alpha$.} }
		\label{10_datos}
	\end{figure}
	\begin{figure}[!h]
		\centering
		\subfigure[]{\includegraphics[width=8cm]{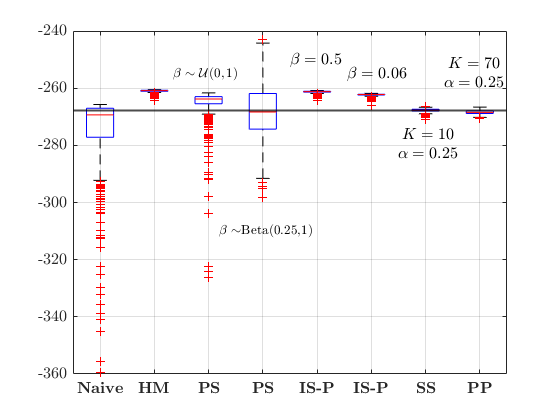}}
		\subfigure[]{\includegraphics[width=8cm]{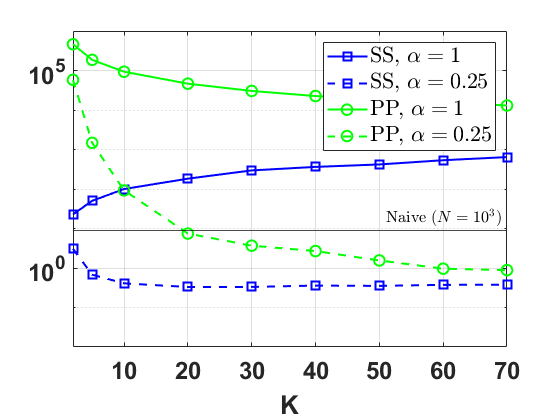}}
		
		\caption{{Simulations when $D_y=100$ and $\Delta=1000$: (a) Estimates of $\log Z$ in 500 independent simulations, (b) MAEs of SS and PP as a function of $K$ for two values of $\alpha$.}}
		\label{100_datos}
	\end{figure}

}

{
\subsection{Third experiment: posterior as mixture of two components}\label{Example mixture} 
We consider a posterior 
which is a mixture of two $D_\theta$-dimensional Gaussian densities. It is a conjugate model where the likelihood is Gaussian and the prior is a mixture of two Gaussian.
Given the observation vector ${\bf y}$, we consider a $D_\theta$-dimensional Gaussian likelihood function 
\begin{align}
\ell({\bf y}|\x) = \mathcal{N}({\bf y}|\x,\bm{\Lambda}),
\end{align}
with covariance $\bm{\Lambda}$, and a $D_\theta$-dimensional Gaussian mixture prior
\begin{align}
g(\x) = \alpha_\text{prior} \mathcal{N}(\x|\bm{\mu}_{\text{pr}}^{(1)}, \bm{\Sigma}_{\text{pr}}^{(1)}) + (1-\alpha_\text{prior})\mathcal{N}(\x|\bm{\mu}_{\text{pr}}^{(2)}, \bm{\Sigma}_{\text{pr}}^{(2)}),
\end{align}
with $\alpha_\text{prior}\in [0,1]$, $\bm{\mu}_{\text{pr}}^{(i)}$ and $\bm{\Sigma}_{\text{pr}}^{(i)}$ being the prior means and covariances of each component of the mixture, respectively. Then, the posterior is also a mixture of two Gaussian densities
\begin{align}\label{NormalizedPosteriorMixGaussExample1}
\post(\x|\y)= \alpha_\text{post} \mathcal{N}(\x|\bm{\mu}_{\text{post}}^{(1)},\bm{\Sigma}_{\text{post}}^{(1)}) + (1-\alpha_\text{post})\mathcal{N}(\x|\bm{\mu}_{\text{post}}^{(2)},\bm{\Sigma}_{\text{post}}^{(2)}),
\end{align}
where the parameters $\alpha_\text{post}\in [0,1]$, $\bm{\mu}_{\text{post}}^{(i)}$ and $\bm{\Sigma}_{\text{post}}^{(i)}$ can be obtained in closed-form from $\alpha_\text{prior}$, $\bm{\mu}_{\text{pr}}^{(i)}$, $\bm{\Sigma}_{\text{pr}}^{(i)}$, ${\bm \Lambda}$ and $\y$. 
Thus, having the analytical expression of the posterior in closed-form allows to compute exactly the marginal likelihood $Z$ (recall $Z = \frac{\pi(\x|\y)}{\post(\x|\y)}$ for any $\x$). In this case, we can also draw samples directly from the posterior. We can interpret this scenario as the use of an ideal MCMC scenario, where the performance is extremely good.  
We compare different estimators of $Z$ changing the Euclidean distance between the means of posterior mixture components,
\begin{align}
{\texttt{dist}}=||\bm{\mu}_{\text{post}}^{(1)}-\bm{\mu}_{\text{post}}^{(2)}||_2,
\end{align}
in $D_\theta=1$ and $D_\theta=5$. 
This distance can be controlled by changing the distance between the prior modes. More specifically,  we choose $\bm{\Lambda}=50{\bf I}_D$, $\bm{\Sigma}_{\text{pr}}^{(1)}=\bm{\Sigma}_{\text{pr}}^{(2)} = 30{\bf I}_D$, where ${\bf I}_D$ denotes the $D$-dimensional identity matrix. The data is a single observation $\y = -0.5 \bm{1}_D$, where ${\bm{1}}_D$ a $D$-dimensional vector of $1$'s. For the prior means we chose $\bm{\mu}_{\text{pr}}^{(1)}=-\bm{\mu}_{\text{pr}}^{(2)}=L \bm{1}_D$, so $\norm{\bm{\mu}_{\text{pr}}^{(1)}-\bm{\mu}_{\text{pr}}^{(2)}}_2=2L\sqrt{D_\theta}$. We can change the distance between the modes of the prior, and hence between the modes of the posterior, by varying $L \in \mathbb{R}^+$.
Specifically, we select $L\in \{1,6,11,16,21,26,31,36,41,46,51\}$ and compare: (i) the Naive-MC estimator, (ii) the HM estimator, (iii) Laplace-Metropolis estimator, (iv) RIS, and (v) CLAIS.
The budget is $10^4$ posterior evaluations. 
In RIS, we set $f(\x)$ to be the mixture in Eq. \eqref{CLAISeq}, that results after applying a clustering algorithm (e.g., k-means algorithm) to the $10^4$ posterior samples. In CLAIS, we use an analogous mixture obtained from $5\cdot10^3$ posterior samples, and then use it to draw other $5\cdot10^3$ samples in the lower layer (hence the total number of posterior evaluations is $10^4$). 
For RIS and CLAIS, we set the number of clusters to $C=4$. 
RIS and  CLAIS also need setting the bandwidth  parameter $h$ (see Eq. \eqref{CLAISeq}). 
We find that  the choices $h=2$ for RIS and $h=10$ for CLAIS show the average performance of both. 
We test the techniques in dimension $D_\theta=1$ and $D_\theta=5$.
We compute the relative Mean Absolute Error (MAE) in the estimation of $Z$, averaged over $200$ independent simulations. 

The results are depicted in Figure \ref{newFig}. They show that RIS and the CLAIS achieve the best overall performances. Their relative error remain small and rather constant for all distances considered, for $D_\theta=1$ and $D_\theta=5$. The RIS estimator performs as well as CLAIS in both $D_\theta=1$ and $D_\theta=5$, and even better for small distances in $D_\theta=1$.   
For the smallest distance, the lowest relative error corresponds to the Naive MC estimator, since prior and posterior are very similar in that case, although it rapidly gets outperformed by  RIS  and CLAIS. The Laplace estimator provides poor results as $\texttt{dist}$ grows, since the posterior becomes bimodal. As one could expect, the estimators that make use of the posterior sample  to adapt its importance density, i.e., RIS and CLAIS, achieve best performances, being almost independent to increasing the distance between the modes. The HM estimator confirms its reputation of relative bad estimator.

\begin{figure}[h!]
	\centering
	\centerline{
	\subfigure{\includegraphics[width=0.45\textwidth]{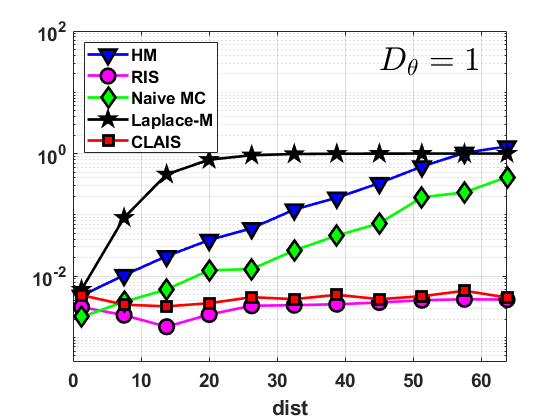}}	
		\hspace{-0.55cm}	
	\subfigure{	\includegraphics[width=0.45\textwidth]{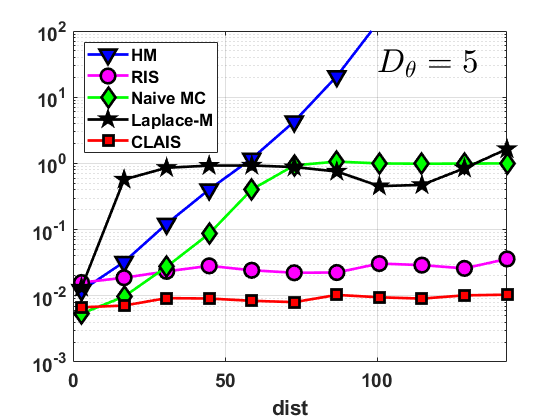}}
	}
	\vspace{-0.25cm}
	\caption{{\small Relative MAE versus $\texttt{dist}$ in  dimension $D_\theta=1$ and dimension $D_\theta=5$.}
	}
	\label{newFig}
\end{figure}

}

\subsection{Experiment with biochemical oxygen demand data}\label{Example Diccicio} 
We consider a numerical experiment studied also in \cite{diciccio1997computing}, that is a nonlinear regression problem modeling data on the biochemical oxygen demand (BOD) in terms of time instants. The outcome variable  $Y_i=\text{BOD}$ (mg/L) is modeled in terms of $t_i=\text{time}$ (days) as
\begin{align}\label{2dNonlinearRegressionModel}
Y_i = \theta_1(1-e^{-\theta_2t_i}) + \epsilon_i, \enskip i=1,\dots,6,
\end{align}
where the $\epsilon_i$'s are independent $\mathcal{N}(0,\sigma^2)$ errors, hence $Y_i \sim \mathcal{N} (\theta_1(1-e^{-\theta_2t_i}),\sigma^2)$. 
The data $\{y_i\}_{i=1}^6$, measured at locations $\{t_i\}_{i=1}^6$, are shown in Table \ref{DiCiccioExTable} below. 
\begin{table}[!h]
\begin{center}
	\caption{Data of the numerical experiment in Section \ref{Example Diccicio}.} \label{DiCiccioExTable}
	\vspace{0.2cm}
	\begin{tabular}{ ll }
		$t_i$ (days)& $y_i$ (mg/L)  \\
		\hline
		1 & 8.3 \\
		2 & 10.3 \\
		3 & 19.0 \\
		4 & 16.0 \\
		5 & 15.6 \\
		7 & 19.8 	
	\end{tabular}
\end{center}
\end{table}

\noindent
The goal is to compute the normalizing constant of the posterior of $\x = [\theta_1,\theta_2]$ given the data $\y=\{(t_i,y_i)\}_{i=1}^6$. 
Following \cite{diciccio1997computing}, we consider uniform priors for  $\theta_1 \sim \mathcal{U}([0,60])$, and  $\theta_2 \sim \mathcal{U}([0,6])$, i.e., $g_1(\theta_1) = \frac{1}{60}$ for  $\theta_1 \in [0,60]$, and $g_2(\theta_2) = \frac{1}{6}$, with $\theta_2 \in [0,6]$. Moreover, we consider an improper prior for $\sigma$,  $g_3(\sigma) \propto \frac{1}{\sigma}$. However, we will integrate out the variable $\sigma$. Indeed, the two-dimensional target $\pi(\x|\y)=\pi(\theta_1,\theta_2|\y)$ results after integrating out $\sigma$ by marginalizing 
$$
\pi(\theta_1,\theta_2,\sigma|\y) = \ell(\y|\theta_1,\theta_2,\sigma)g_1(\theta_1)g_2(\theta_2)g_3(\sigma),
$$
w.r.t. $\sigma$, namely we obtain
\begin{align}
\pi(\x|\y) &= \int \pi(\theta_1,\theta_2,\sigma|\y)d\sigma = \ell\left(\y|\theta_1,\theta_2\right)g_1(\theta_1)g_2(\theta_2) \\
&=\frac{1}{60}\frac{1}{6} \frac{1}{\pi^3}\dfrac{8}{ \left\{ \sum_{i=1}^6[y_i-\theta_1(1-\exp(-\theta_2t_i))]^2 \right\}^3}, \enskip [\theta_1,\theta_2] \in [0,60]\times[0,6],
\end{align}
for which we want to compute its normalizing constant $Z=\int \pi(\x|\y)d\x$.  The derivation is given in the Supplementary Material.
The true value (ground-truth) is $\log Z= -16.208$, considering the data in Table \ref{DiCiccioExTable}. 
\newline
\newline
{{\bf Scenario 1.}}
As in \cite{diciccio1997computing}, we compare the relative MAE,
$
\frac{\E\left[ |\widehat{Z}-Z| \right]}{Z},
$ obtained by different methods: {\b (a)} the naive Monte Carlo estimator; {\bf (b)}  a modified version of the Laplace method (more sophisticated) given in \cite{diciccio1997computing}; {\bf (c)} the  Laplace-Metropolis estimator in Sect. \ref{LaplaceSect} (using sample mean and sample covariance considering MCMC samples from $\post(\x|\y)$); {\bf (d)}  the HM estimator of Eq. \ref{HarmonicMeanEst}; {\bf (f)}  the RIS estimator where $f(\x) = \mathcal{N}(\x|\bm{\mu}, {\bf \Sigma})$, where $\bm{\mu}$ and ${\bf \Sigma}$ are the mean and covariance of the MCMC samples from $\post(\x|\y)$ (it is denoted as RIS in Table \ref{Tabla Ejemplo 2 nuestros results}); {\bf (g)} another RIS scheme where $f(\x)$ is obtained by a clusterized KDE with ${C}=4$ clusters and $h=0$ (in a similar fashion of Eq. \eqref{CLAISeq}); and, finally, a CLAIS scheme with ${C\in\{1,2\}}$, $h=0$  i.e., as in Eq. \eqref{CLAISeq}. 

\begin{table}[!h]
\caption{Relative MAE, and its corresponding standard error, in estimating the marginal likelihood by seven methods} \label{Tabla Ejemplo 2 nuestros results}
\begin{center}
	\begin{tabular}{|c|c|c|c|c|c|c|c|c|}
	\hline
		 Methods & Naive & Laplace (soph)   &  Laplace & HM & RIS & RIS-kde  & {CLAIS} & CLAIS\\
		\hline
		RE & 0.057 & 0.181  &  0.553  &   0.823 &  0.265 &  0.140 & 0.084 &0.082 \\
		std err &  0.001  & 0.013 &  0.003 & 0.018  &  0.006 &   0.004 & 0.015 & 0.014\\
		\hline
		\hline
		 comments &--- & see \cite{diciccio1997computing} &--- &--- & --- & ${C}=4$   &{$C=1$}  & ${C}=2$  \\
		\hline
	\end{tabular}
\end{center}
\end{table}
\noindent All estimators consider 10000 posterior evaluations.
To obtain the samples from the  posterior,  we run $T=10000$ iterations of a Metropolis-Hastings algorithm, using the prior as an independent proposal pdf. 
{The IS estimator employs $5000$ posterior samples to build the normal approximation to the posterior, from which it draws 5000 additional samples.} Similarly, since CLAIS draws additional samples from $\bar{q}(\x)$ in the lower layer, in order to provide a fair comparison, we consider  $N=1$ {(i.e. one chain)}, with $T'=T/2=5000$ iterations and sample $5000$ additional samples in the lower layer.  We averaged the relative MAE over 1000 independent runs.
  Our results are shown in Table \ref{Tabla Ejemplo 2 nuestros results}.

In this example, and with these priors, the results show that the best performing estimator in this case is the Naive Monte Carlo, since prior and likelihood has an ample overlapping region of probability mass. However, the naive Monte Carlo scheme is generally inefficient when there is a small  overlap between likelihood and prior. Note also that IS and CLAIS provide good performance. 
{
RIS-kde performs better than RIS since the choice of $f(\x)$ in the former is probably narrower than in RIS. 
}
The worst performance is provided by the HM estimator.  
\newline
\newline
{
{\bf Scenario 2.}  Now, we consider the following estimators: {\bf (a)} the  Chib's estimator  in Eq. \eqref{eq:ChibsEst}, {\bf (b)} RIS with $f(\x)$ equal to the clusterized KDE in \eqref{CLAISeq} (called RIS-kde in the previous scenario), and ({\bf (c)} CLAIS with clusterized KDE in \eqref{CLAISeq}. We study the effect of the choice of $C$ and $h$ in their performance. 
We test different  numbers of clusters $C\in \{1,2,4,10\}$ and  different values of $h=\{0,1,2,3,4,5\}$. 
\newline
As above, we consider a fair application of CLAIS (using the same budget of posterior evaluations as in the other schemes).
Moreover, in Chib's we need to choose the point $\x^*$. We considered two scenarios: (i) using $\x^* = [19,1]$ that is intentionally located very close to the posterior mode; (ii) using random $\x^*$ drawn from the priors. The first scenario clearly yields more accurate results than the second one, which we refer as a ``fair'' scenario (since, generally, we do not have information about the posterior modes). In summary, we compute the relative MAE of $\widehat{Z}_{\text{chib}}$, $\widehat{Z}_{\text{chib-f}}$ (where the ``f'' stands for ``fair''), $\widehat{Z}_{\text{RIS}}$ and $\widehat{Z}_{\text{CLAIS}}$.  We compute the relative median absolute error of $1000$ independent runs. Figure \ref{FigSIMU} shows the results of the experiment. 
CLAIS and RIS provide results, for all $C$ and $h$, similar to both Chib and Chib-f.  As expected, the error of $\widehat{Z}_\text{chib}$ is lower than $\widehat{Z}_\text{chib-f}$. 
In CLAIS, we note that, for $C=10$, we should not take $h$ too small to avoid the proposal becoming problematic (i.e., narrower than the posterior). Generally, as $C$ increases, $h$ should not be too small since the proposal may not have fatter tails than $\post(\x|\y)$.
The performance of RIS is best when $h=0$, and gets worse as $h$ increases, as expected, since $f(\x)$ may become wider than the posterior. We expect that the results of RIS with $h=0$ would improve further as $C$ increases since the C-KDE pdf, in Eq. \eqref{CLAISeq}, will have lighter tails than the posterior. The Chib's estimator provides also robust and good results. 
Overall, for the choices of $C$ and $h$ considered, CLAIS and RIS (with $f(\x)$ being the clusterized KDE) provide robust results comparable to Chib's estimator. These results are also in line with the theoretical considerations given in the Suppl. Material regarding RIS and IS.

\begin{figure}[h!]
	\centering
	\centerline{
	\subfigure{
		\includegraphics[width=0.35\textwidth]{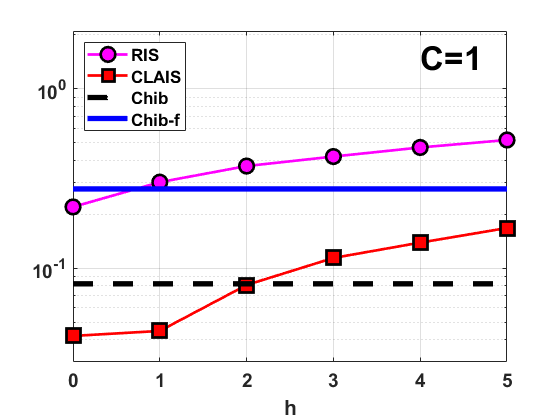}
	}
	\subfigure{	\includegraphics[width=0.35\textwidth]{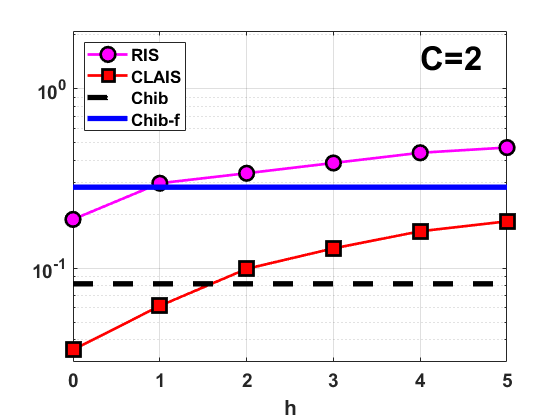}	 }
}
\centerline{
	\subfigure{	\includegraphics[width=0.35\textwidth]{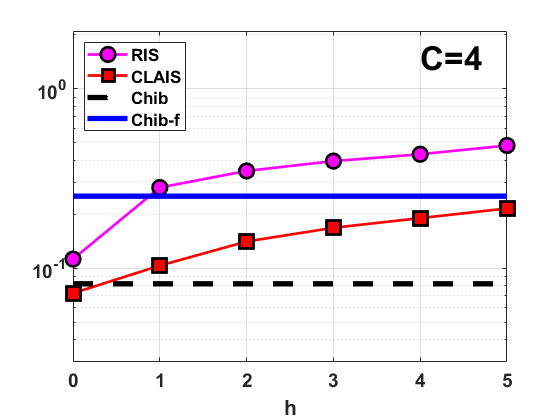}	
}
	\subfigure{	\includegraphics[width=0.35\textwidth]{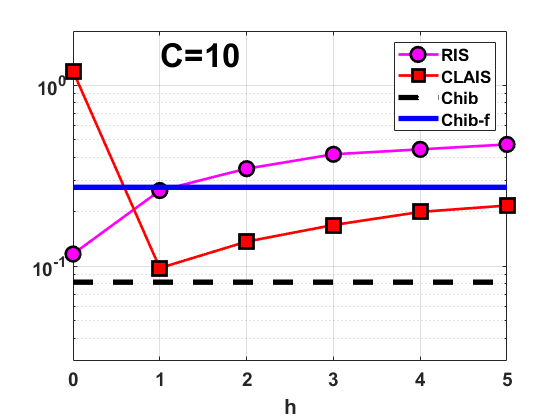}	
}
}
	\vspace{-0.25cm}
	\caption{
		{\small Relative median absolute error of RIS and CLAIS versus $h$ for $C \in \{1,2,4,10\}$. The horizontal lines correspond to Chib's estimator (dashed) and its fair application (solid).
		}
			}
	\label{FigSIMU}
\end{figure}

}

{
	\subsection{Experiment with COVID-19 data}\label{Example covid} 
	
	Let us consider data $\y = [y_1,\dots,y_{D_y}]^\top$ representing the number of daily deaths caused by SAR-CoV-2 in Italy from 18 February 2020 to 6 July 2020. Let $t_i$ denote the  $i$-th day, we model the each observation as 
	$$
	y_i = f(t_i) + e_i, \quad i=1,\dots,D_y=140,
	$$
	where $f$ is the function that we aim to approximate and $e_i$'s are Gaussian perturbations. We consider the approximation of $f$ at some $t$ as a weighted sum of $M$ localized basis functions,
	$$
	{f}(t)=\sum_{m=1}^M { \rho}_m \psi(t|\mu_m,h,\nu),
	$$
	where $\psi(t|\mu_m,h)$ is  $m$-th basis centered at $\mu_m$ with bandwidth $h$. Let also be  $\nu$ an  index denoting the type of basis.  We consider $M\in\{1,....,D_y\}$, then $M\leq D_y$. When $M=D_y$, the model becomes a Relevance Vector Machine (RVM), and the interpolation of all data points (maximum overfitting, with zero fitting error) is possible \cite{MartinoJesse2020}.  
\newline	
We consider $4$ different types of basis (i.e., $ \nu=1,...,4$): Gaussian ($\nu=1$), Laplacian ($\nu=2$), Rectangular ($\nu=3$)  and Triangular-Pyramidal ($\nu=4$). Given $\nu$ and $M$, we select the locations $\{\mu_m\}_{m=1}^M$ as a uniform grid in the interval  $[1,D_y]$ (recall that $D_y=140$).  Hence, knowing $\nu$ and $M$, the locations $\{\mu_m\}_{m=1}^M$ are given.
\newline	
\newline	
{\bf Likelihood and prior of ${\bm \rho}$.} Let ${\bm \Psi}$ be a $D_y\times M$ matrix with elements $[\bm{\Psi}]_{i,m} = \psi(t_i|\mu_m,h)$ for $i=1,\dots,D_y$ and $m=1,\dots,M$, and let $\bm{\rho} = [\rho_1,\dots, \rho_M]^\top$ be the vector of coefficients, where $M$ is the total number of bases. Then, the observation equation in vector form becomes
	$$
	{\bf y}={\bm \Psi} {\bm \rho}+{\bf e},
	$$ 
	where ${\bf e}$ is a $D_y \times 1$ vector of noise. We assume normality ${\bf e} \sim \mathcal{N}({\bf 0},\sigma_e^2 {\bf I}_{D_y})$, where ${\bf I}_{D_y}$ is the $D_y\times D_y$ identity matrix. Therefore, the  likelihood function is  $\ell({\bf y}| {\bm \rho},h,\sigma_e, \nu, M)=\mathcal{N}({\bf y}|{\bm \Psi} {\bm \rho},\sigma_e^2 {\bf I}_N)$. We also consider a Gaussian prior density over the vector of coefficients  ${\bm \rho}$, i.e., $g({\bm \rho}|\lambda) =\mathcal{N}({\bm \rho}|{\bf 0}, {\bm \Sigma}_\rho)$.
	where $ {\bm \Sigma}_\rho= \lambda {\bf I}_M$ and $\lambda>0$. Given $\nu$, $M$, $h$ and $\sigma_e$.
Thus, the complete set of parameters is $\{\bm{\rho}, \nu, M, h, \lambda, \sigma_e\}$.
\newline
\newline
{\bf Posteriors and marginalization.} With our choice of $g({\bm{\rho}}|\lambda)$, the posterior of ${\bm \rho}|\lambda,h,\sigma_e$ is also Gaussian, 
	\begin{eqnarray*}
	\post({\bm \rho}|{\bf y},\lambda,h,\sigma_e,\nu, M) = \frac{\ell({\bf y}| {\bm \rho},h,\sigma_e,\nu, M)g({\bm \rho}|\lambda)}{p({\bf y}|\lambda,h,\sigma_e,\nu, M)} 
	= \mathcal{N}({\bm \rho}|{\bm \mu}_{\rho|y}, {\bm \Sigma}_{\rho|y}),   
	\end{eqnarray*}
	and a likelihood marginalized w.r.t. ${\bm \rho}$ is available in closed-form, 
	\begin{equation}
	p({\bf y}|\lambda,h,\sigma_e,\nu, M) =\mathcal{N}({\bf y}|{\bf 0}, {\bm \Psi} {\bm \Sigma}_\rho{\bm \Psi}^{\top}+\sigma_e^2 {\bf I}_N).
	\end{equation}
	For further details see \cite{MartinoJesse2020}.
	Now, we
	consider priors over $h,\lambda,\sigma_e$, 
	and study the following posterior 
	$$
	\post(\lambda,h,\sigma_e|\y,\nu, M) = \frac{1}{p(\y|\nu, M)} p(\y|\lambda, h,\sigma_e,\nu, M)g_\lambda(\lambda) g_h(h) g_\sigma(\sigma_e),
	$$  
	where $g_\lambda(\lambda)$, $g_h(h)$, $g_\sigma(\sigma_e)$ are folded-Gaussian pdfs defined on $\mathbb{R}_+=(0,\infty)$ with location and scale parameters $\{0,100\}$, $\{0,400\}$ and $\{1.5,9\}$, respectively.
	Finally, we want to compute the marginal likelihood of this posterior, i.e.,
	\begin{align}
		 p(\y|\nu, M) = \int_{\mathbb{R}_+^{3}} p(\y|\lambda,h,\sigma_e,\nu, M)g_\lambda(\lambda) g_h(h) g_\sigma(\sigma_e) d\lambda dh d\sigma_e.
	\end{align}
Furthermore,  assuming a uniform probability mass  $\frac{1}{D_y}$ as prior over $M$, we can also marginalize out $M$, 
	\begin{align}
p(M|\y,\nu) \propto \frac{1}{D_y}p(\y|\nu, M) \quad \mbox{and } \quad p(\y|\nu)= \frac{1}{D_y} \sum_{M=1}^{D_y}p(\y|\nu, M),   \quad \mbox{ for } \quad \nu=1,...,4. 
	\end{align}
	Considering also a uniform prior over $\nu$, we can obtain $
	p(\nu|\y) \propto \frac{1}{4}p(\y|\nu)$.
	\newline
	For approximating $p(\y|\nu, M)$, for $m=1,\dots,D_y$,  we first apply a  Naive Monte Carlo (NMC) method with $N=10^4$ samples. Secondly, we run an MTM algorithm for obtaining the estimator $\widehat{Z}^{(2)}$ (see Table \ref{AIMTM_Table}) 
	and a Markov chain of vectors $\x_t=[\lambda_t,h_t,\sigma_{e,t}]$ for $t=1,\dots,T$. 
	This generated chain $\{\x_t\}_{t=1}^T$ can be also used for obtaining other estimators (e.g., the HM estimator).  
	We consider  the pairs  $T=50$, $N'=1000$, in the MTM scheme. Therefore, $\widehat{Z}^{(2)}$ employs $N'T=5\cdot 10^4$ samples. 
\newline
 {\bf Goal.}  Our purpose is: (a) to make inference regarding the parameters of the model $\{\lambda,h,\sigma_e\}$, (b) approximate $Z=p(\y|\nu, M)$, (c) study the posterior $p(M|\y,\nu)$, and (d) obtain the MAP value, $M_\nu^*$, for $\nu=1,...,4$.  
We also study the marginal posterior $p(\nu|\y)$ of  each of the four candidate bases.
  \newline
	{\bf Results.}  We run once NMC and MTM for all $M=1,...,D_y=140$ different models and approximate the posterior $p(M|\y,\nu)$ for each value of $M$. For illustrative reasons, in Figure \ref{fig_nu_123}, we show the posterior probabilities of $M$ belonging to the intervals $[4\widetilde{M}-3,4\widetilde{M}]$, where $\widetilde{M}$ is an auxiliary index $\widetilde{M}=1,...,\frac{140}{4}=35$. Thus, the first value, $\widetilde{M}=1$ of the curves in Figure \ref{fig_nu_123}, represents the probability of $M\in\{1,2,3,4\}$, the second value represents the probability of $M\in\{5,6,7,8\}$, and so on until the last value, $\widetilde{M}=35$, which represents the probability of $M\in\{137,138,139,140\}$. We can observe that, with both techniques, we obtain that $\widetilde{M}=2$ is the most probable interval,  with a probability generally closer to $0.2$, hence $M_\nu^* \in \{5,6,7,8\}$.  Recall that we have $35$ possible intervals (values of $\widetilde{M}$), so when we compare with a uniform distribution $\frac{1}{35}=0.0286$, the value $0.2$ is quite high. For $\nu=2,3$, the corresponding probabilities are greater than $0.2$, reaching $0.35$ with NMC in $\nu=2$. In Figure \ref{Fig_fit}, we can observe that, with $M=8$ bases, we are already able to obtain a very good fitting to the data. 
	
	Thus, a first conclusion is that the results obtained with models such as RVMs and Gaussian Processes (GPs) (both having $M=140$ \cite{MartinoJesse2020}) can be approximated in a very good way with a much more scalable model, as our model here with  $M \in \{5,6,7,8\}$  \cite{MartinoJesse2020}.  Regarding  the marginal posterior $p(\nu|\y)$, we can observe the results in Table \ref{Locura_ComemeTheDickWithIcreasingStrength}. The basis $\nu=3$ is discarded since is clearly not appropriate, as also shown graphically by Figure \ref{Fig_fit}. With the results provided by NMC, we prefer slightly the Laplacian basis whereas, with the results  of MTM, we have almost $p(\nu=1|\y) \approx p(\nu=2|\y)$. These considerations are reasonable after having a look to Figure \ref{Fig_fit}. 
	As future work, it would be interesting to consider the locations of the bases $\mu_m$, for $m=1,\dots,M$, as additional parameters to be learnt.

\begin{table}[!h]
{
\caption{The approximate marginal posterior $p(\nu|\y)$ with different techniques.} \label{Locura_ComemeTheDickWithIcreasingStrength}
\vspace{-0.4cm}
\begin{center}
	\begin{tabular}{|c|c|c|c|c|c|}
	\hline
		  {\bf Method} & Number of used samples &  $p(\nu=1|\y)$  &  $p(\nu=2|\y)$  & $p(\nu=3|\y)$  & $p(\nu=4|\y)$  \\
		\hline
	         NMC &  $10^4$ & 0.3091  &    0.3307  &  0.0813 &   0.2790\\
	          MTM &$5\cdot 10^4$ & 0.3155   & 0.3100   & 0.0884  &  0.2861 \\
		\hline
	\end{tabular}
\end{center}
}
\end{table}

\begin{figure}[!h]
	\centering
	\centerline{
\includegraphics[width=0.3\textwidth]{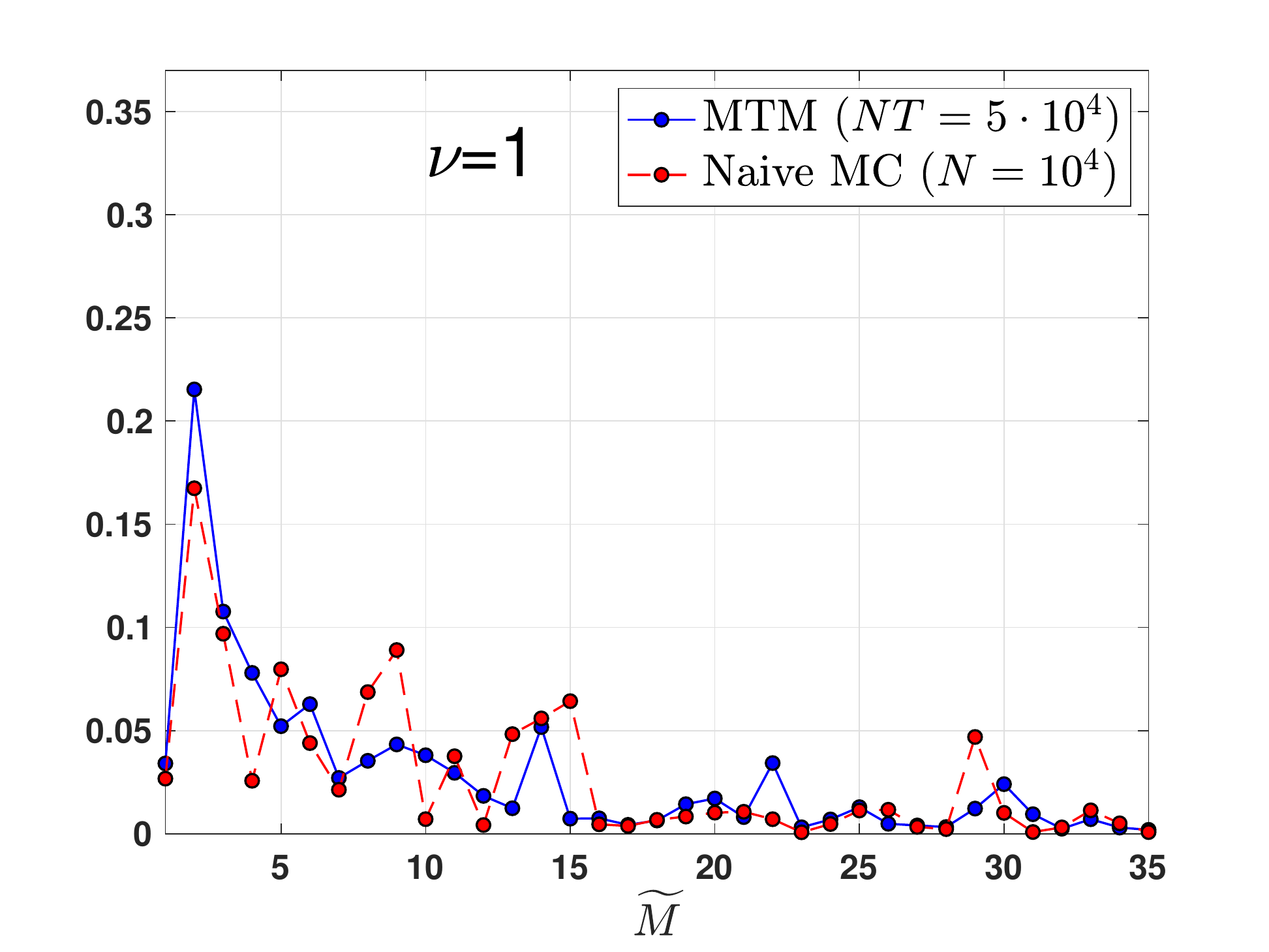}
	\hspace{-0.6cm}
\includegraphics[width=0.3\textwidth]{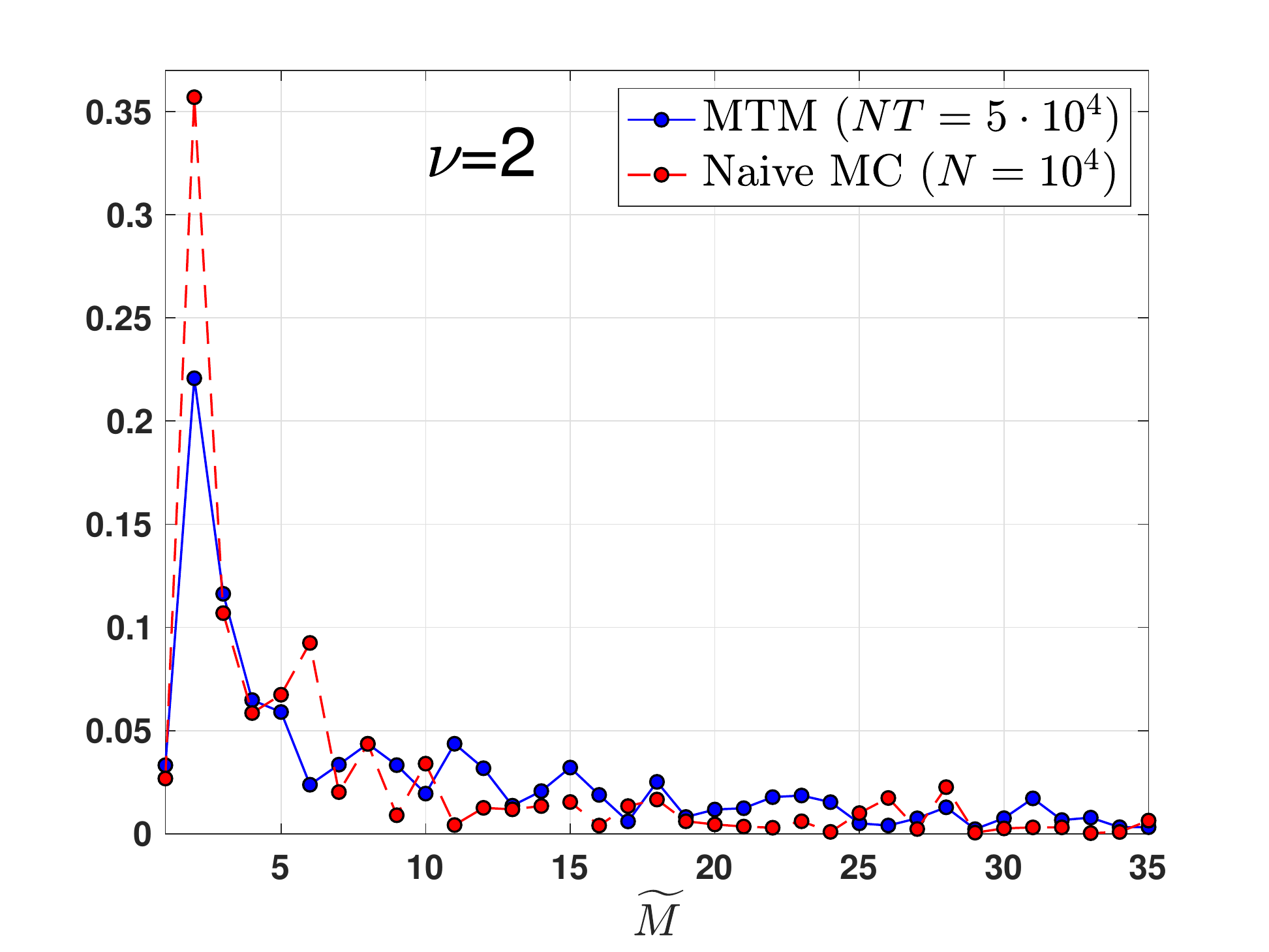}
	\hspace{-0.6cm}
	\includegraphics[width=0.3\textwidth]{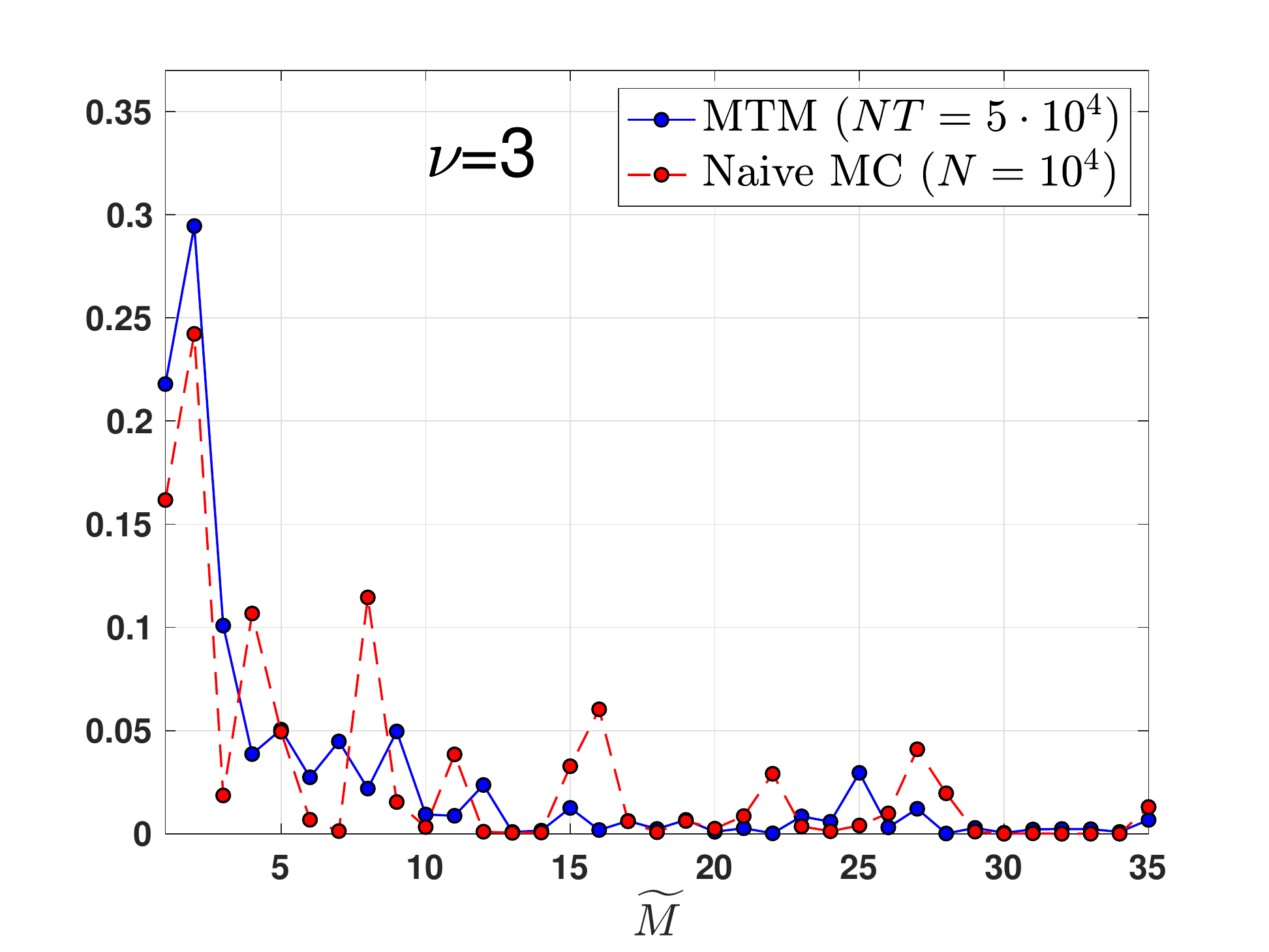}
	\hspace{-0.6cm}
          \includegraphics[width=0.3\textwidth]{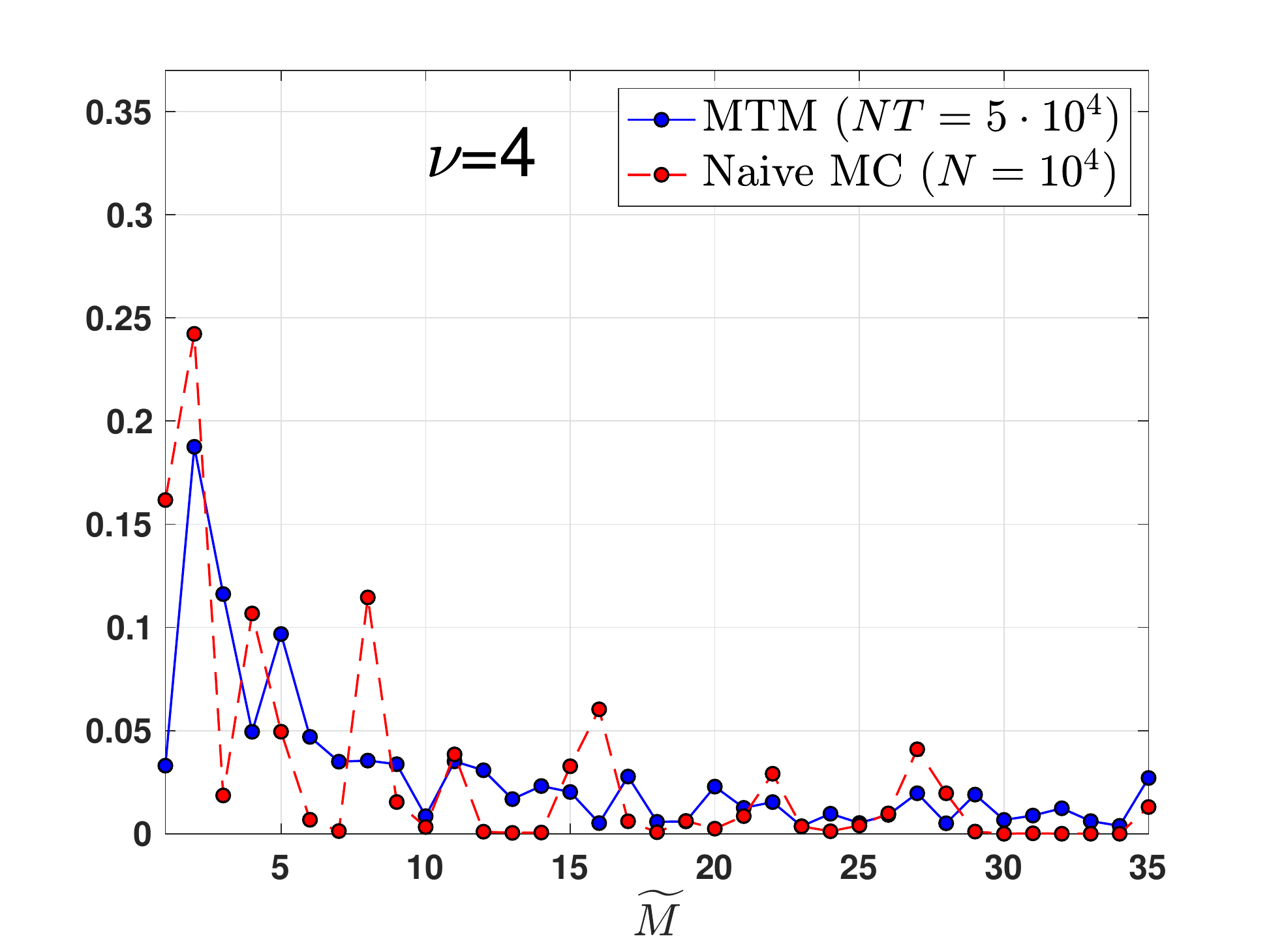}
}
\caption{\footnotesize
	{ Posterior probabilities of the intervals $[4\widetilde{M}-3,4\widetilde{M}]$ with  $\widetilde{M}=1,...,35$, obtained adding $4$ consecutive values of $p(M|\y,\nu)$ with $M\in\{4\widetilde{M}-3,4\widetilde{M}-2,4\widetilde{M}-1,4\widetilde{M}\}$ (and $p(M|\y,\nu)$ is approximated by NMC or MTM). Each figure corresponds to a different type of basis, $\nu=1,2,3,4$.}
}
\label{fig_nu_123}
\end{figure}

\begin{figure}[!h]
	\centering
	\centerline{
\includegraphics[width=0.4\textwidth]{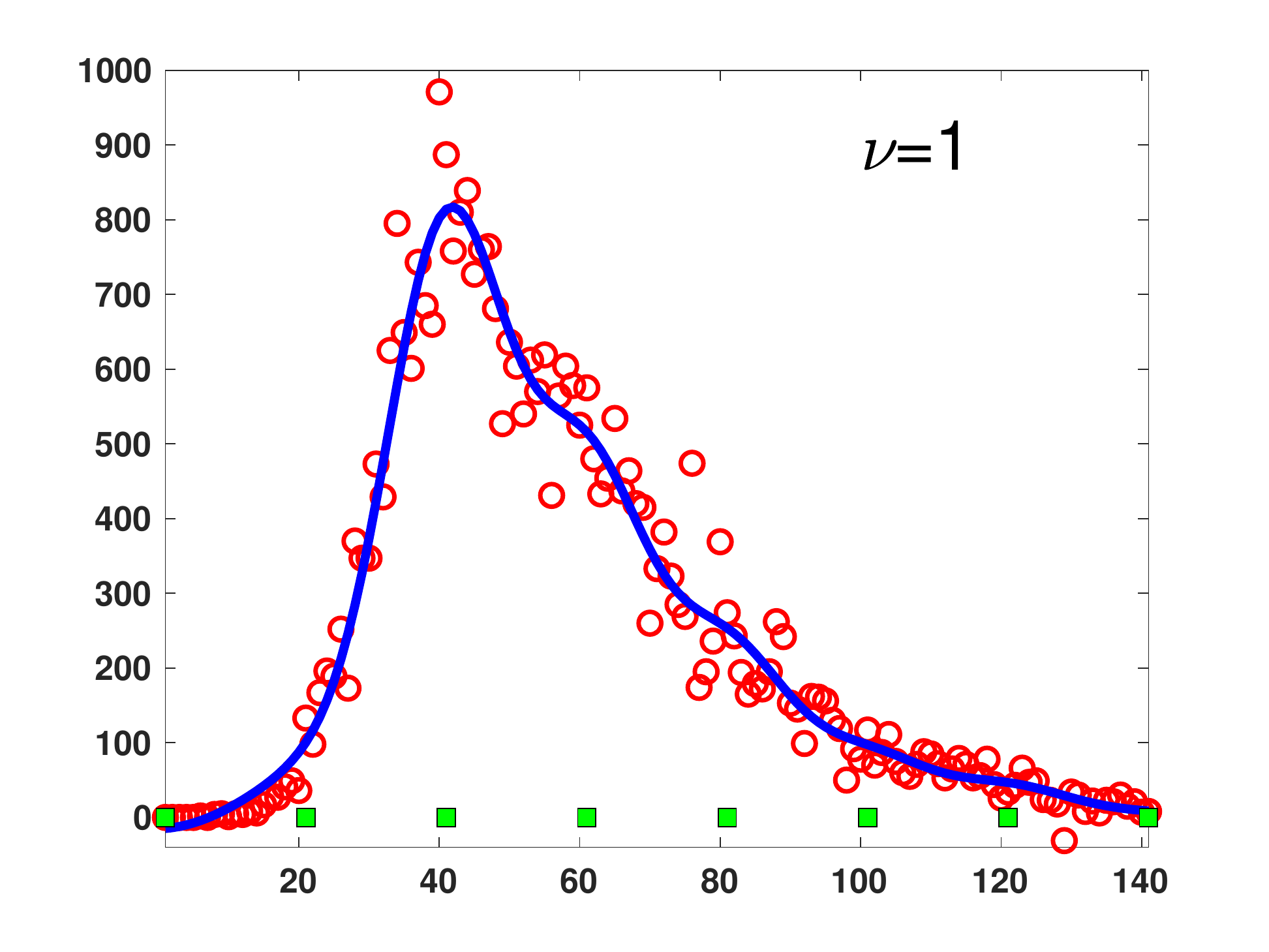}
\includegraphics[width=0.4\textwidth]{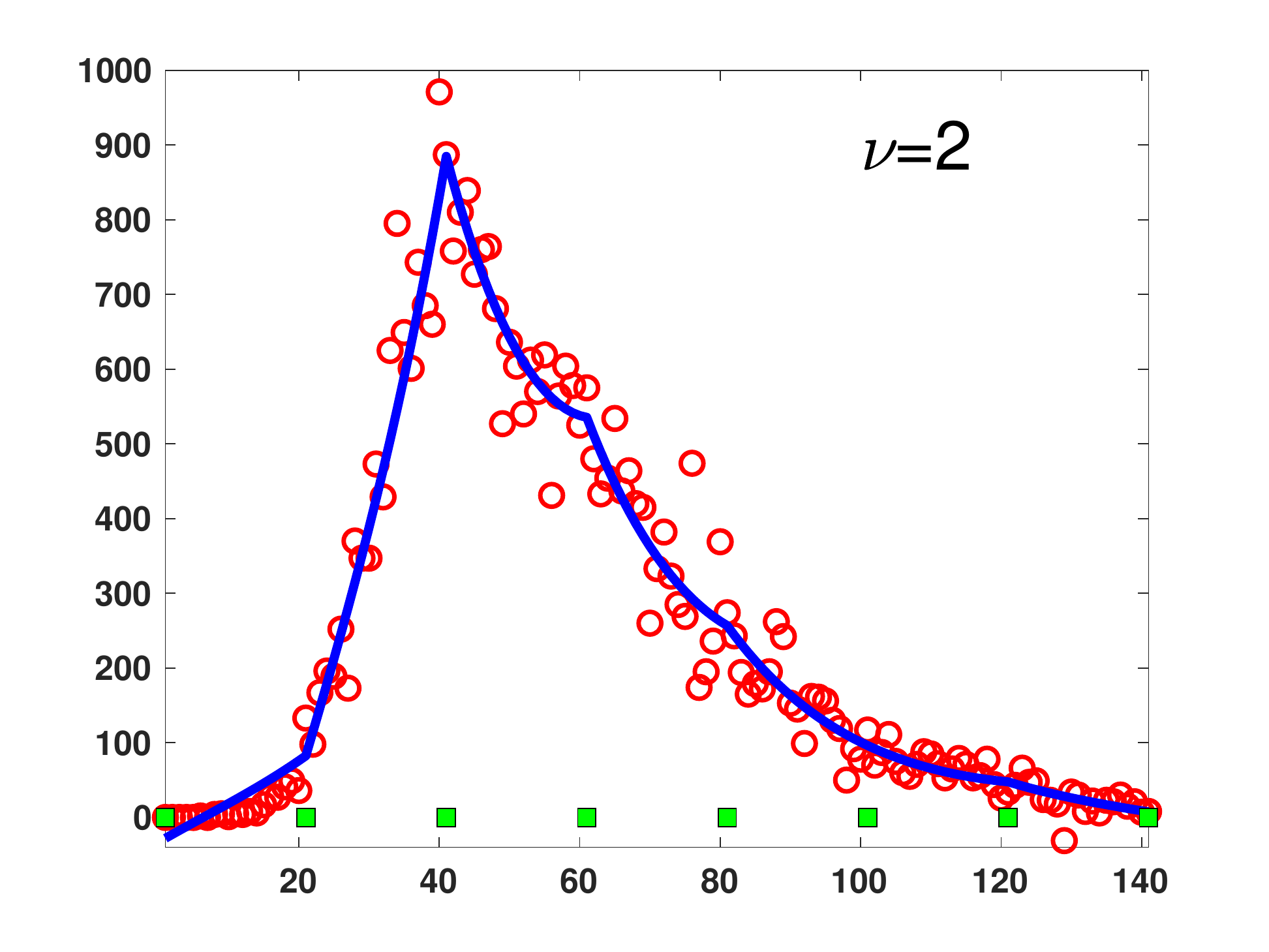}
}
\centerline{
  \includegraphics[width=0.4\textwidth]{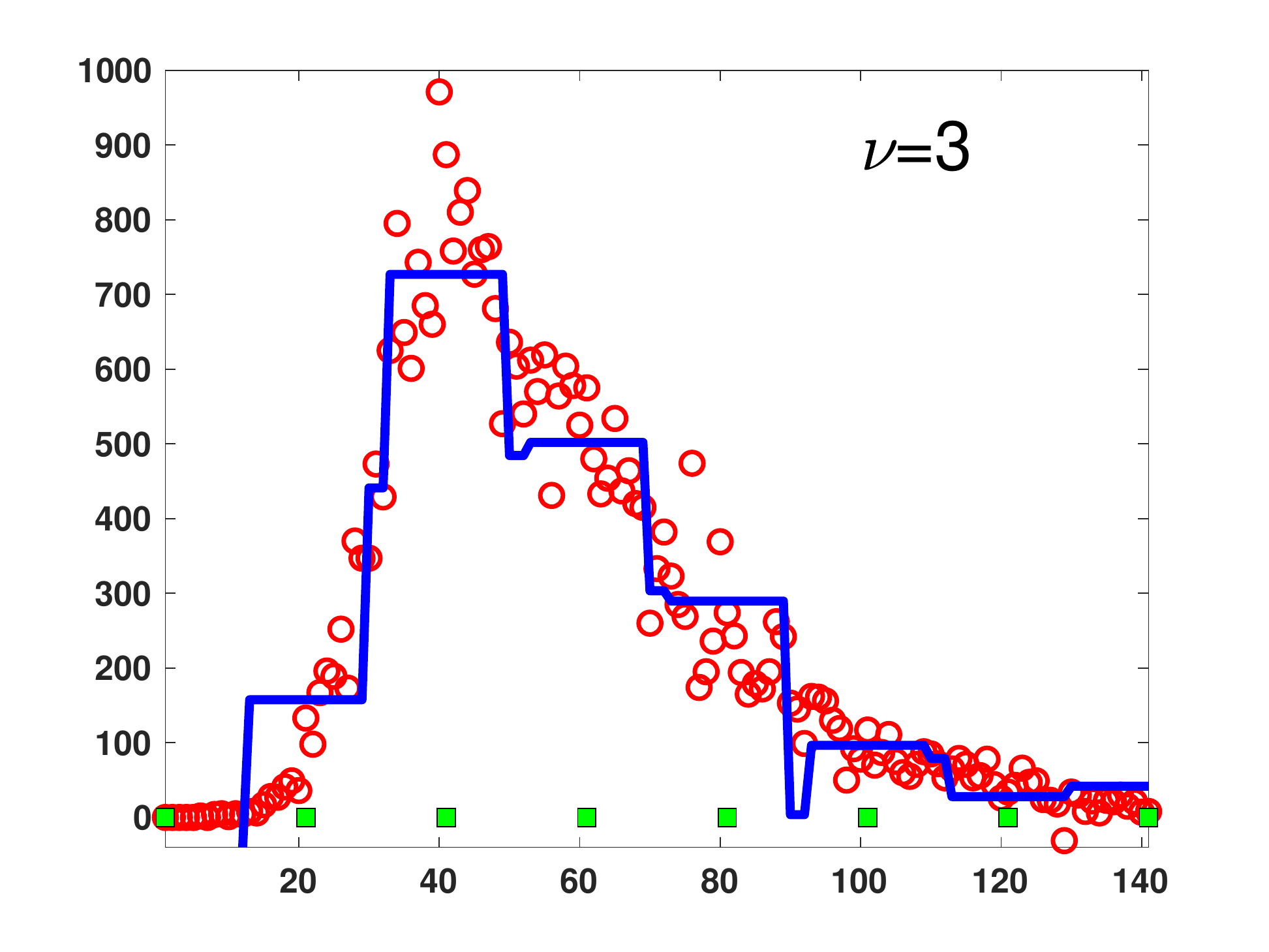}
          \includegraphics[width=0.4\textwidth]{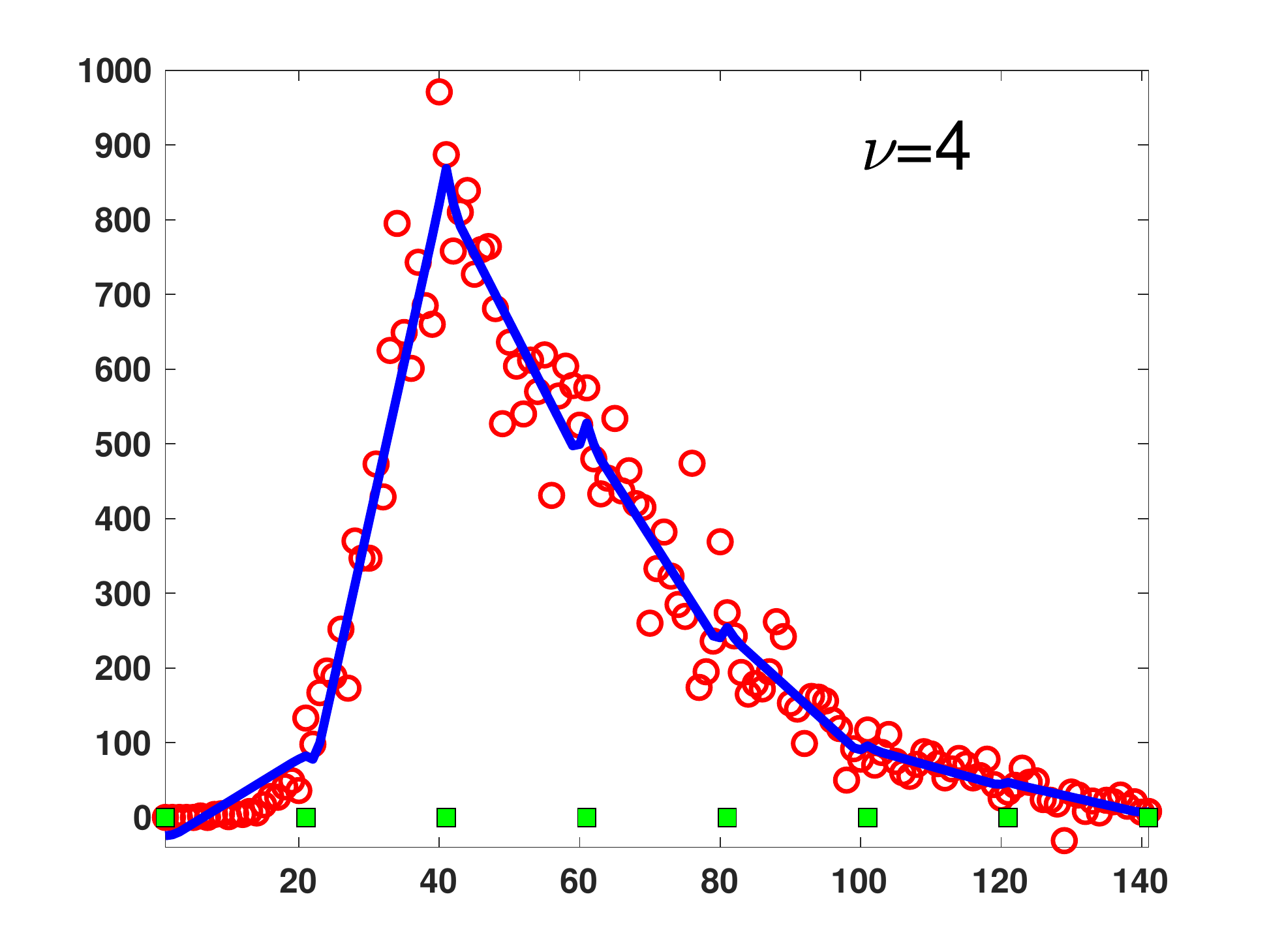}
}
\caption{\footnotesize
	{ Best fit with $8$ bases with different types of basis, $\nu=1,2,3,4$. The circles represent the analyzed data and the squares show the positions of the bases. }
}
\label{Fig_fit}
\end{figure}
}

\section{Final discussion}\label{concluSect}

\begin{figure}[!h]
		\centering
		\includegraphics[width=10cm]{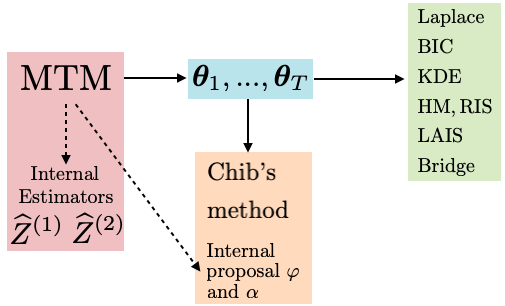}
		
		\caption{{ The application of the MTM algorithm as MCMC provides the generated samples $\{\x_1,...\x_T\}$ and also two possible estimators of $Z$. The generated samples  can employed in other schemes including RIS, LAIS and Bridge sampling. Moreover, considering the proposal and the acceptance function $\alpha$ of the MTM, the Chib's method can be also applied. Indeed, the MTM yields a reversible chain (i.e., fulfills the balance condition). }}
		\label{Fig_MTMlabel}
	\end{figure}

In this work, we have provided an exhaustive review of the techniques for marginal likelihood computation with the purpose of model selection and hypothesis testing. Methods for approximating ratios of normalizing constants have been also described. { The relationships among all of them have been widely described in the text, for instance in Sections \ref{ChibBridgeSect} and \ref{SectpathbridgeSS},  by means of several summary tables (see, as examples, Tables \ref{TablaUmbrella},  \ref{TableBridgeUmbrella}, and  \ref{OnedimRepresentationZ})  and Figures  from \ref{fig1TEO} to \ref{FigTotalRes}.}
 { The careful choice of the prior and} the careful use of the improper priors in the Bayesian setting have been discussed. {A brief description of alternative model selection strategies based on the posterior predictive approach, has been also provided.}
\newline
Most of the presented computational techniques are based on the importance sampling (IS) approach, but also require the use of MCMC algorithms. 
Table \ref{Final_Table} summarizes some methods for estimating $Z$, which involve the generation of the posterior $\post(\x|\y)$ (without using other tempered versions).  This table is devoted to the interested readers which desire to obtain samples $\{\x_n\}_{n=1}^N$ by an MCMC method with invariant pdf $\post(\x|\y)$ (without either any tempering or sequence of densities) and, at the same time, also desire to approximate $Z$. 
Clearly, this table provides only a subset of all the possible techniques. They can be considered the simplest schemes, in the sense that they do not use any tempering strategy or sequence of densities. We also recall that AIC and DIC are commonly used for model comparison, although they do not directly target the actual marginal likelihood. {Table \ref{TablaTemperingMethods} enumerates all the schemes that require the sampling and evaluation of tempered posteriors. For LAIS, the use of tempered posteriors is not strictly required. In PS, one could select a path that does not involve tempered posteriors.   The schemes which provides unbiased estimators of $Z$ or $\log Z$ are given in Table \ref{TablaUnbiasedZ}.  }
\newline
{
We also provide some final advice for practical use. First of all, if informative priors are not available, a very careful choice of the priors must be considered, as remarked in Sections  \ref{SuperSuperIMPSect} and \ref{AppBayeFactImproperPrior}, or alternatively a predictive posterior approach should be applied (see Section \ref{SuperSect_PredPost}).  From a computational point of view, our suggestions are listed below:
\begin{itemize}
\item The use of Naive Monte Carlo (NMC) should be always considered, at least as a first attempt. Moreover, the HM estimator is surely the worst estimator of $Z$, but it could be applied for obtaining an upper bound for $Z$, although it can be very imprecise/loose.
\item The application of an MTM method is a good choice within the MCMC schemes. In fact, as shown in Figure \ref{Fig_MTMlabel}, it also provides two estimators of $Z$, as well as a set of samples. These samples can be employed in other estimators, including Chib, RIS and LAIS, for instance. 
\item 
Regarding the more general task of estimating ratio of constants,  In 
\cite{chen1997monte}, the authors show that (given two unnormalized pdfs) the optimal umbrella estimator provides the best performance theoretically in estimating the ratio of their normalizing constants.
However, the optimal umbrella sampling estimator is difficult and costly to implement (due to the fact sampling from the optimal umbrella proposal is not straightforward), so its best performance may not be achieved in practice.

\item  The Chib's method is  a good alternative, that provide very good performance  as we can observe in Section \ref{Example Diccicio} and also in \cite{marin2009importance,friel2012estimating}. Moreover, the Chib's method is also related to bridge sampling as discussed in Section  \ref{ChibBridgeSect}. However, since it requires internal information regarding the MCMC employed (proposal, acceptance function etc.), it cannot be considered for a possible post-processing scheme after obtaining a Markov chain from a black-box MCMC algorithm. This could be easily done with the HM estimator or LAIS, for instance.
\item 
LAIS can be considered a scheme in between the NMC and HM. NMC draw samples from the prior, which makes it rather inefficient in some setting. The HM estimator uses posterior samples but it is very unstable. 
LAIS uses the posterior samples to build a suitable normalized proposal, so it benefits from localizing samples in regions of high posterior probability (like the HM), while preserving the properties of standard IS (like the Naive MC). In this sense, bridge sampling, the SS method, path sampling, and the rest of techniques based on tempered posteriors, are also schemes  in between the NMC and HM.

\item The methods based on tempered posteriors provide very good performance but the choice of the temperature parameters $\beta_k$ is important. In our opinion, among SS, PS, PP, An-IS,  and SMC,  the more robust to the choice of the $\beta_k$'s is the SS method (that is, perhaps, also the simplest one). Moreover, The SS method does not require the use of several tempered posteriors, unlike PS and PP. The LAIS technique can also be  employed in the upper layer. Since the samples in the upper layer are only used as means of other proposal pdfs and, in the lower layer, the true posterior $\post(\x|\y)$ is always evaluated,  LAIS is also quite robust to the choice of $\beta_k$. More comparisons among SS, An-IS,  and SMC are required, since  these methods are also very related as depicted in Figures \ref{fig_stepping}, \ref{FigAnIS}, \ref{FigSMC} and  \ref{FigTotalRes}.

\item The nested sampling technique has gained attention and is largely applied in the literature. The derivation is complex and several approximations are considered, as discussed in Section \ref{NestedApprox}. The sampling from the truncated priors is the key point and it is not straightforward \cite{chopin2010properties}. In this sense, its success in the literature is surprising. However, the nested sampling includes an implicit optimization of the likelihood. We believe that is an important feature, since the knowledge of high probabilities of the likelihood is a crucial point also to the rest of computational schemes. 
\end{itemize}
}

\begin{table}[!h]
	\caption{Schemes for estimating $Z$, involving  MCMC samples from $\post(\x|\y)$.}
	\label{Final_Table}
	\begin{center}
		\begin{tabular}{|c|c|c|c|}  
			 \hline 
			 \multirow{2}{*}{{\small {\bf Method}}}  &  \multirow{2}{*}{{\small {\bf Section}}}  &  {\small {\bf Need of}} &  \multirow{2}{*}{{\small {\bf Comments}}} \\
	          & 	& {\small {\bf drawing additional samples}}   &  \\
			\hline
			 \multicolumn{4}{c}{\small Below: methods for {\bf post-processing} after generating $N$ MCMC samples from $\post(\x|\y)$.}  \\
			\hline
		   Laplace & \ref{LaplaceSect} & -----  & use MCMC for estimating $\widehat{\x}_{\text{MAP}}$ \\
		   BIC & \ref{BICSect} & -----  &  use MCMC for estimating $\widehat{\x}_{\text{MLE}}$  \\
		   KDE & \ref{KDESect} & -----  &  use MCMC for generating samples\\
		   \hline
 \multirow{2}{*}{Bridge}		 & \multirow{2}{*}{\ref{PUENTE_sect}}   & \multirow{2}{*}{\checkmark} & additional samples are required;   \\
		  & & & see Eq. \eqref{BridgeSamplingIdentityForZ_esy} \\ 
		  \hline
		   RIS  &  \ref{ImportanceSamplingApproaches} & -----  &  the HM estimator is a special case \\
		   	   MTM & \ref{MTMsect} &  -----  & provides  two  estimators of $Z$\\ 
		   LAIS & \ref{LAISsect}   & \checkmark &    with $P(\x|\y)$ in the upper-layer\\
		 \hline
		  \multicolumn{4}{c}{\small Below: methods that require internal information of the MCMC scheme.}  \\
		 \hline
		  \multirow{2}{*}{Chib's method}   & \multirow{2}{*}{\ref{ChibEstSect}}   & \multirow{2}{*}{\checkmark} &      additional samples are required   \\
		  & & & if the proposal is not independent \\
		     MTM & \ref{MTMsect} &  -----  & provides  two  estimators of $Z$\\ 
		 \hline
		  \multicolumn{4}{c}{\small Below: for model selection but do not approximate the marginal likelihood}  \\
		 \hline
		  AIC & \ref{BICSect} & -----  &  use MCMC for estimating $\widehat{\x}_{\text{MLE}}$  \\
                DIC & \ref{BICSect} & -----  &  use MCMC  for estimating $c_p$ and $\bar{\x}$   \\
		 \hline
		 
		\end{tabular}
	\end{center}
\end{table}


\begin{table}[!h]
\caption{Methods using tempered posteriors.}\label{TablaTemperingMethods}
\vspace{-0.3cm}
{
\begin{center}
\footnotesize
	\begin{tabular}{|c||c|c|}
	\hline
		{\bf Method} & {\bf  Section} & {\bf  Use of tempering strictly required}\\
	\hline
	\hline
	IS-P & \ref{TemperedLuca}  & without tempering, it is  HM \\	
	Stepping Stones (SS) & \ref{SSsect}  & \checkmark   \\	
	Path Sampling (PS) &  \ref{PSsect}   & other paths (without tempering) can be used \\	
	Method of Power Posteriors (PP) &  \ref{SectionPP}  & \checkmark  \\
	Annealed Importance Sampling (An-IS) &  \ref{An-ISSect}    &\checkmark  \\
	Sequential Monte Carlo (SMC) &  \ref{SMCsect}    &\checkmark \\
	Layered Adaptive Importance Sampling (LAIS) &  \ref{LAISsect}   & ---  \\ 
         \hline
	\end{tabular}
\end{center}
}
\end{table}

\begin{table}[!h]
\caption{Methods providing unbiased estimators of $Z$ or $\log Z$.} \label{TablaUnbiasedZ}
\vspace{-0.3cm}
{
\footnotesize
\begin{center}
	\begin{tabular}{|c||c|}
	\hline
		{\bf  Method} &  {\bf  Section} \\
	\hline
	 \multicolumn{2}{l}{\small  Unbiased estimators of $Z$:}  \\
	\hline
	IS vers-1 &  \ref{ImportanceSamplingApproaches}  \\	
	Stepping Stones (SS) & \ref{SSsect} \\	
	Annealed Importance Sampling (An-IS) &  \ref{An-ISSect}  \\
	Sequential Monte Carlo (SMC) &  \ref{SMCsect}   \\
	Layered Adaptive Importance Sampling (LAIS) &  \ref{LAISsect}     \\ 
         \hline
	 \multicolumn{2}{l}{\small  Unbiased estimators of $\log Z$:}  \\
	\hline
	 Path Sampling (PS) &  \ref{PSsect}   \\ 
	\hline
	\end{tabular}
\end{center}
}
\end{table}

\bibliographystyle{IEEEtranN}
\bibliography{bibliografia}

\newpage

\begin{appendices}
	
	\section{Table of other reviews}	
	The related literature is rather vast. In this section, we provide a brief summary that intends to be illustrative rather than exhaustive, by means of Table \ref{SurveysTable}.  The most relevant  (in our opinion) and related surveys  are compared according to the topics, material and schemes described in lthe work. The proportion of covering and overlapping with this work is roughly classified as  ``partial'' $\Diamond$, ``complete'' $\surd$, ``remarkable''  or ``more exhaustive''  work with $\bigstar$. From Table \ref{SurveysTable},  we can also notice the completeness of this work. We take into account  also the completeness and the depth of details provided in the different derivations.  The Christian Robert's blog deserves a special mention (\url{https://xianblog.wordpress.com}),
	since Professor C. Robert has devoted several entries of his blog with very interesting comments regarding the marginal likelihood estimation and related topics. 

\begin{landscape}
	\begin{table}[!h]
		{	
		\caption{\footnotesize Covering of the considered topics of other surveys or works ($\Diamond$: partial, $\surd$: complete, $\bigstar$: remarkable or more exhaustive). We take into account  also the completeness and the depth of details provided in the different derivations. To be more precise, in the case of Section 4.1, we have also considered the subsections.}\label{SurveysTable}
		\vspace{-0.2cm}
		\begin{center}
			\begin{tabular}{|c||c||c|c|c||c|c|c|c|c|c||c|c|c||c|c|} 
				\hline
				\multirow{2}{*}{Surveys}   & \multirow{2}{*}{{\footnotesize Families 1--2}}   &  \multicolumn{3}{c||}{IS}&  \multicolumn{6}{c||}{Advanced schemes} &   \multicolumn{3}{c||}{{\footnotesize Vertical likelihood}} & \multirow{2}{*}{{\footnotesize Improper}}  \\
				\cline{3-14}
				&  &  {\footnotesize 1 prop.} & {\footnotesize 2 prop.} &{\footnotesize  Multiple}&  \multicolumn{4}{c|}{{\footnotesize MCMC within IS}}  & {\footnotesize MTM}  & {\footnotesize AIS} &  5.1  & 5.2  & 5.3 &  \\
				\hline
				\hline
				 {\footnotesize Gelfand and Dey (1994)\cite{gelfand1994bayesian}}	&  \multirow{2}{*}{\checkmark}   &  \multirow{2}{*}{$\Diamond$ }  &  & &    &   &  &  &   &  & & &   &  \\
				 {\footnotesize Kass and Raftery (1995)\cite{kass1995bayes}}	&     &     &  & &    &   &  &  &   &  & & &   &  \\
				{\footnotesize Raftery (1995)\cite[Ch. 10]{gilks1995markov}} 	& $\Diamond$   &  \checkmark   & $\Diamond$ & &    &   &  &  &   &  & & &   &  \\
				{\footnotesize Meng and Wong (1996)\cite{meng1996simulating} }	&  &  $\Diamond$  & $\bigstar$ & $\Diamond$ &    &   &  &  &   &  & & &   &  \\
				{\footnotesize DiCiccio et al (1997)\cite{diciccio1997computing}}	& $\bigstar$   &  \checkmark   & \checkmark & &    &   &  &  &   &  & & &   &  \\
				{\footnotesize Chen and Shao (1997)\cite{chen1997monte}}
				& \multirow{2}{*}{$\Diamond$}   & \multirow{2}{*}{$\Diamond$}    & \multirow{2}{*}{\checkmark} & \multirow{2}{*}{\checkmark} &    &   &  &  &   &  & & &   &  \\
				{\footnotesize Chen et al (2012)\cite[Ch. 5]{chen2012monte}}
				&   &    &  &  &    &   &  &  &   &  & & &   &  \\
				{\footnotesize Gelman and Meng (1997)\cite{gelman1998simulating}} &  &  \checkmark  & $\bigstar$ & $\bigstar$ &    &   &  &  &   &  & & &   &  \\
				{\footnotesize Bos (2002)\cite{bos2002comparison}} 	& \checkmark  &   \checkmark  &  & &    &   &  &  &   &  & & &   &  \\
				{\footnotesize Vyshemirsky and Girolami (2007)\cite{vyshemirsky2007bayesian}}	&   & $\Diamond$   &  & $\Diamond$  &    &  $\Diamond$   &  &  &  &  & & &   &  \\
				{\footnotesize Marin and Robert (2009)\cite{marin2009importance}} 	&  $\Diamond$   &  \checkmark  & $\Diamond$ &   &  \mbox{  }\mbox{  }   &   &  &  &   &  & & &   &  \\
				{\footnotesize Robert and Wraith (2009)\cite{robert2009computational}}	&  & \checkmark & \checkmark &   &    &   &  &  &   &  & & \checkmark  &   &  \\
				{\footnotesize Friel and Wyse (2012)\cite{friel2012estimating}}  & $\Diamond$ & $\Diamond$  &  & $\Diamond$ &   & \checkmark & &  &   &  &  & \checkmark  &  & \\
				{\footnotesize Ardia et al (2012)\cite{ardia2012comparative}} 	& $\Diamond$   & $\Diamond$   &  \checkmark &  &  & &  &   &   &  &  &   &  & \\
				{\footnotesize Polson and Scott (2014)\cite{polson2014vertical}}  &   &  $\Diamond$   &  &  $\Diamond$ &    &   &   &  &   &   & \checkmark & $\bigstar$ &  $\bigstar$ & \\ 	
				{\footnotesize Schöniger et al (2014)\cite{schoniger2014model}}  & \checkmark & \checkmark  &  & &    &   &   &  &   &   &  & $\Diamond$ &  & \\ 	
				{\footnotesize Knuth et al (2015)\cite{knuth2015bayesian}} &  $\Diamond$  & $\Diamond$    &  & \checkmark  &   & \checkmark  & &  &   &  &  & \checkmark   &  & \\
				{\footnotesize Liu et al (2016)\cite{liu2016evaluating}} 	& $\Diamond$  & $\Diamond$  &  & \checkmark  &  & &  &  &   &  &  &  $\bigstar$ &  & \\
				{\footnotesize Zhao and Severini (2017)\cite{zhao2017integrated}} 	& $\bigstar$ & \checkmark  &\checkmark  &  &  & &  &  &   &  &  &   &  & \\
				{\footnotesize Martino (2018)\cite{MARTINO_REV_MTM}} &    &     &  & &    &  &   & $\Diamond$  &  $\bigstar$ &  & &   &  & \\ 
				{\footnotesize Bugallo et al (2017)\cite{AIS_SIG_PRO}} &    &     \multirow{2}{*}{$\Diamond$}  &  & &    &   &     &   &   &  \multirow{2}{*}{$\bigstar$} &  &   &  & \\
				{\footnotesize Bugallo et al (2015)\cite{Bugallo15}} &    &   &  & &    &   &     &   &   &  &  &   &  & \\  	
				{\footnotesize Oaks et al (2019)\cite{r2019marginal}}  & & $\Diamond$   &  & \checkmark &   & $\Diamond$  & $\Diamond$  &   &   &  &  & $\Diamond$  &  & \\			
				{\footnotesize O'Hagan (1995)\cite{o1995fractional}}&    &       &  & &    &   &     &   &   &   &  &   &  & \multirow{2}{*}{$\bigstar$} \\
				{\footnotesize Berger and Pericchi (1996)\cite{berger1996intrinsic}}&    &       &  & &    &   &     &   &   &   &  &   &  &  \\  			
				\hline
			\end{tabular}
		\end{center}
	}
	\end{table}
	
\end{landscape}


\end{appendices}

\end{document}